\documentclass[aps,prb,twocolumn,showpacs,eqsecnum]{revtex4}
\usepackage{graphicx}
\usepackage{dcolumn}
\usepackage{bm}
\usepackage[normalem]{ulem}
\usepackage{makeidx}
\usepackage{amsmath}
\usepackage{amsfonts}
\usepackage{amssymb}
\usepackage{color}
\usepackage{colordvi}
\usepackage{dcolumn}

\newcommand{\G}{\Gamma}

\newcommand{\B}{\mathrm{B}}
\newcommand{\D}{\Delta}
\newcommand{\sgn}{\mathrm{sgn}}

\newcommand{\ee}{\mathrm{e}}
\newcommand{\ii}{\mathrm{i}}
\newcommand{\dt}{\mathrm{d}}
\newcommand{\del}{\partial}

\newcommand{\ve}{\varepsilon}

\newcommand{\vp}{\varphi}

\newcommand{\thetax}{{\theta^*}}

\newcommand{\om}{\omega}
\newcommand{\sig}{\sigma}

\newcommand{\komma}{\;\textnormal{,}}
\newcommand{\punkt}{\;\textnormal{.}}
\newcommand{\inte}{\mathrm{int}}
\newcommand{\cS}{\mathcal{S}}
\newcommand{\cA}{\mathcal{A}}

\newcommand{\cW}{\mathcal{W}}
\newcommand{\cD}{\mathcal{D}}

\newcommand{\brperp}{{\mathbf{r}_\perp}}

\newcommand{\bqperp}{{\mathbf{q}_\perp}}
\newcommand{\bp}{\mathbf{p}}
\newcommand{\bn}{\mathbf{n}}
\newcommand{\bnt}{{\tilde{\mathbf{n}}}}
\newcommand{\bnb}{{\bar{\mathbf{n}}}}

\newcommand{\bq}{\mathbf{q}}
\newcommand{\bqbpar}{{\bar{q}_\parallel}}
\newcommand{\bqbperp}{{\bar{\mathbf{q}}_\perp}}
\newcommand{\bk}{\mathbf{k}}

\newcommand{\br}{\mathbf{r}}

\newcommand{\ut}{{\tilde{u}}}

\newcommand{\romI}{\mathrm{I}}
\newcommand{\romII}{\mathrm{II}}

\begin{document}

\title{Low energy excitations and singular contributions\\ in the thermodynamics
of clean Fermi liquids}
\author{Hendrik Meier$^{1}$, Catherine P\'{e}pin$^{2,3}$, and Konstantin B.
Efetov$^{1,3}$}
\affiliation{$^1$ Institut f\"ur Theoretische Physik III, Ruhr-Universit\"at Bochum,
44780 Bochum, Germany \\
$^{2}$ IPhT, CEA-Saclay, L'Orme des Merisiers, 91191 Gif-sur-Yvette, France
\\
$^{3}$ International Institute of Physics, Universidade Federal do Rio
Grande do Norte, 59078-400 Natal-RN, Brazil}
\date{\today }

\begin{abstract}
Using a recently suggested method of bosonization in an arbitrary dimension,
we study the anomalous contribution of the low energy spin and charge
excitations to thermodynamic quantities of a two-dimensional (2D) Fermi
liquid. The method is slightly modified for the present purpose such that
the effective supersymmetric action no longer contains the high energy degrees of freedom
but still accounts for effects of the finite curvature of the Fermi surface.
Calculating the anomalous contribution $\delta c(T)$ to the specific heat,
we show that the leading logarithmic in temperature corrections to $\delta
c(T)/T^{2}$ can be obtained in a scheme combining a summation of ladder diagrams
and renormalization group equations.
The final result is represented as the sum of two
separate terms that can be interpreted as coming from singlet and
triplet superconducting excitations. The latter may diverge in certain regions of the
coupling constants, which should correspond to the formation of
triplet Cooper pairs.
\end{abstract}

\pacs{71.10.Ca, 71.10.Ay, 71.10.Pm}
\maketitle





\section{Introduction}

\label{sec:intro}

At low temperatures, thermodynamic properties of fermions with a repulsive interaction
bear strong resemblance to those of an ideal Fermi gas. This is the quintessence
of the Landau theory of the Fermi liquid\cite{landau}. It is assumed in this theory that
interaction effects merely renormalize quantities such as the fermion mass or the
density of states. In fact, this renormalization can be large for a strong
interaction, thus making perturbative methods inapplicable. However, such obstacles are
always overcome once the renormalized quantities are replaced by phenomenological effective parameters.

Following the similarity to the ideal Fermi gas, one could expect that
quantities like
\begin{align}
g\left( T\right) = \frac{c(T)}{T}\komma  \label{e1}
\end{align}
where $c(T)$ is the Fermi liquid's specific heat, or its spin susceptibility~$\chi (T)$ had to be
analytic functions of $T^{2}/\varepsilon _{F}^{2}$ with $\varepsilon _{F}$
being the Fermi energy and~$T$ the temperature. [In the leading order in~$T$, one should have a finite
$g\left( 0\right) $ considered as a phenomenological Fermi liquid
parameter.] However, several studies revealed for
three-dimensional Fermi liquids the existence of corrections to the
specific heat $c\left( T\right) $ of the order $T^{3}\ln T$, which is incompatible with
ideal Fermi gases.\cite{eliashberg,doniach,brinkman,pethick} In three
dimensions, there are also logarithmic contributions $|\bq|^{2}\ln |\bq|$ to the
non-homogeneous spin susceptibility $\chi (\bq)$, where $\bq$ is the wave vector.
\cite{belitz}

In two dimensions (2D), non-analytic corrections to the quantities $g(T)$, Eq.~(\ref{e1}), and
$\chi (T)$ are stronger and, in the lowest order in interaction, have been
found to be proportional to $T$.\cite{coffey,baranov,chitov,chubukov,betouras}
These anomalous contributions were attributed\cite
{chubukov1,chubukov2} to one-dimensional backscattering processes
imbedded in the two-dimensional momentum space. The linear in~$T$ correction to $g(T)$
has been verified experimentally in a $^3\mathrm{He}$ fluid monolayer.\cite{casey}

The problem of evaluating the anomalous contributions was reconsidered
in Ref.~\onlinecite{aleiner} with the help of a supersymmetric field theory especially
designed to describe those low energy bosonic excitations which are responsible for
the anomalous contributions to the thermodynamics. It was found that earlier
calculations\cite{coffey,baranov,chitov,chubukov,betouras} had not been complete and so far unforeseen
logarithmic contributions to $\delta c(T) /T^{d}$  were discovered for dimensions~$d=2,3$. Similar
anomalous contributions have been found for the spin susceptibility~$\chi(T)$ using either the supersymmetric method mentioned above\cite{schwiete}
or the conventional diagrammatic technique\cite{shekhter}. Both the methods led to identical results for the spin susceptibility in 2D models.

In dimension~$d=1$, the supersymmetric approach to find the spin susceptibility\cite{schwiete}
reproduced the results of earlier theoretical works\cite{dl,lukyanov}.
As to the specific heat of a one-dimensional Fermi gas,
the result of the supersymmetric field theory of Ref.~\onlinecite{aleiner} can be mapped
on known results for the Kondo model\cite{tsvelik} or for the $XXZ$ spin-$\frac{1}{2}$ chain\cite{lukyanov},
showing agreement. A more recent study\cite{cms} confirmed the supersymmetry approach
using the conventional diagrammatic technique.

In spite of the agreement between the results obtained by these different
methods in one dimension, a direct diagrammatic computation of the anomalous specific heat carried out
up to the third order in the fermion-fermion interaction by Chubukov and
Maslov (CM) for the 2D Fermi liquid\cite{chubukov3} led to a result that did not coincide with
the one obtained by the supersymmetric approach in Ref.~\onlinecite{aleiner}. Both of them contained logarithmic
corrections to the anomalous contributions. However, while in the framework of the supersymmetry method of Ref.~\onlinecite{aleiner},
the non-trivial
contributions originated purely from spin excitations, CM obtained
contributions from both spin and charge excitations. They attributed the
difference between the results to the fact that not all effects of the
finite curvature of the Fermi surface had been properly taken into account in
the approach of Ref.~\onlinecite{aleiner}.

Of course, curvature effects are absent in one dimension and charge excitations do
not influence the spin susceptibility~$\chi(T)$ in any dimension in the non-logarithmic lowest order
in the interaction. As a result, no discrepancy could be seen in these cases.
Nevertheless, since evidently certain effects of the finite curvature of the
Fermi surface were neglected in the supersymmetric method of Ref.~\onlinecite{aleiner},
it is important to find the correct way of calculations. At the
same time, approaches based of the conventional diagrammatic expansions
for fermions become inapplicable beyond some low orders of perturbation theory.
Indeed, CM performed a standard perturbation theory to third order and treated
higher orders by plausibility arguments.\cite{chubukov3}

In this paper, we revise the approach of Ref.~\onlinecite{aleiner} for the 2D Fermi liquid taking into
account all the  necessary effects of the curvature of the Fermi surface. As a
result, we are able to sum up all leading logarithms, thus correcting the previous result for the function $g(T)$, Eq.~(\ref{e1}).
To third order in the interaction potential,
our result agrees with the conventional perturbative calculation\cite{chubukov3}.
Moreover, our result in all orders in the large logarithm~$\ln(\ve_F/T)$
shares the same asymptotic behavior as the conjecture suggested by CM.
In principle, our method and results are applicable for both repulsion and attraction
unless one reaches a singularity in the final formulas. We argue that the singularities, if existing, correspond
to the singlet or triplet Cooper superconducting pairing. Remarkably, the
final formula for the function $g\left( T\right) $ contains a sum of separated spin
singlet and spin triplet excitations. It is important to emphasize that
the modification concerns the dimensions $d>1$ only, whereas the method leads for $d=1$
to the same results as those obtained in Ref.~\onlinecite{aleiner}.

The calculations are performed using a modification of a recently suggested bosonization scheme
of Refs.~\onlinecite{epm,epm2}. In contrast to these previous works, we derive an
effective supersymmetric action describing only low-lying modes. This is
achieved by singling out the slowly varying pairs of the fermionic field in the
interaction term. Subsequently, we decouple this interaction by means of Hubbard-Stratonovich auxiliary fields
slowly varying in space and imaginary time --- similarly to what
was done in Ref.~\onlinecite{aleiner}. Here, however, this decoupling is followed by
the derivation of equations of motion using the method of Refs.~\onlinecite{epm,epm2}. In
contrast to the equations of Ref.~\onlinecite{aleiner}, the present equations
preserve all necessary effects of the curvature of the Fermi surface.

The solution of the equations of motion is represented in a form of an integral over
superfields~$\Psi $, which do not only depend on conventional coordinates $\mathbf{r}$ and imaginary time $\tau $
but also on anticommuting variables $\theta,\thetax$. This integral representation allows to average over the auxiliary fields
before we obtain the final effective field theory for the low energy bosonic charge
and spin excitations. Such a representation, suggested in Refs.~\onlinecite{epm,epm2},
differs from the supervector representation used in Ref.~\onlinecite{aleiner} and is
considerably more convenient for explicit calculations.

Although the general calculational scheme based on this superfield
action shares certain similarities with that of Ref.~\onlinecite{aleiner}, the finite curvature
of the Fermi surface suppresses several otherwise logarithmic contributions. Consequently,
different final results are obtained as a result of a different calculational procedure. For instance,
the quartic part of the action can be renormalized by summing ladder diagram series instead of
solving renormalization group equations.

The calculations performed here can be important not only from the point of
view of finding the complete picture about the anomalous contributions to
the thermodynamics of the 2D Fermi liquid but also as a demonstration of how the higher-dimensional
bosonization scheme suggested in Refs.~\onlinecite{epm,epm2}
can be used as a method in analytical studies. The experience gained on this comparably simple example may become
important for attacking more difficult and more interesting problems of strongly correlated
systems.

The paper is organized as follows: In Sec.~\ref{sec:model}, we derive the
effective low energy field theory for the anomalous thermodynamic contributions.
Starting from a general model of repulsive interaction, we discuss and single out
the relevant soft modes and bosonize the microscopic fermion model in the low energy limit.

Section~\ref{sec:renormalization} discusses the leading perturbative corrections
to both the thermodynamic potential and the vertices of the low energy field theory on one-loop level.
We identify the logarithmically divergent one-loop diagrams that are important for the subsequent renormalization group analysis. This analysis is presented in Sec.~\ref{sec:rg}, in which we derive and solve the flow equations for the coupling constants of the low energy field theory.

In Sec.~\ref{sec:c}, we apply the bosonic technique and the results of the renormalization group analysis to evaluate the anomalous contribution
to the specific heat beyond the $T^2$-term obtained from second order perturbation theory.
First performing an explicit perturbation expansion to third order in order to check once more our bosonic approach, we eventually include the completely renormalized vertices and find the non-analytic contribution to the specific heat in all orders in~$\ln(\ve_F/T)$.

Concluding remarks are found in Sec.~\ref{sec:concl}.

\section{Low energy field theory}

\label{sec:model}

In this section, we formulate the microscopic model for the interacting
fermions and derive the low energy field theory that catches the non-trivial
physics of the low-lying bosonic excitations. The derivation does not
require to specify the dimension~$d$ of the system and we assume $d$ to be
arbitrary here.

\subsection{Microscopic fermion model}

We consider a gas of spin-$\frac{1}{2}$ fermions described by the Hamiltonian%
\begin{equation}
\hat{H}=\hat{H}_{0}+\hat{H}_{\mathrm{int}}\komma  \label{e2}
\end{equation}
where $\hat{H}_{0}$ is the kinetic energy,
\begin{equation}
\hat{H}_{0}=\sum_{\sigma }\int c_{\sigma }^{\dagger }(\mathbf{r})\left[
\varepsilon \left( -\ii\mathbf{\nabla }_{\mathbf{r}}\right) -\mu \right] c_{%
\sigma }(\mathbf{r})\ \mathrm{d}^d\mathbf{r} \punkt \label{2a01}
\end{equation}
In Eq.~(\ref{2a01}), $\mathbf{r}$ and $\sigma = \pm 1$ denote the
coordinates and spin, respectively, $c_{\sigma }^{\dagger }\left( \mathbf{r}%
\right) $ $[c_{\sigma }\left( \mathbf{r}\right) ]$ are creation
(annihilation) field operators, and $\mu $ is the chemical potential. In the
simplest case, $\varepsilon \left( \mathbf{p}\right) =\mathbf{p}^{2}/2m$ with
$m$ being the fermion mass. In this case, the Fermi surface is a $(d-1)$%
-dimensional sphere. For a more general spectrum $\varepsilon \left( \mathbf{%
p}\right) $, the Fermi surface has a more complex shape but this does not
lead to a qualitatively different physical picture as long as the Fermi surface remains smooth and
there is no nesting.

The second term in Eq.~(\ref{e2}) stands for the fermion-fermion
interaction and takes the standard form:
\begin{equation}
\hat{H}_{\mathrm{int}}=\frac{1}{2}\sum_{\sigma \sigma ^{\prime }}\int c_{\sigma
}^{\dagger }(\mathbf{r})c_{\sigma ^{\prime }}^{\dagger }({\mathbf{r}^{\prime
}})V(\mathbf{r}-{\mathbf{r}^{\prime }})c_{\sigma ^{\prime }}({\mathbf{r}%
^{\prime }})c_{\sigma }(\mathbf{r})\ \mathrm{d}^d\mathbf{r}\mathrm{d}^d{\mathbf{r}%
^{\prime }}\;\mathnormal{.}  \label{2a02}
\end{equation}
At the moment, we do not specify the form of the function $V\left( \mathbf{%
r-r}^{\prime }\right) $ except for its positivity, guaranteeing repulsive interaction.

Equations~(\ref{e2})--(\ref{2a02}) constitute the model in the Hamiltonian form.
It is more convenient for our purposes to use a functional integral
representation, in which the partition function~$\mathcal{Z}$ is written as
\begin{equation}
\mathcal{Z}=\int \exp \left\{ -\mathcal{S}_{0}-\mathcal{S}_{\mathrm{int}%
}\right\} \ \mathcal{D}(\chi ^{\ast },\chi )\punkt  \label{2a03a}
\end{equation}%
Herein, the Euclidean action is given by
\begin{align}
\mathcal{S}_{0}&=\sum_{\sigma}\int_{0}^{\beta }\!\!\int \chi _{\sigma }^{\ast }(\mathbf{r%
},\tau )
\nonumber\\&\quad\times
\left[ \partial _{\tau }+\varepsilon \left( -\ii\mathbf{\nabla }_{%
\mathbf{r}}\right) -\mu \right] \chi _{\sigma }(\mathbf{r},\tau )\ \mathrm{d}^d%
\mathbf{r}\mathrm{d}\tau \mathnormal{,}  \label{2a03b} \\
\mathcal{S}_{\mathrm{int}}& =\frac{1}{2}\sum_{\sigma \sigma ^{\prime
}}\int_{0}^{\beta }\!\!\int \chi _{\sigma }^{\ast }(\mathbf{r},\tau )\chi
_{\sigma ^{\prime }}^{\ast }({\mathbf{r}^{\prime }},\tau )\chi _{\sigma
^{\prime }}({\mathbf{r}^{\prime }},\tau )\chi _{\sigma }(\mathbf{r},\tau )
\notag \\
& \qquad \qquad \times V(\mathbf{r}-{\mathbf{r}^{\prime }})\ \mathrm{d}^d%
\mathbf{r}\mathrm{d}^d{\mathbf{r}^{\prime }}\mathrm{d}\tau \;\mathnormal{.}
\label{2a03c}
\end{align}%
In Eqs.~(\ref{2a03a})--(\ref{2a03c}), $\beta =1/T$ is the inverse
temperature and $\chi $,$\chi ^{\ast }$ are Grassmann fields which are
antiperiodic in imaginary time~$\tau $, $\chi (\tau +\beta )=-\chi (\tau
)$.

Equations~(\ref{2a03a})--(\ref{2a03c}) are the starting point for our
analysis.

\subsection{Low-lying modes}

\begin{figure}[t]
\includegraphics[width = \linewidth]{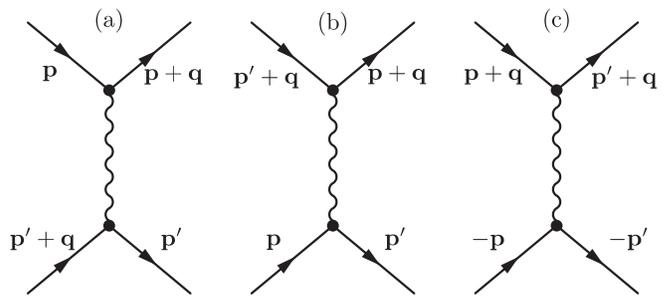}
\caption{Decomposition of the interaction~$\mathcal{S}_\mathrm{int}$, Eq.~(%
\protect\ref{2a03c}), into slow modes, $|\mathbf{q}|\lesssim q_0\ll p_F$.
(a), (b), and (c) are the Hartree, the Fock, and the Cooper vertices,
respectively. In our model, the Cooper vertex should be omitted to avoid double-counting. }
\label{fig: softmodes}
\end{figure}

The spin and charge excitations at low temperatures, which we are interested in,
correspond to the low-lying modes of the microscopic model, Eqs.~(\ref{2a03a})--(\ref%
{2a03c}). Effectively, only those fermions contribute that are energetically
close to the Fermi energy~$\varepsilon _{F}$. From this constraint, we
obtain the three relevant vertices shown in Fig.~\ref{fig: softmodes}
describing scattering processes with momenta $\mathbf{p,p}^{\prime }$ located
near the Fermi surface. We single out these vertices assuming that the momenta $\mathbf{q}$ are small, $|\mathbf{q}|\lesssim
q_{0}$, where $q_{0}$ is a phenomenological momentum cutoff which is much smaller than the Fermi momentum~$p_F$.
Vertices~(a) and (b) describe soft interactions in the particle-hole channel,
whereas (c) is the Cooper vertex. The well-known Hartree-Fock
approximation is obtained using the vertices (a) and (b) and, thus, they can be
referred to as the Hartree and Fock vertex, respectively. The Cooper vertex enters the
ladder diagrams leading to the BCS superconducting instability in case of attractive interaction.

In principle, all three vertices are important when calculating physical
quantities. However, we are here interested in the anomalous contributions to
the thermodynamics, which originate\cite{chubukov1,chubukov2,aleiner} from (quasi-)one-dimensional processes. As we will see later, the main contribution
to the anomalous terms comes from small $\bq$, which is also seen from the
conventional perturbation theory.\cite{chubukov1} Of course, assuming that
all the vertices in Fig.~\ref{fig: softmodes} are different forbids the
regions of the essential momenta attributed to them to overlap.

A quick glance, however, reveals that this is not the case. For example, there is a clear overlap between the
regions of the momenta in the vertices (a) and (b) at small $\left\vert
\mathbf{p}-\mathbf{p}^{\prime }\right\vert \sim \left\vert \mathbf{q}%
\right\vert $. In order to avoid double-counting, one would have to
consider only one of these vertices in this region. In Ref.~\onlinecite{aleiner}, e.g., it was
chosen to remove the region $\left\vert \mathbf{p}-%
\mathbf{p}^{\prime }\right\vert \sim \left\vert \mathbf{q}\right\vert $ from
the Fock vertex. Fortunately, the contribution coming from this region of
the momenta can be neglected at the low temperatures considered here. As a result,
we can consider the vertices Fig.~\ref{fig: softmodes}(a) and~(b) as
practically different vertices for $\left\vert \mathbf{q}\right\vert \lesssim
q_{0}\ll p_F$.

At the same time, the Hartree-Fock (a,b) and the Cooper (c) vertices do
overlap in important momentum regions. As to the role of the Cooper vertex, building ladders out of it,
logarithmic divergencies appear for arbitrary scattering
angles $\widehat{\mathbf{p}{{\mathbf{p}^{\prime }}}}$. The contributions
coming from large angles are not reproduced by using the Fock vertices (b) instead. However, scattering angles essentially different from $0$
or $\pi $ are less important for the anomalous contributions we wish to calculate.
Considering only angles $\widehat{\mathbf{p}{{\mathbf{p}^{\prime }}}}$
close to $0$ or $\pi $ means focusing on almost one-dimensional
scattering and this is where the anomalous contributions emerge.

This region of the momenta attributed to the Cooper vertex Fig.~\ref{fig:
softmodes}(c), however, fully overlaps with that for the Hartree-Fock vertices (a,b). Any diagram containing Cooper loops with small angles $%
\widehat{\mathbf{p}{{\mathbf{p}^{\prime }}}}$ can be represented in an
equivalent way using particle-hole loops built from vertices~(a) and~(b).
Several examples of this equivalence can be found in Ref.~\onlinecite{aleiner}. Therefore, taking into account all the vertices Fig.~\ref{fig:
softmodes}(a,b) and~(c) would imply double-counting.
In order to avoid it, one should choose between either the Hartree-Fock vertices or the
Cooper ones, but not take into account all of them.

In the present study, we choose as in Ref.~\onlinecite{aleiner} the Hartree-Fock route.
This is in contrast to the approach of Ref.~\onlinecite{shekhter} where the Cooper
channel representation was used. Our choice will turn out more convenient for singling
out the anomalous contributions.

As a result, we write the effective interaction describing the low energy
physics as
\begin{align}
\tilde{\mathcal{S}}_{\mathrm{int}}& =\frac{1}{2}\sum_{PP^{\prime }Q,\sigma
\sigma ^{\prime }}\Big\{\chi _{\sigma }^{\ast }(P+Q)\chi _{\sigma }(P)V_{%
\mathbf{q}}\chi _{\sigma ^{\prime }}^{\ast }(P^{\prime })\chi _{\sigma
^{\prime }}(P^{\prime }+Q)  \notag \\
& \qquad -\chi _{\sigma }^{\ast }(P+Q)\chi _{\sigma ^{\prime }}(P)V_{\mathbf{%
p}-{{\mathbf{p}^{\prime }}}}\chi _{\sigma ^{\prime }}^{\ast }(P^{\prime
})\chi _{\sigma }(P^{\prime }+Q)\Big\}  \label{2a03d}
\end{align}%
instead of Eq.~(\ref{2a03c}). In Eq.~(\ref{2a03d}), the fermionic fields are
represented in Fourier space and four-momentum notations
\begin{equation*}
P=(\varepsilon ,\mathbf{p})\mathnormal{,}\quad Q=(\omega ,\mathbf{q})
\end{equation*}%
are used. Herein, $\varepsilon $ and $\omega $ are fermionic and bosonic
Matsubara frequencies, respectively. In the ``fermionic'' summations $\sum_{P^{(\prime
)}}(\ldots)=T\sum_{\varepsilon ^{(\prime) }}\int (\ldots) [\mathrm{d}^d\mathbf{p}^{(\prime )}/(2\pi
)^{d}]$, it is understood that $\mathbf{p}$, ${{\mathbf{p}^{\prime }}}$ are of
order~$p_{F}$. On the contrary in the ``bosonic'' summation~$\sum_Q$,
we introduce a function~$f(\bq)$ which cuts off the momentum~$\bq$ beyond
$q_0\ll p_F$,
\begin{align}
\label{2a03e}
\sum_{Q} (\ldots) =
 T\sum_{\omega }\int
 (\ldots)\ f(\bq)\ \frac{\mathrm{d}^d\mathbf{q}}{(2\pi )^{d}}\punkt
\end{align}
The function~$f(\bq)$ can for instance be modeled as $f(\bq)=\Theta(q_0-|\bq|)$ with
$\Theta$ denoting the Heaviside function. In final formulas, we may choose
a different form which allows to conveniently perform the terminal integrations. For the moment, we
do not specify it more than that it is assumed to fulfill~$f(\bq=0)=1$ and decay fast beyond~$q_0$.

\subsection{Bosonization}

Our choice of the effective interaction, Eq.~(\ref{2a03d}), leads after
bosonization to a field theory representing \emph{particle-hole}-type
bosonic excitations of the Fermi gas. The route to obtain the effective low
energy field theory for these excitations follows the higher-dimensional
bosonization scheme introduced in Refs.~\onlinecite{epm} and \onlinecite{epm2}. It is done in
five steps:
\begin{enumerate}
\item Decouple the effective interaction employing a Hubbard-Stratonovich
transformation.

\item Integrate out the fermionic degrees of freedom.

\item Derive the effective equation of motion for the bosonic field subject
to the random Hubbard-Stratonovich auxiliary field.

\item Write the solution of this equation in form of a functional integral
over superfields thus obtaining a closed supersymmetric field theory.\cite%
{faddeev,brst,parisi,justin}

\item Average over the auxiliary field.
\end{enumerate}
In what follows, each of the steps is presented in a separate section.

\subsubsection{Hubbard-Stratonovich transformation}

In order to decouple the effective interaction $\tilde{\mathcal{S}}_{\mathrm{%
int}}$ by an integration over the auxiliary Hubbard-Stratonovich field, we recast the spin
structure in the Fock channel by means of the relation
\begin{equation}
2\delta _{\sigma _{1}\sigma _{4}}\delta _{\sigma _{2}\sigma _{3}}=\sum_{\mu
=0}^{3}\sigma _{\sigma _{1}\sigma _{2}}^{\mu }\sigma _{\sigma _{3}\sigma
_{4}}^{\mu }\mathnormal{,}  \label{2a04}
\end{equation}%
where $(\sigma ^{\mu })=(\sigma ^{0}, \mbox{\boldmath$\sigma$})$ with $\sigma ^{0}=%
\mathbf{1}$ and $\mbox{\boldmath$\sigma$}=(\sigma ^{1},\sigma ^{2},\sigma ^{3})$ is
the vector of Pauli matrices. In what follows, we use the convention that
Greek upper indices appearing twice imply the summation from $0$ to $3$.

We thus obtain
\begin{align}
\tilde{\mathcal{S}}_{\mathrm{int}}& =\frac{1}{2}\sum_{Q}\Big\{V_{\mathbf{q}%
}n(Q)n(-Q)  \notag \\
& \qquad -\frac{1}{2}\sum_{\mathbf{p}{{\mathbf{p}^{\prime }}}}V_{\mathbf{p}-{%
{\mathbf{p}^{\prime }}}}S_{\mathbf{p}}^{\mu }(Q)S_{\mathbf{p}^{\prime
}}^{\mu }(-Q)\Big\}  \label{2a05}
\end{align}%
where $S_{\mathbf{p}}^{\mu }$ are components of the four-component vector $%
( S_{\mathbf{p}}^{0},S_{\mathbf{p}}^{1},S_{\mathbf{p}}^{2},S_{\mathbf{p}%
}^{3}) $ and
\begin{align}
n(Q)& =\sum_{P,\sigma }\chi _{\sigma }^{\ast }\big(\varepsilon ,\mathbf{p}+%
\frac{\mathbf{q}}{2}\big)\chi _{\sigma }\big(\varepsilon -\omega ,\mathbf{p}-%
\frac{\mathbf{q}}{2}\big)\mathnormal{,}  \label{2a06b} \\
S_{\mathbf{p}}^{\mu }(Q)& =T\sum_{\varepsilon ,\sigma \sigma ^{\prime
}}\chi _{\sigma }^{\ast }\big(\varepsilon ,\mathbf{p}+\frac{\mathbf{q}}{2}%
\big)\sigma _{\sigma {\sigma ^{\prime }}}^{\mu }\chi _{\sigma ^{\prime }}%
\big(\varepsilon -\omega ,\mathbf{p}-\frac{\mathbf{q}}{2}\big).  \notag
\end{align}%
The zero component $S_{\mathbf{p}}^{0}$ is related to the particle density $n(Q)$ as%
\begin{equation}
\sum_{\mathbf{p}}S_{\mathbf{p}}^{0}\left( Q\right) =n\left( Q\right)
\label{e4}
\end{equation}%
whereas the other components form the spin density vector $\mathbf{S}\left( Q\right)$,
\begin{equation}
\sum_{\mathbf{p}}\mathbf{S}_{\mathbf{p}}\left( Q\right) =\mathbf{S}\left(
Q\right)\quad \label{e4a}
\end{equation}
with $\mathbf{S}_{\mathbf{p}}=(S^{1}_{\mathbf{p}},S^{2}_{\mathbf{p}},S^{3}_{\mathbf{p}})$.

Assuming that the interaction $V\left( \mathbf{r}\right) $ decays
sufficiently fast, the interaction amplitudes $V$ entering Eq.~(\ref{2a05})
can be written as
\begin{equation}
V_{\mathbf{q}}\simeq V_{0}\;\mathnormal{,}\quad V_{\mathbf{p}-{{\mathbf{p}%
^{\prime }}}}\simeq \tilde{V}\Big(2p_{F}\sin \big(\frac{\widehat{\mathbf{p}%
\mathbf{p}^{\prime }}}{2}\big)\Big)  \label{2a07}
\end{equation}%
with $\widehat{\mathbf{p}\mathbf{p}^{\prime }}$ denoting the angle between~$%
\mathbf{p}$ and~$\mathbf{p}^{\prime }$. In Eq.~(\ref{2a07}), $V_{0}$ is a
positive constant and $\tilde{V}$ is a function of the angle between the
vectors $\mathbf{p}$ and $\mathbf{p}^{\prime }$ located at the Fermi
surface. As the main contribution will come from momenta $\mathbf{p}$ near
the Fermi surface and small $\bq$, Eq.~(\ref{2a07}) is a good approximation.
For contact interaction, $\tilde{V}\equiv V_{0}$.

In principle, we are now ready to decouple the quartic interaction, Eq.~(\ref%
{2a05}), by means of a Hubbard-Stratonovich (HS) transformation. Considering
the low energy theory, we have not said anything about high energies so far.
Actually, one can first integrate out the high energy degrees of freedom.
Following the philosophy of the Landau Fermi liquid theory, integrating out
the high energies results in a renormalization of the original spectrum, Eq.~(\ref{2a01}), of the interaction, Eq.~(\ref{2a02}),
and of the ground state energy. Neglecting fluctuations, one would obtain a partition
function containing the renormalized constants to be considered as
phenomenological parameters.

As a result of the integration over the high energies, one obtains instead
of the quadratic forms $nn$ and $S^{\mu }S^{\mu }$ written in Eq.~(\ref{2a05})
expressions of the type $(n-\langle n\rangle )(n-\langle n\rangle )$ and $%
(S^{\mu }-\langle S^{\mu }\rangle )(S^{\mu }-\langle S^{\mu }\rangle )$,
where $\langle \ldots \rangle $ denotes the averages with respect to the high energy
part of the Hamiltonian. Applying the HS transformation to this effective
interaction yields
\begin{widetext}
\begin{align}
\label{2a08}
\exp\big\{\!-\!\tilde{\cS}_\inte\!\big\} &= \int \exp
 \Big\{
   \ii \sum_Q \vp(-Q)\big[n(Q)\!-\!\langle n(Q)\rangle\big] + \sum_Q\sum_\bp h_{\bp}^\mu(-Q)\big[S_{\bp}^\mu(Q)\!-\!\langle S_{\bp}^\mu(Q)\rangle\big]
 \Big\}\
 \cW_h[\vp]\cW_f[h^\mu] \cD \vp\cD h^\mu\punkt
\end{align}
\end{widetext}The Gaussian weights entering Eq.~(\ref{2a08}) are given by
\begin{align}
\mathcal{W}_{h}[\varphi ]& =\exp \Big\{-\frac{1}{2V_{0}}\sum_{Q}\varphi
(Q)\varphi (-Q)\Big\}\mathnormal{,}  \label{2a09b} \\
\mathcal{W}_{f}[h^{\mu }]& =\exp \Big\{-\sum_{Q}\sum_{\mathbf{p}{{\mathbf{p}%
^{\prime }}}}h_{\mathbf{p}}^{\mu }(Q)\big[\hat{V}^{-1}\big]_{\mathbf{p}{{%
\mathbf{p}^{\prime }}}}h_{\mathbf{p}^{\prime }}^{\mu }(-Q)\Big\}\mathnormal{,%
}  \notag
\end{align}%
and $\mathcal{D}\varphi \mathcal{D}h^\mu$ is the measure normalized such that
$\int \mathcal{W}_{h}[\varphi ]\mathcal{W}_{f}[h^{\mu }]%
\mathcal{D}\varphi \mathcal{D}h^{\mu }=1$. The HS fields~$\varphi (Q)$ and~$%
h^{\mu }(Q)$ depend
on bosonic Matsubara frequencies $\omega $, corresponding in imaginary time representation to
periodicity, $\varphi (\tau +\beta ;%
\mathbf{q})=\varphi (\tau ;\mathbf{q})$ and $h_{\mathbf{p}}^{\mu }(\tau
+\beta ;\mathbf{q})=h_{\mathbf{p}}^{\mu }(\tau ;\mathbf{q})$. The constant~$%
V_{0}$ has been introduced in Eq.~(\ref{2a07}) and $\hat{V}^{-1}$ is the
inverse of the operator $\hat{V}_{\mathbf{p}}$ which acts on a function~$g(\bp)$ as
\begin{equation*}
[\hat{V}g](\mathbf{p})=\sum_{{\mathbf{p}^{\prime }}}V_{\mathbf{p}-%
{{\mathbf{p}^{\prime }}}}g({{\mathbf{p}^{\prime }}})\;\mathnormal{.}
\end{equation*}
Actually, $V_{\mathbf{p}-{{\mathbf{p}^{\prime }}}}$ is essentially the
function~$\tilde{V}$ from Eq.~(\ref{2a07}) for momenta~$\mathbf{p}$ close to
the Fermi surface. By construction, the momentum $\mathbf{q}$ in the HS
fields~$\varphi (Q)$ and~$h_{\mathbf{p}}^{\mu }(Q)$ is restricted by the
cutoff $q_{0}\ll p_{F}$ and the coupling of the HS fields to the fermions
displaces the fermion momentum $\mathbf{p}$ of
order~$p_{F}$ only locally.

\subsubsection{Integration over the fermionic fields}

After the HS transformation, Eq.~(\ref{2a08}), the partition function is
given by
\begin{equation}
\mathcal{Z}=\mathcal{Z}_{0}\int Z[\Phi ]\ \mathcal{W}[\Phi ]\mathcal{D}\Phi
\label{2a11}
\end{equation}%
where $\Phi $ is the $2\times 2$-matrix
\begin{equation}
\Phi _{\mathbf{p}}(Q)=\mathrm{i}\varphi (Q)\ +h_{\mathbf{p}}^{\mu }(Q)\
\sigma ^{\mu },  \label{2a12}
\end{equation}%
while $\mathcal{W}[\Phi ]=\mathcal{W}_{h}[\varphi ]\mathcal{W}_{f}[h^{\mu
}]$ and $\mathcal{D}\Phi =\mathcal{D}\varphi \mathcal{D}h^{\mu }$. The functional $Z[\Phi ]$ is essentially the formal partition function of
non-interacting fermions in the field~$\Phi $,
\begin{equation}
Z[\Phi ]=\mathcal{Z}_{0}^{-1}\exp \big\{\!-\!\mathcal{F}[\Phi ]\big\}\int
\exp \big\{-\tilde{\mathcal{S}}_{0}[\Phi ]\big\}\ \mathcal{D}(\chi ^{\ast
},\chi ) \label{2a13}
\end{equation}%
with the action $\tilde{\mathcal{S}}_{0}[\Phi ]$ given by
\begin{equation}
\tilde{\mathcal{S}}_{0}[\Phi ]=\mathcal{S}_{0}-\int_{0}^{\beta }\sum_{%
\mathbf{p}\mathbf{q}}\chi^{\dagger }\big(\tau ,\mathbf{p}\!-\!%
\frac{\mathbf{q}}{2}\big)\Phi _{\mathbf{p}}(\tau ,\mathbf{q})\chi %
\big(\tau ,\mathbf{p}\!+\!\frac{\mathbf{q}}{2}\big)\mathrm{d}\tau
\mathnormal{.}  \label{2a14}
\end{equation}%
Herein, the bare action $\mathcal{S}_{0}$ has been defined in Eq.~(\ref{2a03b}), the
field $\Phi $ is written in the imaginary time $\tau $ representation, and
the spinor notation $\chi=(\chi _{\uparrow },\chi _{\downarrow })$ is
used.
The functional~$\mathcal{F}[\Phi ]$ in Eq.~(\ref{2a13}) is given by
\begin{align}
\mathcal{F}[\Phi ]& =\int_{0}^{\beta }\sum_{\mathbf{p}%
\mathbf{q}}\Big\langle\chi ^{\dagger }\big(\tau ,\mathbf{p}\!-\!%
\frac{\mathbf{q}}{2}\big)\Phi _{\mathbf{p}}\big(\tau ,\mathbf{q}\big)
\chi \big(\tau ,\mathbf{p}\!+\!\frac{\mathbf{q}}{2}\big)\Big\rangle\mathrm{d%
}\tau  \label{2a15} \\
& =\int_{0}^{\beta }\sum_{\mathbf{p}\mathbf{q}}\mathrm{tr}\big[\Phi _{%
\mathbf{p}}(\tau ,\mathbf{q})
\mathcal{G}_{\mathbf{p}\!+\!\frac{\mathbf{q}}{2}%
,\mathbf{p}\!-\!\frac{\mathbf{q}}{2}}^{[0]}(\tau ,\tau \!+\!0)\big]\mathrm{d}%
\tau\punkt  \notag
\end{align}%
Herein, $\mathcal{G}^{[0]}$ is the Green's function of the ideal Fermi gas
described by the action $\mathcal{S}_{0}$, Eq.~(\ref{2a03b}), and $\mathrm{tr%
}$ denotes the trace over spins.

In order to integrate out the fermion fields in Eq.~(\ref{2a13}), we follow the route suggested in Refs.~\onlinecite{aleiner,epm}
and recast the fermion determinant as $\det \equiv \exp \mathrm{Tr}\ln $.
The logarithm is then replaced by an inverse using an additional integration
over a variable~$u$ of the form
\begin{equation*}
\ln (x+\phi )=\ln x+\int_{0}^{1}\phi (x+u\phi )^{-1}\mathrm{d}u\punkt
\end{equation*}%
This yields
\begin{align}
& Z[\Phi ]=\exp \big\{\!-\!\mathcal{F}[\Phi ]\big\}
\label{2a16} \\
& \times \exp \Big\{\int_{0}^{1}\!\!\int_{0}^{\beta }\sum_{\mathbf{p}\mathbf{%
q}}\mathrm{tr}\big[\Phi _{\mathbf{p}}(\tau ,\mathbf{q})\mathcal{G}_{\mathbf{p%
}\!+\!\frac{\mathbf{q}}{2},\mathbf{p}\!-\!\frac{\mathbf{q}}{2}}^{[u\Phi
]}(\tau ,\tau \!+\!0)\big]\mathrm{d}\tau \mathrm{d}u\Big\}  \notag
\end{align}%
where $\mathcal{G}^{[u\Phi ]}$ is the $2\times 2$-matrix Green's function of
non-interacting fermions in the external field~$u\Phi $, i.e. the propagator
of the action~$\tilde{\mathcal{S}}_{0}[u\Phi]$, Eq.~(\ref{2a14}). It satisfies the
equation
\begin{align}
\label{2a16a}
-\big\{\partial_\tau+&\ve\big(\bp+\frac{\bq}{2}\big)-\mu -u\Phi_\bp(\tau,\bq)\big\}\
 \mathcal{G}_{\mathbf{p%
}\!+\!\frac{\mathbf{q}}{2},\mathbf{p}\!-\!\frac{\mathbf{q}}{2}}^{[u\Phi
]}(\tau ,\tau') \nonumber\\
&= \delta(\tau-\tau')\delta_{\bq,0}\punkt
\end{align}
Thus, in order to study the thermodynamics of the low-lying excitations, we
need to determine the Green's function of non-interacting fermions in a random
external HS field at equal times. This quantity $\mathcal{G}^{[u\Phi ]}(\tau
,\tau \!+\!0)$ is a complex matrix function being \emph{periodic} in
imaginary time~$\tau $, $\mathcal{G}^{[u\Phi ]}(\tau \!+\!\beta ,\tau
\!+\!\beta \!+\!0)=\mathcal{G}^{[u\Phi ]}(\tau ,\tau \!+\!0)$. This
observation motivates the introduction of the complex $2\times 2$-matrix
field $\mathcal{A}_{\mathbf{p}{{\mathbf{p}^{\prime }}}}=\mathcal{A}_{\mathbf{%
p}{{\mathbf{p}^{\prime }}}}^{[u\Phi ]}$ as
\begin{equation}
\mathcal{A}_{\mathbf{p}{{\mathbf{p}^{\prime }}}}(\tau )=\mathcal{G}_{\mathbf{%
p}{{\mathbf{p}^{\prime }}}}^{[0]}(\tau ,\tau \!+\!0)-\mathcal{G}_{\mathbf{p}{{%
\mathbf{p}^{\prime }}}}^{[u\Phi ]}(\tau ,\tau \!+\!0)\;\mathnormal{.}
\label{2a17}
\end{equation}%
The field~$\mathcal{A}_{\mathbf{p}{{\mathbf{p}^{\prime }}}}(\tau )$ obeys
\emph{bosonic} periodicity in time $\tau $, $\mathcal{A}_{\mathbf{p}{%
{\mathbf{p}^{\prime }}}}(\tau +\beta )=\mathcal{A}_{\mathbf{p}{{\mathbf{p}%
^{\prime }}}}(\tau )$, and captures the entire physics of the low-lying
excitations.

By this definition, the functional $Z[\Phi ]$, Eq.~(\ref{2a16}), takes the
form of a functional of the boson field~$\mathcal{A}_{\mathbf{p}{{\mathbf{p}%
^{\prime }}}}(\tau )$, Eq.~(\ref{2a17}),
\begin{equation}
Z[\Phi ]=\exp \Big\{-\int_{0}^{1}\!\!\int_{0}^{\beta }\sum_{\mathbf{p}%
\mathbf{q}}\mathrm{tr}\big[\Phi _{\mathbf{p}}(\tau ,\mathbf{q})\mathcal{A}_{%
\mathbf{p}\!+\!\frac{\mathbf{q}}{2},\mathbf{p}\!-\!\frac{\mathbf{q}}{2}%
}\left( \tau \right) \big]\ \mathrm{d}\tau \mathrm{d}u\Big\}\mathnormal{.}
\label{2a18}
\end{equation}%
Using Eqs.~(\ref{2a11}) and~(\ref{2a18}) one can compute the partition function $%
\mathcal{Z}$ provided the function $\mathcal{A}_{\bp\bp'}\left( \tau \right) $ is known for any configuration of~$\Phi$. Of course, one could solve Eq.~(%
\ref{2a16a}) for the Green's function and find $\mathcal{A}_{\bp\bp'}\left( \tau \right) $ from Eq.~(\ref{2a17}) but this
would lead to the conventional diagrammatic expansion for the fermions. Here,
we follow a different route deriving a closed equation for the field~$\mathcal{A}%
_{\bp\bp'}\left( \tau \right)$.

\subsubsection{Dynamics of excitations}

\label{ssec:dynamics}

The derivation of the equation of motion for the boson field~$\mathcal{A}_{%
\mathbf{p}{{\mathbf{p}^{\prime }}}}$ follows the route presented in detail
in Refs.~\onlinecite{epm} and~\onlinecite{epm2}: We start with the differential
equations for the Green's functions~$\mathcal{G}^{[0]}(\tau ,\tau ^{\prime })$
and~$\mathcal{G}^{[u\Phi ]}(\tau ,\tau ^{\prime })$, Eq.~(\ref{2a16a}).
Subtracting them according to the definition of~$\mathcal{A}_{\mathbf{p}{{%
\mathbf{p}^{\prime }}}}$, Eq.~(\ref{2a17}), yields a differential equation
as an intermediate result. Next, we repeat the same steps with the equations for~$\mathcal{G}^{[0]}(\tau ,\tau ^{\prime })$ and~$\mathcal{G}%
^{[u\Phi ]}(\tau ,\tau ^{\prime })$ while this time the operator $\partial_\tau+\hat{H}_0-u\Phi$ acts from the right.
The equation obtained in this way is then subtracted from the intermediate result. Eventually putting~$\tau
^{\prime }=\tau +0$, we find that the dynamics of~$\mathcal{A}_{\mathbf{p}%
\!+\!\frac{\mathbf{k}}{2},\mathbf{p}\!-\!\frac{\mathbf{k}}{2}}$ is governed
by the equation
\begin{widetext}
\begin{align}
\label{2a19}
 \Big[\del_\tau
 + \frac{(\bp\!\cdot\!\bk)}{m}\Big]\cA_{\bp\!+\!\frac{\bk}{2},\bp\!-\!\frac{\bk}{2}}
 - u \!\sum_\bq\!
  \Big[
   \Phi_{\bp\!+\!\frac{\bk\!+\!\bq}{2}}(\bq)\cA_{\bp\!+\!\frac{\bk}{2}\!+\!\bq,\bp\!-\!\frac{\bk}{2}}
   -\cA_{\bp\!+\!\frac{\bk}{2},\bp\!-\!\frac{\bk}{2}\!-\!\bq}\Phi_{\bp\!-\!\frac{\bk\!+\!\bq}{2}}(\bq)
  \Big]
 &=-u \Phi_{\bp}\big(\!-\!\bk\big)
   \big[
     n_{\bp\!-\!\frac{\bk}{2}}-n_{\bp\!+\!\frac{\bk}{2}}
   \big]\punkt
\end{align}
\end{widetext}Herein, $n_{\mathbf{p}}=[\exp (\xi _{\mathbf{p}}/T)+1]^{-1}$
with $\xi _{\mathbf{p}}=\varepsilon(\mathbf{p})-\varepsilon _{F}$ is the
Fermi distribution function of the ideal Fermi gas.

In Eq.~(\ref{2a19}), the relevant momenta~$\mathbf{k}$ in~$\mathcal{A}_{\mathbf{p}\!+\!\frac{\mathbf{k}}{2},\mathbf{p}%
\!-\!\frac{\mathbf{k}}{2}}$ are small and do not exceed the cutoff momentum $%
q_{0}\ll p_{F}$. This can easily be seen from Eq.~(\ref{2a18}) because the fields~$\Phi _{\mathbf{p}%
}(\tau ,\mathbf{k})$ are by construction non-zero only if the momentum~$\mathbf{k}$ is small.
The fact that the two momenta $\mathbf{p+k/}2$ and $\mathbf{p-k/}2$ entering~%
$\mathcal{A}_{\mathbf{p}\!+\!\frac{\mathbf{k}}{2},\mathbf{p}\!-\!\frac{%
\mathbf{k}}{2}}$ are essentially close to each other significantly simplifies the
analytical study as compared to the general formulation studied
perturbatively in Ref.~\onlinecite{epm2} as a check of the bosonization
procedure.

In order to obtain a feasible low energy theory, the still present degrees
of freedom on the scale~$p_{F}$ should be integrated out from Eqs.~(\ref{2a18}) and~(\ref{2a19}).
Since the right-hand side of Eq.~(\ref{2a19}) contains the
combination $n_{\mathbf{p}-\frac{\mathbf{k}}{2}}-n_{\mathbf{p}+\frac{\mathbf{k}}{2}}$, the dependence of the function $\mathcal{A}_{\mathbf{p}\!+\!\frac{\mathbf{k}%
}{2},\mathbf{p}\!-\!\frac{\mathbf{k}}{2}}$ on $\left\vert \mathbf{p}%
\right\vert $ near $p_{F}$ is for small $\mathbf{k}$ sharper than the dependence
of all other functions entering the left-hand side of Eq.~(\ref{2a19}) or the exponent in Eq.~(\ref{2a18}). As a result,
when integrating both sides of Eq.~(\ref{2a19}) or the integrand in the exponent of Eq.~(\ref{2a18}) over the variable $\xi _{\mathbf{p}}$,
it is justified to approximate $\mathbf{p} \simeq p_{F}\mathbf{n}$ with $\mathbf{n}
^{2}=1$ in all other functions depending smoothly on~$\left\vert
\mathbf{p}\right\vert$.

Proceeding in this way, we rewrite the exponent in the expression for $%
Z[\Phi ]$, Eq.~(\ref{2a18}), as
\begin{align}
& \sum_{\mathbf{p}\mathbf{q}}\Phi _{\mathbf{p}}(\mathbf{q})\mathcal{A}_{%
\mathbf{p}\!+\!\frac{\mathbf{q}}{2},\mathbf{p}\!-\!\frac{\mathbf{q}}{2}}
\label{2a20} \\
\simeq & \sum_{\mathbf{q}}\int \Phi _{\mathbf{n}}(\mathbf{q})\left( \nu \int
\mathcal{A}_{\mathbf{p}\!+\!\frac{\mathbf{q}}{2},\mathbf{p}\!-\!\frac{%
\mathbf{q}}{2}}\ \mathrm{d}\xi _{\mathbf{p}}\right) \mathrm{d}\mathbf{n}\;%
\mathnormal{,}  \notag
\end{align}%
thus separating the radial and angular integration of the momentum vector~$%
\mathbf{p}=[2m(\varepsilon _{F}+\xi _{\mathbf{p}})]^{1/2}\mathbf{n}$. In Eq.~(\ref{2a20}), $\nu $ is the density of states at the Fermi surface and the
integration $\int \mathrm{d}\mathbf{n}$ is done over the $(d-1)$-dimensional
sphere~$S^{d-1}$. We normalize the integration over $\mathbf{n}$ by the
convention
\begin{equation}
\int_{S^{d-1}}\mathrm{d}\mathbf{n}=1\mathnormal{,}  \label{2a20a}
\end{equation}%
and write the function $\Phi _{\mathbf{p}}\left( \mathbf{q}\right) $ as $%
\Phi _{\mathbf{n}}\left( \mathbf{q}\right) $ since $\mathbf{p}\simeq p_{F}\mathbf{n}$%
.

Equation~(\ref{2a20}) shows us that we need the integral over $\xi _{%
\mathbf{p}}$ of~$\mathcal{A}_{\mathbf{p}\!+\!\frac{\mathbf{k}}{2},\mathbf{p}%
\!-\!\frac{\mathbf{k}}{2}}$ rather than this function itself. This observation
motivates us to introduce the \emph{quasiclassical} field
\begin{equation}
a_{\mathbf{n}}(\mathbf{k})=\nu \int \mathcal{A}_{\mathbf{p}\!+\!\frac{%
\mathbf{k}}{2},\mathbf{p}\!-\!\frac{\mathbf{k}}{2}}\ \mathrm{d}\xi _{\mathbf{%
p}}  \label{2a21}
\end{equation}%
and to construct the low energy theory for this field.

For this purpose, we integrate the equation of motion for~$\mathcal{A}_{%
\mathbf{p}\!+\!\frac{\mathbf{k}}{2},\mathbf{p}\!-\!\frac{\mathbf{k}}{2}}$,
Eq.~(\ref{2a19}), in a similar manner term by term. For the first term of
Eq.~(\ref{2a19}) we find
\begin{equation}
\nu \int \Big[\partial _{\tau }+\frac{(\mathbf{p}\!\cdot \!\mathbf{k})}{m}%
\Big]\mathcal{A}_{\mathbf{p}\!+\!\frac{\mathbf{k}}{2},\mathbf{p}\!-\!\frac{%
\mathbf{k}}{2}}\mathrm{d}\xi _{\mathbf{p}}\simeq \Big[\partial _{\tau
}+v_{F}(\mathbf{n}\!\cdot \!\mathbf{k})\Big]a_{\mathbf{n}}(\mathbf{k}),
\label{2a22}
\end{equation}%
where $v_{F}$ is the Fermi velocity.

For the first expression in the second term, we obtain
\begin{align}
& \nu \int
 \Phi _{\mathbf{p}\!+\frac{\mathbf{k}\!+\!\mathbf{q}}{2}}(\mathbf{q})
   \mathcal{A}_{\mathbf{p}\!+\!\frac{\mathbf{k}+\mathbf{q}}{2}\!
               +\!\frac{\mathbf{q}}{2},\mathbf{p}\!-\!\frac{\mathbf{k}+\mathbf{q}}{2}+\!\frac{\mathbf{q}}{2}}
   \dt\xi_{\bp}  \label{2a23} \\
  &\qquad\qquad\simeq
  \Phi_{\mathbf{n}\!+\!\frac{\mathbf{k}_{\perp }\!+\!\mathbf{q}_{\perp }}{2p_{F}}}(\bq)
  a_{\bn+\frac{\bq_{\perp}}{2p_{F}}} (\mathbf{k}+\mathbf{q})\komma  \notag
\end{align}%
where $\mathbf{k}_{\perp }$ is the component of the vector $\mathbf{k}$
perpendicular to the vector $\mathbf{n,}$
\begin{equation}
\mathbf{k}_{\perp }=\mathbf{k}-(\mathbf{n}\cdot \mathbf{k})\bn\punkt
\label{2a24}
\end{equation}%
In Eq.~(\ref{2a23}), we made the same approximation as in Eq.~(\ref{2a20})
using the fact that the field $\Phi $ is a slow function of the variable~$%
\left\vert \mathbf{p}\right\vert $.
The radial integration in Eq.~(\ref{2a23}) absorbs the normal component of~$%
\mathbf{q}$ and that is why only the tangent component~$\mathbf{q}_{\perp }$
enters the result of the integration. The combination $\mathbf{n}+\mathbf{q}%
_{\perp }/(2p_{F})$ in Eq.~(\ref{2a23}) for small vectors~$\mathbf{q}_{\perp
}$ corresponds to a rotation of~$\mathbf{n}$. Accounting for these rotations
is very important because they capture the essential effects of the curvature of
the Fermi surface arising in $d>1$.

The remaining terms of Eq.~(\ref{2a19}) are treated similarly. Integrating
the right-hand side is especially simple because $n_{\mathbf{p}\!-\!\frac{\mathbf{k}}{%
2}}-n_{\mathbf{p}\!+\!\frac{\mathbf{k}}{2}}\simeq m^{-1}(\mathbf{p}\!\cdot \!%
\mathbf{k})\delta (\xi _{\mathbf{p}})$. Finally, we reduce the functional~$%
Z[\Phi ]$, Eq.~(\ref{2a11}), to the form
\begin{equation}
Z[\Phi ]=\exp \Big\{-\int_{0}^{1}\!\!\int_{0}^{\beta }\!\!\int \sum_{\mathbf{%
q}}\mathrm{tr}\big[\Phi _{\mathbf{n}}(\mathbf{q})a_{\mathbf{n}}(\mathbf{q})%
\big]\ \mathrm{d}\mathbf{n}\mathrm{d}\tau \mathrm{d}u\Big\}\;\mathnormal{,}
\label{2a25a}
\end{equation}%
where~$a_{\mathbf{n}}(\mathbf{k})$ is the solution of the equation
\begin{widetext}
\begin{align}
\label{2a25b}
 \Big[\del_\tau
 \!+\! v_F(\bn\!\cdot\!\bk)\Big]a_\bn(\bk)
 - u \!\sum_\bq\!
  \Big[
   \Phi_{\bn\!+\!\frac{\bk_\perp\!+\!\bq_\perp}{2p_F}}(\bq)a_{\bn+\frac{\bq_\perp}{2p_F}}(\bk\!+\!\bq)
   \!-\!a_{\bn-\frac{\bq_\perp}{2p_F}}(\bk\!+\!\bq)\Phi_{\bn\!-\!\frac{\bk_\perp\!+\!\bq_\perp}{2p_F}}(\bq)
  \Big]
 \!=\!-u \nu v_F (\bn\!\cdot\!\bk) \Phi_{\bn}(-\bk)\punkt
\end{align}
\end{widetext}

Let us discuss the analytical properties of the field~$a_{\mathbf{n}}(%
\mathbf{k})=a_{\mathbf{n}}(u,\tau ,\mathbf{k})$ and its equation of motion
Eq.~(\ref{2a25b}). By construction, the field $a_{\mathbf{n}}(\tau ,\mathbf{k%
})$ describing the low energy excitations is bosonic: It is a complex $%
2\times 2$-matrix field periodic in time, $a_{\mathbf{n}}(\tau +\beta ,%
\mathbf{k})=a_{\mathbf{n}}(\tau ,\mathbf{k})$. The momentum~$\mathbf{k}$
entering the field~$a$, $\left\vert \mathbf{k}\right\vert \lesssim q_{0}\ll
p_{F}$, determines the scale of the bosonic excitations while the argument $%
\mathbf{n}$ of the field~$a_{\mathbf{n}}(\mathbf{k})$ determines the
position on the Fermi surface. The dependence of $a_{\mathbf{n}}(\mathbf{k})$
on~$\mathbf{n}$ and~$\mathbf{k}$ contains the full
information needed to describe the spin and charge excitations in a
higher-dimensional Fermi gas.

In dimension~$d=1$, transverse momenta~$\mathbf{q}_{\perp }$, $\mathbf{k}%
_{\perp }$ do not exist. If we neglected these momenta in higher dimensions,
we would come to the low energy model of Ref.~\onlinecite{aleiner}. In this
approximation, Eq.~(\ref{2a25b}) describes one-dimensional processes and all
effects of the Fermi surface curvature in~$d>1$ are lost. Writing the field $%
a_{\mathbf{n}}\left( \mathbf{k}\right) $ in the form $a_{\mathbf{n}}(\mathbf{%
k})=\varrho _{\mathbf{n}}(\mathbf{k})\sig^0+\mathbf{s}_{\mathbf{n}}(\mathbf{k}%
)\cdot \mbox{\boldmath$\sigma$}$, one can see that, as a result of this
approximation, the charge $\varrho _{\mathbf{n}}\left( \mathbf{k}\right) $ and
spin $\mathbf{s}_{\mathbf{n}}\left( \mathbf{k}\right) $ parts of the field
decouple from each other. Eventually, one obtains a model of non-interacting
charge excitations~$\varrho _{\mathbf{k}}(\mathbf{n})$ and a non-trivial
field theory for the interacting spin excitations~$\mathbf{s}_{\mathbf{n}}(%
\mathbf{k})$. All the results of Ref.~\onlinecite{aleiner} have been
obtained in this way, implying their validity in one dimension but requesting
reconsideration for $d>1$.

Taking into account the curvature effects in dimensions $d>1$, Eq.~(\ref{2a25b}) shows
that the spin and charge excitations cannot be treated separately but
interact with each other. Moreover, the charge modes also interact themselves similarly to the spin modes. The interaction of spin and charge modes
is what constitutes the major difference between the physics of the fermion
gases in $d=1$ and $d>1$. Furthermore, we will see that there are classes of diagrams in
a perturbative analysis of the final boson theory that are logarithmic
in $d=1$, but due to the presence
of the $\mathbf{q}_{\perp }$-terms become regular in dimensions $d>1$.

In principle, one can solve Eq.~(\ref{2a25b}) employing a perturbation theory in
the HS field~$\Phi $. In the zero-order approximation, one should neglect the
terms containing the HS field~$\Phi $ in the left-hand side of Eq.~(\ref{2a25b}),
yielding $a_{\mathbf{n}}(\omega,\mathbf{k})\simeq a_{\mathbf{n}}^{(0)}(\omega,\mathbf{k})$ with
\begin{equation}
a_{\mathbf{n}}^{\left( 0\right) }(\omega ,\mathbf{k})= \nu u\ \frac{%
v_{F}(\mathbf{n}\!\cdot \!\mathbf{k})}{\mathrm{i}\omega -v_{F}(\mathbf{n}%
\!\cdot \!\mathbf{k})}\ \Phi _{\mathbf{n}}(\omega ,-\mathbf{k})\;\mathnormal{%
.}  \label{e5}
\end{equation}%
Inserting this zero order approximation $a_{\mathbf{n}}^{\left( 0\right)
}(\omega ,\mathbf{k})$ into the functional~$Z[\Phi ]$, Eq.~(\ref{2a25a}), we
reduce the partition function~$\mathcal{Z}$, Eq.~(\ref{2a11}), to a Gaussian
integral over the field~$\Phi =\mathrm{i}\varphi +h^{\mu }\sigma ^{\mu }$,
Eq.~(\ref{2a12}). It is not difficult to understand that this limit yields
the contributions obtained by summing certain ladder series in the
conventional fermion diagrammatics.\cite{agd} Considering only the HS field~$%
\varphi $ while neglecting~$h^{\mu }$, we obtain the contribution of the
rings built from polarization bubbles, i.e. reproduce the random phase
approximation (RPA). Alternatively, keeping only~$h^{\mu }$ the contribution
of the particle-hole ladder ring is reproduced. Keeping both~$\varphi $ and~$%
h^{\mu }$, we obtain the contribution of all particle-hole ring diagrams.

However, the interesting logarithmic contributions to the non-analytic terms
arise from the fluctuations on top of these ladders. Considering the ladders
as propagators of elementary bosonic excitations, one can say that the
logarithmic contributions arise as a result of interaction between these
excitations. In order to treat the fluctuations properly, we need a more
efficient route of solving the equation of motion for~$a_{\mathbf{n}}(\omega
,\mathbf{k})$, Eq.~(\ref{2a25b}), including the $\Phi $-term in the
left-hand side.

\subsubsection{Superfield representation}

\label{ssec: superfield}

In this section, we represent the solution of the equation of motion for the bosonic field~$a_{\mathbf{n}}(\mathbf{k})$,
Eq.~(\ref{2a25b}),  for an arbitrary $\Phi $ in the form of a functional integral over superfields. We begin by
noting a remarkable symmetry in the left-hand side of Eq.~(\ref{2a25b}).

Being linear in~$a_{\mathbf{n}}(\mathbf{k})$, the left-hand side
can be formally represented as $[\mathcal{L}a]_{\mathbf{n}}(\mathbf{k})$. We observe that the
operator~$\mathcal{L}$ is \emph{antisymmetric} with respect to the inner
product given by
\begin{align}
\big(f^{\dagger },a\big)=\!\int_{0}^{1}\!\!\int_{0}^{\beta }\!\!\int \!\sum_{%
\mathbf{k}}\mathrm{tr}\big[f_{\mathbf{n}}^{\dagger }(-\mathbf{k})a_{\mathbf{n%
}}(\mathbf{k})\big]\ \mathrm{d}\mathbf{n}\mathrm{d}\tau \mathrm{d}u
\label{2a26}\punkt
\end{align}
Herein, $f^{\dagger }$ is a field having the same structure like~$a$. The
antisymmetry condition $(f^{\dagger },\mathcal{L}a)=-(a,\mathcal{L}%
f^{\dagger })$ is straightforwardly checked using the definitions of the
operator~$\mathcal{L}$ and the inner product Eq.~(\ref{2a26}). We will see
shortly that this antisymmetry of~$\mathcal{L%
}$ leads to an important simplification of the theory.

Since the remaining of the derivation of the low energy field theory is purely formal,
we will use short-hand notations in this section. We define
\begin{equation}
\mathcal{R}\equiv \mathcal{R}_{\mathbf{n}}(u,\tau ,\mathbf{k})=-u\nu v_{F}(%
\mathbf{n}\!\cdot \!\mathbf{k})\Phi _{\mathbf{n}}(-\mathbf{k})  \label{2a27}
\end{equation}%
as short-hand notation for the right-hand side of Eq.~(\ref{2a25b}) and use the
notation $\mathcal{L}$ introduced above for the antisymmetric operator in the left-hand side.
As a result, Eq.~(\ref{2a25b}) can be written in the compact form
\begin{equation}
\mathcal{L}a=\mathcal{R}\punkt  \label{2a28}
\end{equation}
Both the operator~$\mathcal{L}$ and the inhomogeneity term~$\mathcal{R}$
depend (linearly) on the HS field~$\Phi $, Eq.~(\ref{2a12}). With $\Phi $
being a Gauss-distributed random field, Eq.~(\ref{2a28}) is technically a
\emph{stochastic differential equation} for the boson field~$a_{\mathbf{n}%
}(u,\tau ,\mathbf{k})$. In the context of stochastic field
equations, a well-known method of analysis is the Becchi-Rouet-Stora-Tyutin
(BRST) transformation\cite{faddeev,brst,parisi,justin} which brings the
problem of solving the stochastic equation into the form of a supersymmetric
field theory. The latter formulation allows for a study by means of standard field
theory techniques.

We now apply the BRST map on our problem. First, we rewrite Eq.~(\ref{2a25a})
with $a$ satisfying Eq.~(\ref{2a25b}) in a form of a functional integral
over fields $a$
\begin{equation}
Z[\Phi ]=\int \delta \lbrack \mathcal{L}a-\mathcal{R}]\Big|\det \frac{\delta
\mathcal{L}}{\delta a}\Big|\ Z[\Phi ;a]\ \mathcal{D}a^{\dagger }\mathcal{D}a
\;\mathnormal{.}  \label{2a29}
\end{equation}%
Herein, the integration with respect to the measure $\mathcal{D}a^{\dagger }\mathcal{D}a$ is performed over all complex fields $a$
which do not necessarily satisfy Eq.~(\ref{2a25b}). The functional $Z[\Phi ;a]$ in
Eq.~(\ref{2a29}) denotes formally the functional from Eq.~(\ref{2a25a}), yet
the field~$a$ is here an unspecified complex field and not the solution to
the constraint equation Eq.~(\ref{2a25b}). The equivalence of the
representation by Eq.~(\ref{2a29}) and the original one, Eqs.~(\ref{2a25a}) and~(\ref{2a25b}),
for $Z\left[\Phi \right] $ is easily seen as the constraint~(\ref{2a25b}) is enforced
by the integration over the functional $\delta $-function in the
integrand of Eq.~(\ref{2a29}). The determinant in Eq.~(\ref{2a29}) arises as a consequence of
changing variables from~$a$ to~$\mathcal{L}a$.

Our goal is to integrate the functional $Z\left[ \Phi \right] $, Eq.~(\ref{2a29}),
over the fields $\Phi$ and obtain a field theory for the interacting
bosonic excitations. This can be achieved representing the $\delta$-function
as a Fourier integral,
\begin{align}
& \delta \lbrack \mathcal{L}a-\mathcal{R}]  \label{2a30} \\
\propto & \int \exp \Big\{\frac{\mathrm{i}}{2}\left( f^{\dagger },\mathcal{L}a-%
\mathcal{R}\right) +\frac{\mathrm{i}}{2}\left( [\mathcal{L}a-\mathcal{R}%
]^{\dagger },f\right) \Big\}\ \mathcal{D}f^{\dagger }\mathcal{D}f\;%
\mathnormal{,}  \notag
\end{align}%
and the determinant as an integral over Grassmann variables,
\begin{align}
& \det \frac{\delta \mathcal{L}}{\delta a}  \label{2a31} \\
\propto & \int \exp \Big\{\frac{\mathrm{i}}{2}\left( \rho ^{\dagger },\mathcal{L}%
\sigma \right) +\frac{\mathrm{i}}{2}\left( [\mathcal{L}\sigma ]^{\dagger
},\rho \right) \Big\}\ \mathcal{D}\sigma ^{\dagger }\mathcal{D}\sigma
\mathcal{D}\rho ^{\dagger }\mathcal{D}\rho \;\mathnormal{.}  \notag
\end{align}%
In Eqs.~(\ref{2a30}) and~(\ref{2a31}), $f$ is a complex field, while $\sigma $
and $\rho $ are Grassmann fields of the same structure as $a$. The
brackets~$\left( \ldots ,\ldots \right) $ denote the inner product defined
in Eq.~(\ref{2a26}). All the fields in Eqs.~(\ref{2a30}) and~(\ref{2a31}) are
assumed to be periodic in imaginary time $\tau $ in order to reproduce
the bosonic boundary condition of the solution~$a$ of Eq.~(\ref{2a25b}).

Substituting Eqs.~(\ref{2a30}) and~(\ref{2a31}) into Eq.~(\ref{2a29}), we come to
the representation of the functional $Z\left[ \Phi \right] $ in terms of a
Gaussian integral over the fields $a$,$a^{\dagger }$,$f$,$f^{\dagger }$,$%
\sigma $,$\sigma ^{\dagger }$,$\rho $,$\rho ^{\dagger }$. However, as the
functional $Z[\Phi ;a]$ contains only the field $a$ and not $a^{\dagger }$,
the integral over the fields $a^{\dagger },f,\sigma ^{\dagger },\rho $ can
immediately be calculated giving unity. Then, we are left with an integral
only over the fields $a,f^{\dagger },\sigma ,\rho ^{\dagger }$.

Instead of writing all these fields separately we unify them into one \emph{
superfield}~$\Psi$ which we define as
\begin{equation}
\Psi (\theta ,{\theta ^{\ast }})=a\theta +f^{\dagger }{\theta ^{\ast }}%
+\sigma +\rho ^{\dagger }{\theta ^{\ast }}\theta \;\mathnormal{.}
\label{2a32}
\end{equation}%
$\theta $ and ${\theta ^{\ast }}$ are additional Grassmann anticommuting
variables. By construction, $\Psi $ is an \emph{anticommuting} field. This,
however, does not mean that it describes fermions as the periodicity in
imaginary time,
\begin{equation}
\Psi \left( \tau \right) =\Psi \left( \tau +\beta \right) \komma \label{e6}
\end{equation}%
guarantees the boson statistics.

The antisymmetry of the operator~$\mathcal{L}$, Eq.~(\ref{2a28}), implies
the remarkable and important relation
\begin{equation}
\int \left( \Psi ,\mathcal{L}\Psi \right) \ \mathrm{d}\theta \mathrm{d}{%
\theta ^{\ast }}=2\left( f^{\dagger },\mathcal{L}a\right) +2\left( \rho
^{\dagger },\mathcal{L}\sigma \right) \punkt  \label{2a33}
\end{equation}%
Using additionally the relations
\begin{equation}
a=-\int \Psi {\theta ^{\ast }}\ \mathrm{d}\theta \mathrm{d}{\theta ^{\ast }}%
\;\mathnormal{,}\quad f^{\dagger }=\int \Psi \theta \ \mathrm{d}\theta
\mathrm{d}{\theta ^{\ast }}\;\mathnormal{,}  \label{e7}
\end{equation}%
we can express the entire field theory solely in terms of the
superfield~$\Psi$.

As a result, we write the partition
function~$\mathcal{Z}$ of the low energy field theory for the excitation
modes of the interacting Fermi gas, Eq.~(\ref{2a11}), in
the form of a functional integral over the superfield~$\Psi$ and the auxiliary field~$\Phi$,
\begin{equation}
\mathcal{Z}=\mathcal{Z}_{0}\int \exp \big\{-\mathcal{S}_{S}-\mathcal{S}_{B}%
\big\}\ \mathcal{W}[\Phi ]\mathcal{D}\Phi \mathcal{D}\Psi \punkt \label{2a35}
\end{equation}%
The action~$\cS_S$ with
\begin{equation}
\mathcal{S}_{S}=\frac{\mathrm{i}}{2}\int \Big[-\frac{1}{2}\left( \Psi ,%
\mathcal{L}\Psi \right) +\left( \Psi ,\mathcal{R}\right) \theta \Big]\
\mathrm{d}\theta \mathrm{d}{\theta ^{\ast }} \label{2a36a}
\end{equation}
is invariant under the BRST symmetry transformation $\Psi\mapsto\Psi+\delta\Psi$ with the
variation given by
\begin{align}
\label{2a36c}
\delta\Psi &= \eta\ \frac{\partial \Psi}{\partial \theta^*}\punkt
\end{align}
Herein, the transformation parameter~$\eta$ is a Grassmann variable.
The action~$\mathcal{S}_{B}$, which derives from
the functional~$Z[\Phi;a]$, takes the form
\begin{equation}
\mathcal{S}_{B}=-\int_{0}^{1}\!\!\int_{0}^{\beta }\!\!\int \sum_{\mathbf{q}}%
\mathrm{tr}\big[\Phi _{\mathbf{n}}(\mathbf{q})\Psi _{\mathbf{n}}(\mathbf{q})%
\big]{\theta ^{\ast }}\ \mathrm{d}\theta \mathrm{d}{\theta ^{\ast }}\mathrm{d%
}\mathbf{n}\mathrm{d}\tau \mathrm{d}u\punkt  \label{2a36b}
\end{equation}
In contrast to~$\cS_S$, Eq.~(\ref{2a36a}), the action~$\cS_B$ is not invariant when varying
the superfield~$\Psi$ according to Eq.~(\ref{2a36c}) and, thus, breaks the BRST symmetry.\cite{footnote00}

\subsubsection{Final form of the low energy model}

Since the action~$\mathcal{S}_{S}+\mathcal{S}_{B}$ is linear in the
auxiliary field~$\Phi $, the integral over~$\Phi $ in Eq.~(\ref{2a35}) is
Gaussian and can easily be performed. This yields the final form of the low
energy field theory for the bosonic excitations of the interacting Fermi
gas. All the interesting physics is described by the $2\times 2$-matrix
superfield~$\Psi $ only. The partition function can be written after the integration
over the HS field $\Phi $ in the form
\begin{equation}
\mathcal{Z}=\mathcal{Z}_{0}\int \exp \big\{-\mathcal{S}_{\mathrm{bare}}-%
\mathcal{S}_{2}-\mathcal{S}_{3}-\mathcal{S}_{4}\big\}\ \mathcal{D}\Psi\punkt
\label{2a37}
\end{equation}%
Herein, the bare action~$\mathcal{S}_{\mathrm{bare}}$ is in Fourier
representation given by
\begin{align}
\mathcal{S}_{\mathrm{bare}}& =-\frac{\mathrm{i}}{4}\!\int \sum_{K}\mathrm{tr}%
\big[\Psi _{\mathbf{n}}(-K,\kappa)  \label{2a37b} \\
& \quad \times \left\{ -\mathrm{i}\varepsilon +v_{F}(\mathbf{n}\!\cdot \!%
\mathbf{k})\right\} \Psi _{\mathbf{n}}(K,\kappa)\big]\ \mathrm{d}%
\mathbf{n}\mathrm{d}\kappa \;\mathnormal{.}  \notag
\end{align}%
For the sake of compact notations, we use here and in the following
the four-momentum notations
\begin{align*}
K =(\varepsilon ,\mathbf{k})\;\mathnormal{,}\;
\sum_{K}(\ldots) &=T\sum_{\varepsilon
}\int(\ldots)\ \frac{\dt^d\bk}{(2\pi)^d}
\;\mathnormal{,}\;
\nonumber \\
\delta _{K,-K^{\prime }}&=\frac{\delta
_{\varepsilon ,-\varepsilon ^{\prime }}}{T}\delta _{\mathbf{k},-\mathbf{k}%
^{\prime }}\;\mathnormal{,} \nonumber \\
Q =(\omega ,\mathbf{q})\;\mathnormal{,}\;
\sum_{Q}(\ldots) &=T\sum_{\omega }\int(\ldots)\ \frac{\dt^d\bq}{(2\pi)^d} \;\mathnormal{,}
\end{align*}%
where $\varepsilon $ and $\omega $ denote bosonic Matsubara frequencies.
Also, we use the following short-hand notations in the remaining of our
analysis:
\begin{align}
\kappa & =(u,\theta ,{\theta ^{\ast }})\;\mathnormal{,}  \label{2a38} \\
\mathrm{d}\kappa & =\mathrm{d}\theta \mathrm{d}{\theta ^{\ast }}\mathrm{d}u\;%
\mathnormal{,}  \notag \\
\delta (\kappa -\kappa ^{\prime })& =\delta (u-u^{\prime })({\theta ^{\ast }}%
-\theta ^{\prime \ast })(\theta -\theta ^{\prime })\;\mathnormal{.}  \notag
\end{align}%
Whenever we integrate over~$u$, integration over the interval~$(0,1)$
is implied.

Averaging quadratic forms with respect to the bare action~$\mathcal{S}_{\mathrm{bare}}$,
Eq.~(\ref{2a37}), is done as
\begin{align}
\left\langle \Psi _{\mathbf{n}}^{\sigma {\tilde{\sigma}}}(K,\kappa)\Psi _{%
\mathbf{n}^{\prime }}^{\sigma ^{\prime }{\tilde{\sigma}}^{\prime }}(K^{\prime},\kappa
^{\prime })\right\rangle & =2\mathrm{i}\ g_{\mathbf{n}}(K)
\label{2a39} \\
\times \delta _{\sigma {\tilde{\sigma}}^{\prime }}\delta _{{\tilde{\sigma}}%
\sigma ^{\prime }}\delta (\kappa -\kappa ^{\prime })\delta (\mathbf{n}-&
\mathbf{n}^{\prime })\delta _{K,-K^{\prime }}\;\mathnormal{.}  \notag
\end{align}%
In Eq.~(\ref{2a39}),
\begin{align}
\label{2a40}
g_{\mathbf{n}}(K)=\frac{1}{\mathrm{i}\varepsilon -v_{F}(\mathbf{n}\!\cdot \!%
\mathbf{k})}
\end{align}%
is the bare Green's function for the bosonic modes. Higher moments of the
field~$\Psi $ are reduced to second moments, Eq.~(\ref{2a39}), using Wick's
theorem.

\begin{figure}[t]
\includegraphics{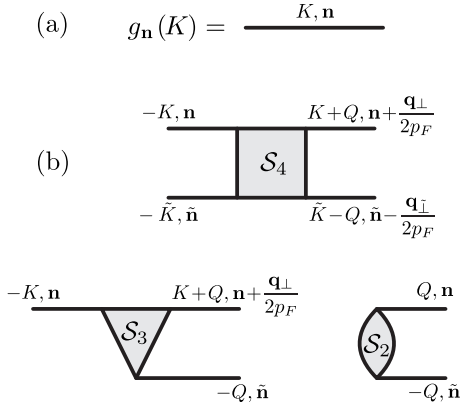}
\caption{Diagrammatic building blocks of our low energy field theory: (a)
the propagator $g_{\mathbf{n}}(K)$, Eq.~(\protect\ref{2a40}), and (b) the
interaction vertices~$\mathcal{S}_{4}$, $\mathcal{S}_{3}$, and~$\mathcal{S}%
_{2}$ from Eqs.~(\protect\ref{2a41}), (\protect\ref{2a43}), and~(\protect\ref%
{2a44}). }
\label{fig: blocks}
\end{figure}

Let us now have a look at the interaction vertices in Eq.~(\ref{2a37}).
The quartic interaction term~$\mathcal{S}_{4}$ is given by
\begin{align}
\mathcal{S}_{4}& =-\frac{1}{4\nu }\int \mathrm{d}\mathbf{n}\mathrm{d}{\tilde{%
\mathbf{n}}}\mathrm{d}\kappa \mathrm{d}{\tilde{\kappa}}\sum_{K{\tilde{K}}Q}u{%
\tilde{u}}\ \gamma _{\widehat{\mathbf{n}{\tilde{\mathbf{n}}}}}^{c}f(\bq)
\label{2a41} \\
& \quad \times \mathrm{tr}\left[ \Psi _{\mathbf{n}}(-K,\kappa )\Psi _{%
\mathbf{n}+\frac{\mathbf{q}_{\perp }}{2p_{F}}}\big(K+Q,\kappa \big)\right]
\notag \\
& \quad \times \mathrm{tr}\left[ \Psi _{{\tilde{\mathbf{n}}}}(-\tilde{K},%
\tilde{\kappa})\Psi _{{\tilde{\mathbf{n}}}-\frac{\mathbf{q}_{\tilde{\perp}}}{%
2p_{F}}}\big({\tilde{K}}-Q,\tilde{\kappa}\big)\right]  \notag \\
& \quad -\frac{1}{4\nu }\int \mathrm{d}\mathbf{n}\mathrm{d}{\tilde{\mathbf{n}%
}}\mathrm{d}\kappa \mathrm{d\tilde{\kappa}}\sum_{K\tilde{K}Q}u{\tilde{u}}\
\gamma _{\widehat{\mathbf{n}{\tilde{\mathbf{n}}}}}^{s}f(\bq)  \notag \\
& \quad \times \mathrm{tr}\left[ \Psi _{\mathbf{n}}(-K,\kappa )\sigma
^{k}\Psi _{\mathbf{n}+\frac{\mathbf{q}_{\perp }}{2p_{F}}}\big(K+Q,\kappa %
\big)\right]  \notag \\
& \quad \times \mathrm{tr}\left[ \Psi _{{\tilde{\mathbf{n}}}}(-{\tilde{K}},{%
\tilde{\kappa}})\sigma ^{k}\Psi _{{\tilde{\mathbf{n}}}-\frac{\mathbf{q}_{%
\tilde{\perp}}}{2p_{F}}}\big({\tilde{K}}-Q,{\tilde{\kappa}}\big)\right] \;%
\mathnormal{,}  \notag
\end{align}%
where the vector~$\mathbf{q}_{\tilde{\perp}}$ is the projection of~$\mathbf{q%
}$ onto the plane perpendicular to~${\tilde{\mathbf{n}}}$, i.e. $\mathbf{q}_{%
\tilde{\perp}}=\mathbf{q}-{\tilde{\mathbf{n}}}({\tilde{\mathbf{n}}}\!\cdot \!%
\mathbf{q})$. The amplitudes for the spin $\gamma _{\widehat{\mathbf{n}{%
\tilde{\mathbf{n}}}}}^{s}$ and charge $\gamma _{\widehat{\mathbf{n}{\tilde{%
\mathbf{n}}}}}^{c}$ channel are expressed at weak interaction in terms of
the interaction potential~$V$, Eq.~(\ref{2a07}), as
\begin{equation}
\gamma _{\widehat{\mathbf{n}{\tilde{\mathbf{n}}}}}^{s}=-\frac{\nu }{4}\tilde{%
V}(2p_{F}\sin \frac{\widehat{\mathbf{n}{\tilde{\mathbf{n}}}}}{2})\;%
\mathnormal{,}\quad \gamma _{\widehat{\mathbf{n}{\tilde{\mathbf{n}}}}}^{c}=%
\frac{\nu }{2}V_{0}+\gamma _{\widehat{\mathbf{n}{\tilde{\mathbf{n}}}}}^{s}\;%
\mathnormal{,}  \label{2a42}
\end{equation}%
respectively. If the interaction is not weak these amplitudes can be
considered as effective coupling constants of the Fermi liquid. The cutoff function~$f(\bq)$
introduced for the soft modes in Eq.~(\ref{2a03e}) is from now on written explicitly
in the formulas.
The cubic interaction~$\mathcal{S}_{3}$ reads
\begin{align}
\mathcal{S}_{3}& =\frac{1}{2\nu }\int \mathrm{d}\mathbf{n}\mathrm{d}{\tilde{%
\mathbf{n}}}\mathrm{d}\kappa \mathrm{d}{\tilde{\kappa}}\sum_{KQ}u\ \gamma _{%
\widehat{\mathbf{n}{\tilde{\mathbf{n}}}}}^{c}f(\bq)  \label{2a43} \\
& \quad \times \mathrm{tr}\left[ \Psi _{\mathbf{n}}(-K,\kappa )\Psi _{%
\mathbf{n}+\frac{\mathbf{q}_{\perp }}{2p_{F}}}\big(K+Q,\kappa \big)\right]
\notag \\
& \quad \times \left\{ -{\tilde{u}}\nu v_{F}({\tilde{\mathbf{n}}}\!\cdot \!%
\mathbf{q}){\tilde{\theta}}+2\mathrm{i}{\tilde{\theta}^{\ast }}\right\}
\mathrm{tr}\left[ \Psi _{{\tilde{\mathbf{n}}}}(-Q,\tilde{\kappa})\right]
\notag \\
& \quad +\frac{1}{2\nu }\int \mathrm{d}\mathbf{n}\mathrm{d}{\tilde{\mathbf{n}%
}}\mathrm{d}\kappa \mathrm{d}{\tilde{\kappa}}\sum_{KQ}u\ \gamma _{\widehat{%
\mathbf{n}{\tilde{\mathbf{n}}}}}^{s}f(\bq)  \notag \\
& \quad \times \mathrm{tr}\left[ \Psi _{\mathbf{n}}(-K,\kappa )\sigma
^{k}\Psi _{\mathbf{n}+\frac{\mathbf{q}_{\perp }}{2p_{F}}}\big(K+Q,\kappa %
\big)\right]  \notag \\
& \quad \times \left\{ -{\tilde{u}}\nu v_{F}({\tilde{\mathbf{n}}}\!\cdot \!%
\mathbf{q}){\tilde{\theta}}+2\mathrm{i}{\tilde{\theta}^{\ast }}\right\}
\mathrm{tr}\left[ \sigma ^{k}\Psi _{{\tilde{\mathbf{n}}}}(-Q,\tilde{\kappa})%
\right]  \notag
\end{align}%
and, finally, the quadratic action~$\mathcal{S}_{2}$ takes the form
\begin{align}
\mathcal{S}_{2}& =-\frac{1}{4\nu }\int \mathrm{d}\mathbf{n}\mathrm{d}{\tilde{%
\mathbf{n}}}\mathrm{d}\kappa \mathrm{d}{\tilde{\kappa}}\sum_{Q}\ \gamma _{%
\widehat{\mathbf{n}{\tilde{\mathbf{n}}}}}^{c}f(\bq)  \label{2a44} \\
& \quad \times \left\{ u\nu v_{F}(\mathbf{n}\!\cdot \!\mathbf{q})\theta +2%
\mathrm{i}{\theta ^{\ast }}\right\} \mathrm{tr}\left[ \Psi _{\mathbf{n}%
}(Q,\kappa )\right]  \notag \\
& \quad \times \left\{ -{\tilde{u}}\nu v_{F}({\tilde{\mathbf{n}}}\!\cdot \!%
\mathbf{q}){\tilde{\theta}}+2\mathrm{i}{\tilde{\theta}^{\ast }}\right\}
\mathrm{tr}\left[ \Psi _{{\tilde{\mathbf{n}}}}(-Q,{\tilde{\kappa}})\right]
\notag \\
& \quad -\frac{1}{4\nu }\int \mathrm{d}\mathbf{n}\mathrm{d}{\tilde{\mathbf{n}%
}}\mathrm{d}\kappa \mathrm{d}{\tilde{\kappa}}\sum_{Q}\ \gamma _{\widehat{%
\mathbf{n}{\tilde{\mathbf{n}}}}}^{s}f(\bq)  \notag \\
& \quad \times \left\{ u\nu v_{F}(\mathbf{n}\!\cdot \!\mathbf{q})\theta +2%
\mathrm{i}{\theta ^{\ast }}\right\} \mathrm{tr}\left[ \sigma ^{k}\Psi _{%
\mathbf{n}}(Q,\kappa )\right]  \notag \\
& \quad \times \left\{ -{\tilde{u}}\nu v_{F}({\tilde{\mathbf{n}}}\!\cdot \!%
\mathbf{q}){\tilde{\theta}}+2\mathrm{i}{\tilde{\theta}^{\ast }}\right\}
\mathrm{tr}\left[ \sigma ^{k}\Psi _{{\tilde{\mathbf{n}}}}(-Q,{\tilde{\kappa}}%
)\right] \;\mathnormal{.}  \notag
\end{align}%
Diagrammatically, perturbative calculations within the low energy boson
model can be conveniently represented using the building blocks shown in
Fig.~\ref{fig: blocks}. In the diagrammatic representation and explicitly in
Eqs.~(\ref{2a41}) and~(\ref{2a44}), it is evident that the building blocks
constituted by~$\mathcal{S}_{4}$ and~$\mathcal{S}_{2}$ are invariant under
vertical reflection. On the other hand, it is important to note that for~$%
d>1 $, neither~$\mathcal{S}_{4}$ nor~$\mathcal{S}_{3}$ are symmetric with
respect to horizontal flipping.

The quartic interaction~$\mathcal{S}_{4}$, Eq.~(\ref{2a41}), is fully obtained from the $\Phi$-average
of the BRST-symmetric action~$\mathcal{S}_{S}$, Eq.~(\ref{2a36a}).
Since $\mathcal{S}_{4}$ inherits this BRST symmetry, there is
no (perturbative) contribution to thermodynamics originating purely from~$\mathcal{S}_{4}$.
In contrast, the cubic and quadratic interactions~$\mathcal{S}_{3}$ and~$\mathcal{S}_{2}$, Eqs.~(\ref{2a43}) and~(\ref{2a44}),
are formed using also the symmetry breaking action~$\mathcal{S}_{B}$, Eq.~(\ref{2a36b}),
when averaging over the auxiliary field~$\Phi $. That is why diagrams contributing
to a physical thermodynamic quantity necessarily contain $\mathcal{S}_{2}$
or $\mathcal{S}_{3}$ among their building blocks.
For example, considering only the terms $\mathcal{S}_{\mathrm{bare}}+
\mathcal{S}_{2}$ we reproduce the RPA and particle-hole ladder rings because
this is equivalent to neglecting the auxiliary field~$\Phi$ in the left-hand side
of Eq.~(\ref{2a25b}), cf. the discussion at the end of Sec.~\ref%
{ssec:dynamics}. Blocks of $\mathcal{S}_{4}$ may additionally decorate
diagrams built from~$\mathcal{S}_{2}$ or $\mathcal{S}_{3}$ and the
contribution may consequently acquire logarithmic renormalizations.

In summary, Eqs.~(\ref{2a37})--(\ref{2a44})
specify our effective low energy model for the interacting Fermi gas
in $d>1$ dimensions. It is a field theory for the anticommuting
superfield~$\Psi $ which describes the bosonic excitations. The interaction
between these excitations appears as sum of the quadratic term~$\mathcal{S}_{2}$,
the cubic term~$\mathcal{S}_{3}$ and the quartic term~$\mathcal{S}_{4}$. The bare
coupling constants are written in Eq.~(\ref{2a42}). In principle, one
can immediately start perturbative studies of the model using the
contraction rule, Eq.~(\ref{2a39}), and Wick's theorem. A possible
diagrammatic representation is shown in Fig.~\ref{fig: blocks}. Although the
effective field theory may look somewhat complex, it allows to conveniently
treat the low energy limit, identifying the interesting logarithms and
summing them. This is what the next sections are devoted to.

\section{Perturbation theory}

\label{sec:renormalization}

The bosonized model, Eqs.~(\ref{2a37})--(\ref{2a44}), is not trivial and the
perturbation theory in the coupling constants $\gamma _{\widehat{\mathbf{n}{%
\tilde{\mathbf{n}}}}}^{s}$, $\gamma _{\widehat{\mathbf{n}{\tilde{\mathbf{n}}}%
}}^{c}$, Eq.~(\ref{2a42}), yields logarithmic contributions diverging in the limit $T\rightarrow 0$.
In this section, we identify the relevant classes of logarithmic one-loop diagrams.
Later in Sec.~\ref{sec:rg}, these logarithmic contributions will be summed up to infinite order by means of
a one-loop renormalization group scheme.

In one dimension, such a procedure would essentially repeat the steps from Ref.~\onlinecite{aleiner}.
The peculiarity of higher dimensions, $d>1$, appears in form of the ``rotations'' $\mathbf{n}+\mathbf{q}_{\perp }/2p_{F}$
of the angular arguments in the interacting superfields, cf. Eqs.~(\ref{2a41})--(\ref{2a44}).
Consequently, the running momentum~$Q$ in a one-loop diagram affects at the same time
the (actual) momentum~$K$ and the direction~$\bn$ of the propagators. As a result, we will find
that logarithms which certain classes of diagrams feature in $d=1$ dimension are suppressed
in dimensions~$d>1$ because of transverse fluctuations~$\mathbf{q}_{\perp }$ along the Fermi surface.
Eventually, the effects of the finite Fermi surface curvature lead to renormalization group
equations different from the ones obtained\cite{aleiner} in one dimension.

Before studying the one-loop vertex corrections,
we begin the perturbative analysis of this section as we discuss the relevant diagrams for the thermodynamic potential. These
diagrams describe physical \emph{backscattering processes}.

While the boson model, Eqs.~(\ref{2a37})--(\ref{2a44}), has been derived for an arbitrary dimension~$d$,
we consider from now on the most interesting case of a two-dimensional Fermi liquid, $d=2$.

\subsection{Backscattering diagrams}

\label{ssec:backscattering}

\begin{figure}[t]
\includegraphics{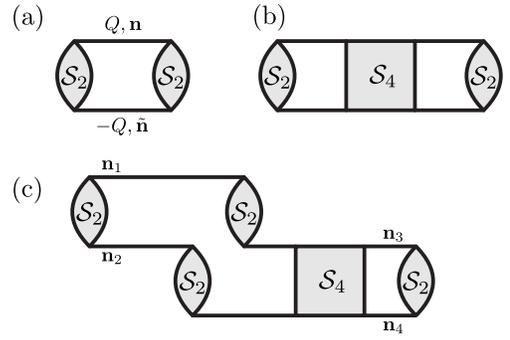}
\caption{Backscattering diagrams for the thermodynamic potential~$\Omega $:
Diagram~(a) is the second order diagram containing the leading
backscattering contribution for $\mathbf{n}\sim -{\tilde{\mathbf{n}}}$ while
diagram~(b) represents an exemplary logarithmic renormalization to
diagram~(a), cf. Sec.~\protect\ref{sssec:S2S3}. Finally, diagram~(c)
gives for $\mathbf{n}_{1}\sim -\mathbf{n}_{2}\sim -\mathbf{n}_{3}\sim
\mathbf{n}_{4}$ a backscattering contribution of higher order in the
interaction and also includes a renormalizing building block~$\mathcal{S}%
_{4} $. For weakly interacting fermions, diagram~(c) can be neglected.}
\label{fig: back}
\end{figure}
In the second order in the interaction, only diagram Fig.~\ref{fig: back}(a)
describes a contribution to the thermodynamic potential~$\Omega $
relevant for studying the backscattering effects. All other second order
diagrams cannot contain two boson propagators~$g_{\mathbf{n}}(K)$ and~$g_{%
\mathbf{\tilde{n}}}(K^{\prime })$ with $\mathbf{n}\sim -{\tilde{\mathbf{n}}}$%
. Figure~\ref{fig: back}(b) shows an exemplary diagram that renormalizes the
bare diagram Fig.~\ref{fig: back}(a) while Fig.~\ref{fig: back}(c)
represents a backscattering contribution of higher order in the interaction.
Considering the limit of weak interaction, we are safe to neglect such
higher order diagrams because they do only describe high energy renormalizations of
the coupling constants.

Working with the effective low energy theory, we have to be sure that the
main contribution to the physical quantities of interest indeed comes from
the low energies not exceeding $T$. Whether this is the case or not, it should be
checked for each quantity under investigation. In fact, the low energy contributions
are not the most important for a perturbative correction $\Delta \Omega \left(
T\right)$ to the thermodynamic potential and, thus, we cannot compute $\Delta \Omega \left(
T\right)$ using the low energy limit only. However, the main
contribution to the difference
\begin{equation}
\delta \Omega (T)=\Delta \Omega (T)-\Delta \Omega (T=0)
\label{3b000}
\end{equation}%
does come from the low energies. In order to determine such quantities as
the specific heat, the quantity $\delta \Omega \left( T\right) $ contains
all the necessary information and the low energy bosonized
model, Eqs.~(\ref{2a37})--(\ref{2a44}), becomes useful.

Formally, the quantity $\Delta \Omega \left( T\right) $ will be
represented in terms of sums over Matsubara frequencies such as $T\sum_{\omega _{n}}\psi(\omega
_{n})$. The corresponding expression for $\delta \Omega(T) $ consequently takes the form
\begin{equation}
T\sum_{\omega _{n}}\psi(\omega _{n})-\int \psi(\omega )\ \frac{\mathrm{d}\omega }{%
2\pi }\simeq \sum_{l\neq 0}\int \psi(\omega )\mathrm{e}^{-\mathrm{i}\omega l/T}%
\frac{\mathrm{d}\omega }{2\pi }\;\mathnormal{.}  \label{3b00}
\end{equation}%
Equation~(\ref{3b00}) follows from the Poisson summation formula. It shows
that, when calculating $\delta \Omega \left( T\right) $, essential $\omega $
in the function $\psi\left( \omega \right) $ are of order $T$ provided the
function $\psi\left( \omega \right) $ decays sufficiently at $|\omega| \rightarrow \infty $.

Using the developed formalism, we can start the calculation of
thermodynamic quantities. As a first example, we are going to compute the correction $\delta
\Omega ^{\left( 2\right) }\left( T\right) $ in the second order in the
coupling constants $\gamma _{\widehat{\mathbf{n}{\tilde{\mathbf{n}}}}}^{s}$ and
$\gamma _{\widehat{\mathbf{n}{\tilde{\mathbf{n}}}}}^{c}$. Logarithmic contributions are taken into account later
by replacing the bare coupling constants $\gamma _{\widehat{\mathbf{n}{\tilde{\mathbf{n}}}}}^{s}$ and $\gamma
_{\widehat{\mathbf{n}{\tilde{\mathbf{n}}}}}^{c}$ with effective amplitudes obtained
from the summation of logarithmic contributions. This computational procedure is
justified by the fact that logarithmic contributions come from energies
exceeding $T$ with the logarithms cut from below by $\max (2\pi
T,v_{F}q_{0}|\mathbf{n}+{\tilde{\mathbf{n}}}|)$. Therefore with
logarithmic accuracy, one may replace the energies in the effective
amplitudes by the temperature $T$ and treat them as constants when
calculating Matsubara sums.

The second order contribution is given by the diagram Fig.~\ref%
{fig: back}(a). Analytically, we obtain for $\Delta \Omega
^{(2)}(T)=-(T/2)\left\langle \mathcal{S}_{2}^{2}\right\rangle $ the expression
\begin{align}
& \Delta \Omega ^{(2)}(T)=-\frac{T}{16\nu ^{2}}\int \mathrm{d}\mathbf{n}%
\mathrm{d}{\tilde{\mathbf{n}}}\mathrm{d}\kappa \mathrm{d}{\tilde{\kappa}}%
\sum_{Q}f^2(\bq) \notag  \\
\times & \left\{ 4\big[\gamma _{\widehat{\mathbf{n}{\tilde{\mathbf{n}}}}}^{c}%
\big]^{2}+\big[\gamma _{\widehat{\mathbf{n}{\tilde{\mathbf{n}}}}}^{s}\big]%
^{2}\mathrm{tr}[\sigma ^{k}\sigma ^{l}]\mathrm{tr}[\sigma ^{k}\sigma
^{l}]\right\}  \notag \\
\times & \left\{ u\nu v_{F}(\mathbf{n}\!\cdot \!\mathbf{q})\theta +2\mathrm{i%
}{\theta ^{\ast }}\right\} \left\{ -u\nu v_{F}(\mathbf{n}\!\cdot \!\mathbf{q}%
)\theta +2\mathrm{i}{\theta ^{\ast }}\right\}  \notag \\
\times & \left\{ -{\tilde{u}}\nu v_{F}({\tilde{\mathbf{n}}}\!\cdot \!\mathbf{%
q}){\tilde{\theta}}+2\mathrm{i}{\tilde{\theta}^{\ast }}\right\} \left\{ {%
\tilde{u}}\nu v_{F}({\tilde{\mathbf{n}}}\!\cdot \!\mathbf{q}){\tilde{\theta}}%
+2\mathrm{i}{\tilde{\theta}^{\ast }}\right\}  \notag \\
\times & \frac{(2\mathrm{i})^{2}}{T}\ g_{\mathbf{n}}(Q)g_{{\tilde{\mathbf{n}}%
}}(-Q)\;\mathnormal{,} \label{3b001}
\end{align}%
where $g_{\mathbf{n}}(Q)$ is the bosonic Green's function, Eq.~(\ref{2a40}).
Performing the remaining integrations over the Grassmann variables, evaluating the spin
traces, and using Eq.~(\ref{3b000}), we obtain the relevant low energy second order
correction~$\delta \Omega ^{(2)}\left( T\right) $ as
\begin{align}
\delta \Omega ^{(2)}(T)& =
 \left(T\sum_\om-\int\frac{\dt\omega}{2\pi}\right)
 \int \frac{\mathrm{d}^{2}\mathbf{q}}{(2\pi )^{2}}
      \mathrm{d}\mathbf{n}\mathrm{d}{\tilde{\mathbf{n}}}\ f^2(\bq)
      \nonumber\\
 \times & \gamma _{\widehat{\mathbf{n}{\tilde{\mathbf{n}}}}}^{2}
 \frac{v_{F}(\mathbf{n}\!\cdot \!\mathbf{q})\
 v_{F}({\tilde{\mathbf{n}}}\!\cdot \!\mathbf{q})}
 {[\mathrm{i}\omega -v_{F}(\mathbf{n}\!\cdot \!\mathbf{q})]
  [-\mathrm{i}\omega +v_{F}({\tilde{\mathbf{n}}}\!\cdot \!\mathbf{q})]}\punkt  \label{3b01}
\end{align}%
The scattering amplitudes for the charge and spin channels,
$\gamma^c_{\widehat{\mathbf{n}{\tilde{\mathbf{n}}}}}$
and~$\gamma^s_{\widehat{\mathbf{n}{\tilde{\mathbf{n}}}}}$, enter Eq.~(\ref{3b01}) as
\begin{equation}
\label{3b01a}
\gamma _{\widehat{\mathbf{n}{\tilde{\mathbf{n}}}}}^{2}=
4([\gamma_{\widehat{\mathbf{n}{\tilde{\mathbf{n}}}}}^{c}]^{2}+3[\gamma _{\widehat{\mathbf{n}{%
\tilde{\mathbf{n}}}}}^{s}]^{2})\;\mathnormal{.}
\end{equation}%
The evaluation of the integral in Eq.~(\ref{3b01}) follows with some minor
deviations the steps of a similar calculation in Ref.~\onlinecite{aleiner}.
This calculation is not trivial and we present a possible route of how to
carry it out in Appendix~\ref{App_1}. As a result, we find
\begin{equation}
\delta \Omega ^{(2)}(T)=\frac{\zeta (3)}{\pi v_{F}^{2}}\ \gamma _{\pi
}^{2}T^{3}\;\mathnormal{.}  \label{3b11}
\end{equation}%
By virtue of the thermodynamic relation~$\delta c=-T\partial ^{2}\delta
\Omega /\partial T^{2}$, the anomalous correction to the specific heat~$%
\delta c$ is in the second order in the interaction obtained as
\begin{equation}
\delta c^{(2)}(T)=-\frac{6\zeta (3)}{\pi v_{F}^{2}}\ \gamma _{\pi
}^{2}T^{2}\;\mathnormal{.}  \label{3b11a}
\end{equation}%
We see that only the backscattering amplitude [$\widehat{\mathbf{n}{\tilde{%
\mathbf{n}}}}=\pi $] enters $\delta c^{(2)}(T)$, Eq.~(\ref{3b11a}).

Equation~(\ref{3b11a}) gives the well-known anomalous lowest order specific
heat contribution.\cite{chubukov1,chubukov2} It is quadratic in~$T$, which
contrasts what one would expect from the Sommerfeld expansion for the Fermi
gas of weakly-interacting quasiparticles. Thus, Eq.~(\ref{3b11a})
confirms the equivalence of the perturbative calculations in both
the conventional approach and the bosonization one which we are studying here. In
the remaining of this paper, we investigate the logarithmic renormalizations to
the backscattering contribution~$\delta c^{(2)}(T)$, Eq.~(\ref{3b11a}), and thus
refine this intermediate result.

\subsection{One-loop corrections to $\mathcal{S}_{4}$}

\label{ssec:log_diagrams}

The second order result~Eq.~(\ref{3b11}) for the thermodynamic potential
cannot provide the full qualitative picture of the non-analytic corrections
because logarithmic contributions arise in higher orders in the coupling
constants. At sufficiently low temperatures, they become large for an
arbitrarily weak interaction. In this and the next section, we study the logarithmic
divergencies in the leading one-loop order.

\begin{figure*}[t]
\includegraphics[width=\linewidth]{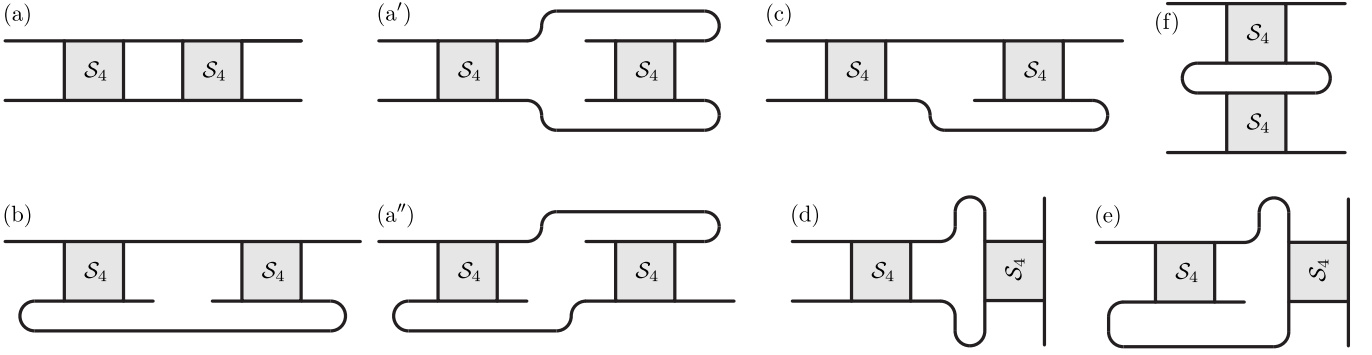}
\caption{One-loop diagrams for the quartic action~$\mathcal{S}_{4}$:
Diagram~(a) is logarithmic independently from the dimensionality
of the system whereas diagrams~(b) and~(c) are negligible in dimensions~$d>1$
due to curvature effects. Diagrams~(d) and~(e) vanish after~${\bar{q}%
_{\parallel }}$-integration for any~$d$ and, finally, diagram~(f) vanishes
due to supersymmetry. The contributions of diagrams~(a') and~(a''), each of which is logarithmic on its own, cancel each other.}
\label{fig: 1loopS4}
\end{figure*}

Considering first the quartic action~$\mathcal{S}_{4}$, Eq.~(\ref{2a41}),
the one-loop order of the expansion in the coupling constants yields the
diagrams shown in Fig.~\ref{fig: 1loopS4}. We want to show that
only diagram~(a) is important and leads to
the logarithmic divergency in the limit $T\rightarrow 0$, $\mathbf{%
n\rightarrow -\tilde{n}}$, whereas the contribution of the other diagrams remains finite in this
limit and does not contain large logarithms. For this purpose, we need to
focus on the momentum structure only while the spin structure present in $\mathcal{S}_{4}$ in the
$\gamma ^{s}$-term has nothing to say about the \emph{existence} of a logarithmic divergency.
Therefore for the sake of a simpler presentation, we only consider the $\gamma ^{c}$-term for the moment.

The logarithmic contributions come from
running bosonic frequencies with~$|\omega |\gg 2\pi T$, which are in the focus
of the following considerations. Contributions from the region~$|\omega
|\lesssim 2\pi T$ produce terms of higher order in $\gamma^{c}$ and,
therefore, only give perturbative corrections to the coefficients in front of a large logarithm.

The standard diagrammatic technique based on the contraction rule, Eq.~(\ref%
{2a39}), yields for diagram Fig.~\ref{fig: 1loopS4}(a) the one-loop vertex
correction
\begin{align}
\delta \mathcal{S}_{4}^{\text{(a)}}
 &=-\frac{1}{4\nu }\int \mathrm{d}\mathbf{n}\mathrm{d}{\tilde{\mathbf{n}}}\mathrm{d}\kappa \mathrm{d}{\tilde{\kappa}}u{\tilde{u}}%
 \sum_{KQ_{1}Q_{2}} \frac{-\delta\gamma_{\widehat{\mathbf{n}{\tilde{\mathbf{n}}}}}^{c}}{2}
                              \nonumber \\
\quad \times \mathrm{tr}&\left[ \Psi _{\mathbf{n}+\frac{\mathbf{q}_{1,\perp
}}{2p_{F}}}(-K-Q_{1},\kappa )\Psi _{\mathbf{n}+\frac{\mathbf{q}_{2,\perp }}{%
2p_{F}}}\big(K+Q_{2},\kappa \big)\right]  \notag \\
\quad \times \mathrm{tr}&\left[ \Psi _{{\tilde{\mathbf{n}}}-\frac{\mathbf{q}%
_{1,{\tilde{\perp}}}}{2p_{F}}}(-K+Q_{1},\tilde{\kappa})\Psi _{{\tilde{%
\mathbf{n}}}-\frac{\mathbf{q}_{2,{\tilde{\perp}}}}{2p_{F}}}\big(K-Q_{2},{%
\tilde{\kappa}}\big)\right] \;\mathnormal{.}  \label{3c01}
\end{align}%
This vertex~$\delta \mathcal{S}_{4}^{\text{(a)}}$ reproduces the analytical
form of the quartic action~$\mathcal{S}_{4}$, Eq.~(\ref{2a41}), where a
correction $-\delta\gamma_{\widehat{\mathbf{n}{\tilde{\mathbf{n}}}}}^{c}/2$ should now be
added to the bare quantity~$\gamma _{\widehat{\mathbf{n}{\tilde{\mathbf{n}}}}}^{c}f(\bq_{1}\!-\!\bq_{2})$. This correction --- a function of~$u$, ${\tilde{u}}$, $\mathbf{n}$%
, ${\tilde{\mathbf{n}}}$, $\bq_{1}\!-\!\bq_{2}$, and~$K$ --- is determined by the integral over the
running four-momentum.

We find
\begin{align}
\delta\gamma_{\widehat{\mathbf{n}{\tilde{\mathbf{n}}}}}^{c} &=
  \frac{4u{\tilde{u}}}{\nu}\ [\gamma_{\widehat{\mathbf{n}{\tilde{\mathbf{n}}}}}^{c}]^2
  \sum_{Q} f(\bq\!-\!\bq_1)f(\bq\!-\!\bq_2)\nonumber\\
 &\quad\times g_{\mathbf{n}+\frac{\mathbf{q}_{\perp }}{2p_{F}}}\big(K+Q\big)g_{{%
   \tilde{\mathbf{n}}}-\frac{\mathbf{q}_{\tilde{\perp}}}{2p_{F}}}\big(K-Q\big)\punkt  \label{3c02}
\end{align}%
This integral is conveniently calculated introducing the angular variables
\begin{align}
{\bar{\mathbf{n}}}& =\frac{1}{2}(\mathbf{n}-{\tilde{\mathbf{n}}})\;%
\mathnormal{,}  \label{3c03} \\
\delta \mathbf{n}& =\frac{1}{2}(\mathbf{n}+{\tilde{\mathbf{n}}})\;%
\mathnormal{,}  \notag
\end{align}%
and their projections of the momentum vector~$\mathbf{q}$,
\begin{align}
{\bar{q}_{\parallel}}& =({\bar{\mathbf{n}}}\!\cdot \!\mathbf{q})\;%
\mathnormal{,}  \label{3c04} \\
{\bar{\mathbf{q}}_{\perp}}& =\mathbf{q}-{\bar{q}_{\parallel }}{\bar{\mathbf{%
n}}}\;\mathnormal{.}  \notag
\end{align}%
Calculating the integral in Eq.~(\ref{3c02}) with logarithmic accuracy
and keeping in mind that the essential $K$ in the final integration, e.g. in
the diagram in Fig.~\ref{fig: back}(b), will be of order $T$, which is the
lower cutoff of the logarithms, we can safely put $K=0$ in Eq.~(\ref{3c02}).
Thus, $\delta\gamma_{\widehat{\mathbf{n}{\tilde{\mathbf{n}}}}}^{c}$ can be written as
\begin{align}
\delta\gamma_{\widehat{\mathbf{n}{\tilde{\mathbf{n}}}}}^{c}
 =&\frac{4u{\tilde{u}}}{\nu }\ [\gamma_{\widehat{\mathbf{n}{\tilde{\mathbf{n}}}}}^{c}]^2\ \int\frac{\mathrm{d}{\bar{\mathbf{q}}_{\perp }}}{2\pi}
   f(\bqbperp\!-\!\bq_{1})f(\bq_{2}\!-\!\bqbperp)\nonumber \\
\times\  T\sum_{\omega } &\int \frac{\mathrm{d}{\bar{q}_{\parallel }}}{2\pi } \frac{1}{\mathrm{i}\omega -v_{F}{\bar{q}_{\parallel }}-v_{F}(\delta
\mathbf{n}\!\cdot \!{\bar{\mathbf{q}}_{\perp }})-\frac{1}{2m}{\bar{\mathbf{q}%
}_{\perp }}^{2}}  \notag \\
\times & \frac{1}{-\mathrm{i}\omega -v_{F}{\bar{q}_{\parallel }}%
+v_{F}(\delta \mathbf{n}\!\cdot \!{\bar{\mathbf{q}}_{\perp }})-\frac{1}{2m}{%
\bar{\mathbf{q}}_{\perp }}^{2}}\;\mathnormal{.}  \label{3c05}
\end{align}%
As to the radial component~${\bar{q}_{\parallel }}$, it will be sufficient
to know the cutoff's order of magnitude while the precise form of the cutoff
function~$f(\bq)$ is irrelevant to it. As a result, it is justified to restrict
its dependence to the transverse momenta~$\bqbperp$ --- as has been done in Eq.~(\ref{3c05}) --- while keeping in mind
that~$v_F{\bar{q}_{\parallel}}\lesssim \Lambda$ with~$\Lambda$ being the upper boundary of
the bosonic spectrum.

Integrating over the radial component~${\bar{q}_{\parallel }}$, we find
\begin{align}
\delta\gamma_{\widehat{\mathbf{n}{\tilde{\mathbf{n}}}}}^{c}
&=
 [\gamma_{\widehat{\mathbf{n}{\tilde{\mathbf{n}}}}}^{c}]^2
  \int\frac{\mathrm{d}{\bar{\mathbf{q}}_{\perp }}}{2q_0}
   f(\bqbperp\!-\!\bq_{1})f(\bq_{2}\!-\!\bqbperp)\ \lambda _{\widehat{\mathbf{n}{\tilde{\mathbf{n}}}}}\label{3c06a}
\end{align}
with
\begin{align}
\lambda_{\widehat{\mathbf{n}{\tilde{\mathbf{n}}}}} &=
 \frac{4u{\tilde{u}}}{\nu v_{F}}\frac{q_0}{\pi}\
 \int_{2\pi T}^{\Lambda }
 \frac{\mathrm{d}\omega }{2\pi }
 \frac{\omega}{\omega^2 -[v_{F}(\delta \mathbf{n}\!\cdot \!{\bar{\mathbf{q}}_{\perp }})]^2}\punkt  \label{3c06}
\end{align}
Remarkably after the integration over ${\bar{q}_{\parallel }}$, the term~${%
\bar{\mathbf{q}}_{\perp }}^{2}/(2m)$, which is of order~$\Lambda$ and, thus, in principle large, has dropped
out. Therefore, the subsequent integration over the frequencies~$\omega $
leads in the limit $T,\left\vert \delta \mathbf{n}\right\vert \rightarrow 0$
to a logarithmic divergency. The important observation is that the divergency of
diagram Fig.~\ref{fig: 1loopS4}(a) is not sensitive to the question whether
we are in dimension~$d=1$ or~$d>1$ and thus the logarithm appears in \emph{%
any} dimension~$d$. The term~$v_{F}(\delta \mathbf{n}\!\cdot \!{\bar{\mathbf{%
q}}_{\perp }})\sim v_{F}q_{0}|\delta \mathbf{n}|$, though being dependent on
the transverse momenta, vanishes in the limit $\delta \mathbf{n\rightarrow }%
0$ independently from~$d$. This however is exactly the limit
in which~$\lambda _{\widehat{\mathbf{n}{\tilde{\mathbf{n}}}}}$ eventually
enters physical quantities such as the thermodynamic potential~$\delta\Omega $,
Eq.~(\ref{3b11}). Explicitly, we find
\begin{align}
\lambda _{\widehat{\mathbf{n}{\tilde{\mathbf{n}}}}}&= 4u{\tilde{u}}\
\frac{\nu ^{\ast }}{\nu }\ln \left( \frac{\Lambda }{\max (2\pi
T,v_{F}q_{0}|\delta \mathbf{n}|)}\right)\label{3c07}
\end{align}
with the constant~$\nu ^{\ast }$ defined as
\begin{equation}
\nu ^{\ast }= \frac{1}{2\pi ^{2}}\frac{q_{0}}{v_{F}}\punkt
\label{3c08}
\end{equation}%
By construction, the transverse momentum only
varies on a small arc with a length of order $2q_0\ll p_F$ on the Fermi circle, but in final results this arc should extend
to a semicircle, corresponding to~$q_{0}\sim (\pi/2) p_F$ or~$\nu^*\sim \nu/2$.

Still the integration over~$\bar{\mathbf{q}}_{\perp }$ in Eq.~(\ref{3c06a}) remains to be done.
Since $v_F \bar{q}_{\parallel } \sim {\bar{\mathbf{q}}_{\perp }}^{2}/(2m)\sim\Lambda $ in the region of the logarithm,
the transverse momenta are much larger than the parallel ones, $\bar{\mathbf{q}}_{\perp }\gg\bar{q}_{\parallel }$.
As this statement remains true for the leading terms in all logarithmic orders, we may neglect in the relevant limit of
$\delta \mathbf{n\rightarrow }0$ the parallel momenta~$\bar{q}_{1,2;\parallel}$ in the cutoff functions in Eq.~(\ref{3c06a}).
Then, the remaining integral becomes
\begin{align}
\label{3c08a0}
\int\frac{\mathrm{d}{\bar{\mathbf{q}}_{\perp }}}{2q_0}
   f(\bqbperp\!-\!\bq_{1,\perp})f(\bq_{2,\perp}\!-\!\bqbperp)\punkt
\end{align}
This is nothing but a convolution $[f\!\ast\! f](\bq_{2,\perp}\!-\!\bq_{1,\perp})$, which becomes
a product after employing a Fourier transformation,
\begin{align}
\label{3c08a}
f(\brperp) &= \int f(\bqperp)\ \ee^{\ii\brperp\bqperp/q_0}\frac{\dt\bqperp}{2q_0}\punkt
\end{align}
The form of the Fourier transform in Eq.~(\ref{3c08a}) has been chosen such that both~$\brperp$ and~$f(\brperp)$
are dimensionless. The value $\brperp = 1$ corresponds to the minimal length of the theory, which is given by~$1/q_0$.
In Fourier representation, the vertex correction~$\delta\gamma_{\widehat{\mathbf{n}{\tilde{\mathbf{n}}}}}^{c} $
takes the final form
\begin{align}
\delta\gamma_{\widehat{\mathbf{n}{\tilde{\mathbf{n}}}}}^{c}(\brperp)&=
 [\gamma_{\widehat{\mathbf{n}{\tilde{\mathbf{n}}}}}^{c} f(\brperp)]^2\ \lambda _{\widehat{\mathbf{n}{\tilde{\mathbf{n}}}}}
\label{3c08b}
\end{align}
with the function~$\lambda _{\widehat{\mathbf{n}{\tilde{\mathbf{n}}}}}$, Eq.~(\ref{3c07}), containing the logarithm.
From Eq.~(\ref{3c08b}), we understand that it is actually the quantity $\gamma _{\widehat{\mathbf{n}{\tilde{\mathbf{n}}
}}}^{c}f(\brperp)$ which flows during a renormalization procedure.

In conclusion, the diagram Fig.~\ref{fig: 1loopS4}(a) logarithmically renormalizes the
backward scattering amplitude~$\gamma _{\pi }^{c}$ of the quartic action~$%
\mathcal{S}_{4}$ and this logarithmic contribution comes independently of
the dimension~$d$. Diagram~(a) corresponds in conventional fermion
diagrammatics to a rung of the particle-particle ladder. We discuss this correspondence
in Appendix~\ref{App_3}.

Let us now turn our attention to diagram Fig.~\ref{fig: 1loopS4}(b). After the
integration over the internal momenta and frequencies, this diagram also
reproduces the structure of~$\mathcal{S}_{4}$ but in contrast to diagram~(a),
it is not logarithmic for dimensions~$d>1$ . In order to support this
statement, we consider the vertex correction~$\delta \mathcal{S}_{4}^{\text{%
(b)}}$, which has the same form as~$\delta \mathcal{S}_{4}^{\text{(a)}}$,
Eq.~(\ref{3c01}), except that $\lambda _{\widehat{\mathbf{n}{\tilde{\mathbf{n}
}}}}$, Eq.~(\ref{3c06}), is replaced by the function
\begin{align}
& \lambda _{\widehat{\mathbf{n}{\tilde{\mathbf{n}}}}}^{\text{(b)}}\propto
\sum_{Q}g_{\mathbf{n}+\frac{\mathbf{q}_{\perp }}{2p_{F}}}\big(K+Q\big)
\label{3c09} \\
\times & g_{{\tilde{\mathbf{n}}}+\frac{\mathbf{q}_{\tilde{\perp}}}{2p_{F}}-%
\frac{\mathbf{q}_{1,{\tilde{\perp}}}+\mathbf{q}_{2,{\tilde{\perp}}}}{2p_{F}}}%
\big(K+Q-(Q_{1}+Q_{2})\big)\;\mathnormal{.}  \notag
\end{align}%
In order to estimate the integral over $Q$ in Eq.~(\ref{3c09}), we put all
external momenta and frequencies equal to zero. Also, the precise form of the cutoff functions~$f(\bq)$
is not needed for this estimate and therefore, we do not write them here for simplicity.
Then, in the frame of the angular coordinates from Eqs.~(\ref{3c03}) and~(\ref{3c04}) we obtain
\begin{align}
& \lambda _{\widehat{\mathbf{n}{\tilde{\mathbf{n}}}}}^{\text{(b)}}\propto
\sum_{Q}g_{\mathbf{n}+\frac{\mathbf{q}_{\perp }}{2p_{F}}}\big(Q\big)g_{{%
\tilde{\mathbf{n}}}+\frac{\mathbf{q}_{\tilde{\perp}}}{2p_{F}}}\big(Q\big)
\label{3c10} \\
& \simeq T\sum_{\omega }\!\int \frac{\mathrm{d}{\bar{q}_{\parallel }}}{2\pi }%
\frac{\mathrm{d}{\bar{\mathbf{q}}_{\perp }}}{2\pi }\frac{1}{\mathrm{i}\omega
\!-\!v_{F}{\bar{q}_{\parallel }}\!-\!\frac{1}{2m}{\bar{\mathbf{q}}_{\perp
}^{2}}}\frac{1}{\mathrm{i}\omega \!+\!v_{F}{\bar{q}_{\parallel }}\!-\!\frac{1%
}{2m}{\bar{\mathbf{q}}_{\perp }^{2}}}\;\mathnormal{.}  \notag
\end{align}%
Similar to the case of diagram Fig.~\ref{fig: 1loopS4}(a), the unit vectors~$%
\mathbf{n}$ and~${\tilde{\mathbf{n}}}$ need to be close to anticollinearity
if we want to achieve the largest value of the integral. Passing from the first to the second line in Eq.~(\ref{3c10}),
this has already been assumed. Integrating over~${\bar{q}_{\parallel }}$ yields
\begin{equation}
\lambda _{\widehat{\mathbf{n}{\tilde{\mathbf{n}}}}}^{\text{(b)}}\propto
T\sum_{\omega }\int \frac{1}{|\omega |+\mathrm{i}\ \frac{\mathrm{sgn}\omega
}{2m}{\bar{\mathbf{q}}_{\perp }^{2}}}\frac{\mathrm{d}{\bar{\mathbf{q}}%
_{\perp }}}{2\pi }\;\mathnormal{.}  \label{3c11}
\end{equation}%
This intermediate result for~$\lambda _{\widehat{\mathbf{n}{\tilde{\mathbf{n}%
}}}}^{\text{(b)}}$ already demonstrates what makes diagram Fig.~\ref{fig:
1loopS4}(b) essentially different from diagram~(a): Here, the transverse term~${\bar{\mathbf{q}}_{\perp }}%
^{2}/(2m) $ does not drop out. Moreover, nothing prevents the momentum $\bar{%
\mathbf{q}}$ from being large and the energy ${\bar{\mathbf{q}}_{\perp }}%
^{2}/(2m) $ from taking values of the order of the cutoff~$\Lambda$.
As a result, logarithms analogous to those that appeared from the
diagram Fig.~\ref{fig: 1loopS4}(a) are suppressed by the presence of the
transverse term. Since the latter exists for dimensions~$d>1$ only, the
vanishing of the divergent contribution of diagram~(b) is
clearly due to the higher dimensionality of the system. In one dimension, diagram~(b)
would give the same logarithmic contribution as
diagram~(a). Thus, one can interpret the cancellation of the diagram's
logarithmic contribution as an effect of
the finite curvature of the Fermi surface in~$d>1$.

Studying the correspondence of diagram Fig.~\ref{fig: 1loopS4}(b) to the
conventional fermion diagrammatic technique --- cf. the discussion of Fig.~\ref{fig: 1loopS4_conv} in Appendix~\ref{App_3} ---, it is possible to identify
this contribution with the particle-hole ladder and polarization bubble
diagrams. As is well-known, both these contributions are logarithmic in~$d=1$%
, but not in higher dimensions~$d>1$.

Diagrams Fig.~\ref{fig: 1loopS4}(d) and (e) trivially vanish because the
angular variables~$\mathbf{n}$ in both the internal propagators are
necessarily close to each other. Therefore, the integration contour for ${%
\bar{q}_{\parallel }}$ can be closed without residues inside and the
integral equals zero. Diagram~(f) contains a closed loop of bosonic propagators. Since this diagrammatic substructure is
fully supersymmetric in the sense of Eq.~(\ref{2a36c}), the contribution of the entire diagram vanishes.

For the study of the remaining diagrams, the discussions of the diagrams
Fig.~\ref{fig: 1loopS4}(a) and~(b) can rather easily be extended. The
reasons why a logarithmic divergency appears in diagrams Fig.~\ref{fig:
1loopS4}(a') and~(a'') while it is suppressed in diagram~(c)
in dimensions~$d>1$ are similar to those used for diagrams~(a) and~(b).

Considering the diagrams Fig.~\ref{fig: 1loopS4}(a'), (a''), and (c), one encounters
an, at first glance, very unpleasant property: Neither of them reproduces
the formal structure of~$\mathcal{S}_{4}$ for fixed external momenta.
Once more neglecting the cutoff functions for the moment, we obtain for~(a') and~(c)
the analytical expressions
\begin{align}
& \delta \mathcal{S}_{4}^{\text{(a')}}=-\frac{1}{4\nu }\int \mathrm{d}%
\mathbf{n}\mathrm{d}{\tilde{\mathbf{n}}}\mathrm{d}\kappa \mathrm{d}{\tilde{%
\kappa}}\sum_{KQ_{1}Q_{2}}u{\tilde{u}}\int \frac{\mathrm{d}\mathbf{q}_{%
\tilde{\perp}}}{4\pi q_{0}}\frac{-\big[\gamma _{\widehat{\mathbf{n}{\tilde{%
\mathbf{n}}}}}^{c}\big]^{2}}{2}\lambda _{\widehat{\mathbf{n}{\tilde{\mathbf{n%
}}}}}  \notag \\
& \times \mathrm{tr}\left[ \Psi _{\mathbf{n}+\frac{\mathbf{q}_{1,\perp }}{%
2p_{F}}}(-K-Q_{1},\kappa )\Psi _{\mathbf{n}-\frac{\mathbf{q}_{2,\perp }}{%
2p_{F}}+\frac{\mathbf{q}_{{\tilde{\perp}}}}{p_{F}}}\big(K+Q_{2},\kappa \big)%
\right]  \notag \\
& \times \mathrm{tr}\left[ \Psi _{{\tilde{\mathbf{n}}}-\frac{\mathbf{q}_{1,{%
\tilde{\perp}}}}{2p_{F}}}(-K+Q_{1},\tilde{\kappa})\Psi _{{\tilde{\mathbf{n}}}%
+\frac{\mathbf{q}_{2,{\tilde{\perp}}}}{2p_{F}}-\frac{\mathbf{q}_{{\tilde{%
\perp}}}}{p_{F}}}\big(K-Q_{2},{\tilde{\kappa}}\big)\right] \label{3c12}
\end{align}
and
\begin{align}
& \delta \mathcal{S}_{4}^{\text{(c)}}=-\frac{1}{4\nu }\int \mathrm{d}%
\mathbf{n}\mathrm{d}{\tilde{\mathbf{n}}}\mathrm{d}\kappa \mathrm{d}{\tilde{%
\kappa}}\sum_{KQ_{1}Q_{2}}u{\tilde{u}}\int \frac{\mathrm{d}\mathbf{q}_{%
\tilde{\perp}}}{4\pi q_{0}}\frac{-\big[\gamma _{\widehat{\mathbf{n}{\tilde{%
\mathbf{n}}}}}^{c}\big]^{2}}{2}\lambda _{\widehat{\mathbf{n}{\tilde{\mathbf{n%
}}}}}  \notag \\
& \times \mathrm{tr}\left[ \Psi _{\mathbf{n}+\frac{\mathbf{q}_{1,\perp }}{%
2p_{F}}}(-K-Q_{1},\kappa )\Psi _{\mathbf{n}-\frac{\mathbf{q}_{2,\perp }}{%
2p_{F}}}\big(K+Q_{2},\kappa \big)\right]  \notag \\
& \times \mathrm{tr}\left[ \Psi _{{\tilde{\mathbf{n}}}-\frac{\mathbf{q}_{1,{%
\tilde{\perp}}}}{2p_{F}}}(-K+Q_{1},{\tilde{\kappa}})\Psi _{{\tilde{\mathbf{n}%
}}+\frac{\mathbf{q}_{2,{\tilde{\perp}}}}{2p_{F}}-\frac{\mathbf{q}_{{\tilde{%
\perp}}}}{p_{F}}}\big(K-Q_{2},{\tilde{\kappa}}\big)\right] \;\mathnormal{.}
 \label{3c13}
\end{align}%
The contributions $\delta \mathcal{S}_{4}^{\text{(a')}}$ and~$\delta \mathcal{S}%
_{4}^{\text{(c)}}$, Eqs.~(\ref{3c12}) and~(\ref{3c13}), contain the logarithmic function $\lambda _{\widehat{\mathbf{n}{\tilde{%
\mathbf{n}}}}}$, Eq.~(\ref{3c07}). This function~$\lambda _{\widehat{\mathbf{%
n}{\tilde{\mathbf{n}}}}}$ is sensitive to deviations of $\mathbf{n}$ from $-{%
\tilde{\mathbf{n}}}$ with $\mathbf{n}\sim -\tilde{\mathbf{n}}$ being the only region of importance in our
consideration. The structure of the action~$\mathcal{S}_{4}$ is
formally not reproduced by $\delta \mathcal{S}_{4}^{\text{(a')}}$ and~$%
\delta \mathcal{S}_{4}^{\text{(c)}}$ because of the presence of the momentum
$\mathbf{q}_{{\tilde{\perp}}}$ in Eqs.~(\ref{3c12}) and~(\ref{3c13}).

However, one can easily see that the momentum $\mathbf{q}_{{\tilde{\perp}}}$
enters the vertex corrections~$\delta \mathcal{S}_{4}^{\text{(a')}}$ and~$%
\delta \mathcal{S}_{4}^{\text{(c)}}$ quite differently. As concerns~$\delta
\mathcal{S}_{4}^{\text{(a')}}$, the additional rotation of the vectors $%
\mathbf{n}$ and $\mathbf{\tilde{n}}$ represented by~$\mathbf{q}_{{\tilde{%
\perp}}}$ does not change the direction of~$\mathbf{n}$ and~${\tilde{\mathbf{%
n}}}$ with respect to each other. In other words, if $\mathbf{n}=-{\tilde{%
\mathbf{n}}}$, then $\mathbf{n}+\mathbf{q}_{{\tilde{\perp}}}/p_{F}=-[{\tilde{%
\mathbf{n}}}-\mathbf{q}_{{\tilde{\perp}}}/p_{F}]+\mathcal{O}[(\mathbf{q}_{{%
\tilde{\perp}}}/p_{F})^{2}]$. This makes the presence of the momentum $%
\mathbf{q}_{{\tilde{\perp}}}$ in $\delta \mathcal{S}_{4}^{\text{(a')}}$
unimportant. One should simply keep in mind that eventually we are to calculate the
thermodynamic potential correction~$\delta \Omega $. Here, $\delta \mathcal{S}_{4}^{%
\text{(a')}}$ can enter in the leading order correction as in Fig.~\ref{fig: back}(b)
or be a part of a larger ladder containing other $\mathcal{S}_{4}$%
-blocks. It will turn out that the terms including $\mathbf{q}_{{\tilde{\perp}}}$ drop out after the integration
over the parallel momenta~$\bar{q}_{\parallel}$ in the additional loops.
Formally, it is therefore legitimate in such diagrams to simply put $\mathbf{q}_{{\tilde{\perp}%
}}=0$, thus making the diagram~(a') give
the same logarithmic contribution as $\delta \mathcal{S}_{4}^{\text{(a)}}$, Eq.~(\ref{3c01}).

In contrast to~$\delta \mathcal{S}_{4}^{\text{(a')}}$, the vertex~$\delta \mathcal{S}_{4}^{\text{(c)}}$ is entered by~$%
\mathbf{q}_{{\tilde{\perp}}}$ in an asymmetric way changing the mutual
direction of~$\mathbf{n}$ and~${\tilde{\mathbf{n}}}$. As a consequence, if
we attach another~$\mathcal{S}_{4}$-block to the right of~$\delta \mathcal{S}%
_{4}^{\text{(c)}}$, a curvature term of order~$\Lambda $ containing $\mathbf{%
q}_{{\tilde{\perp}}}$ will necessarily survive the parallel momentum
integration, resulting similarly to the scenario of diagram~(b) in the
cancellation of the otherwise emerging logarithm. Inserting~$\delta \mathcal{S%
}_{4}^{\text{(c)}}$ into the perturbation series for the thermodynamic potential, Fig.~%
\ref{fig: back}(b), smears --- due to the presence of~$\mathbf{q}_{{\tilde{%
\perp}}}$ in \emph{one} of the propagators --- the important region around $%
\mathbf{n}=-{\tilde{\mathbf{n}}}$. The expression of the form Eq.~(\ref{3b01})
in Sec.~\ref{ssec:backscattering} will accordingly be no longer
sufficiently sensitive to the backscattering limit, thus resulting in the
suppression of all backscattering logarithms. For these reasons, the vertex~$\delta \mathcal{S}_{4}^{\text{(c)}}$ can be excluded from the class of the important one-loop diagrams.

The final one-loop contribution to be dealt with is the diagram Fig.~\ref{fig: 1loopS4}(a'').
Its evaluation at small external momenta yields the exactly same analytical form as~$\delta\cS^{\text{(a')}}$, Eq.~(\ref{3c12}),
but due to a necessary transposition of the Grassmann fields \emph{with opposite sign},
\begin{align}
\delta\cS_4^{\text{(a'')}}=-\delta\cS_4^{\text{(a')}}\punkt
\end{align}
Thus renormalizing the quartic action using the renormalization group, diagrams~(a') and~(a'') cancel each other. One straightforwardly
checks that this cancellation still prevails when we consider both charge and spin channel of the quartic action~$\cS_4$, Eq.~(\ref{2a41}), at
the same time.

As a result, the first-loop analysis of the quartic action~$\cS_4$ clearly
demonstrates the existence of logarithmic divergencies arising due to the interaction
of the collective excitations of the Fermi gas. These divergencies originate
from just one of the various one-loop diagrams shown Fig.~\ref{fig: 1loopS4}, namely diagram~(a).

\subsection{One-loop corrections to $\mathcal{S}_{2}$ and~$\mathcal{S}_{3}$}

\label{sssec:S2S3}

In the previous section, we have identified the relevant logarithmic one-loop corrections to the quartic interaction
$\mathcal{S}_{4}$. In a general leading logarithmic diagram of order~$\ln^{n}(\Lambda /T)$ with~$n$ being a large integer,
nearly all logarithmic factors will be due to $\cS_4\cS_4$-loops. The quadratic and cubic parts~$\mathcal{S}_{2}$ and~$\mathcal{S}_{3}$ of the
action, Eqs.~(\ref{2a44}) and~(\ref{2a43}), that are needed to break the
BRST symmetry in diagrams for a thermodynamic quantity, serve as the
``abutments'' of the big $\cS_4\cS_4$-loop structure.

\begin{figure}[t]
\includegraphics{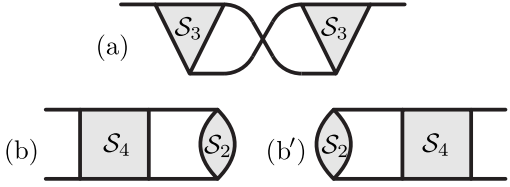}
\caption{Logarithmic one-loop corrections to the quadratic action~$\mathcal{S%
}_{2}$, Eq.~(\protect\ref{2a44}). }
\label{fig: 1loopS2}
\end{figure}
In principle, the analysis of the one-loop diagrams for the
quadratic and cubic vertices is very similar to that performed for~$\mathcal{S}_{4}$
in Sec.~\ref{ssec:log_diagrams}. Let us begin with the quadratic action~$\cS_2$: One-loop
corrections to~$\mathcal{S}_{2}$ come from diagrams built from~$\cS_3\cS_3$ and $\mathcal{S}_{2}\mathcal{S}_{4}$.
As in the renormalization
of the quartic action, most diagrams are negligible since they vanish due to
supersymmetry [like diagram Fig.~\ref{fig: 1loopS4}(f)], as a result of integration over ${\bar{q}_{\parallel }}$ in
cases when the integrand is an odd function of ${\bar{q}_{\parallel }}$ [like diagram Fig.~\ref{fig: 1loopS4}(d)], or due to
higher-dimensional curvature effects as in the case of diagrams Fig.~\ref{fig: 1loopS4}(b) and~(c).
The only diagrams yielding in fact a logarithmic
contribution proportional to $\lambda _{\widehat{\mathbf{n}{\tilde{\mathbf{n}}}}}$, Eq.~(\ref{3c07}), in dimensions~$d>1$ are those shown in Fig.~\ref{fig: 1loopS2}.

\begin{figure}[t]
\includegraphics{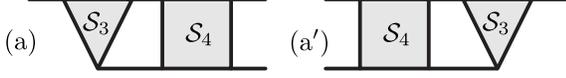}
\caption{Diagrams (a) and (a') represent the logarithmic one-loop corrections to
the cubic action~$\mathcal{S}_{3}$, Eq.~(\protect\ref{2a43}).}
\label{fig: 1loopS3}
\end{figure}
Corrections to the cubic action~$\mathcal{S}_{3}$ appear in form of the one-loop
diagrams built from~$\mathcal{S}_{3}\mathcal{S}_{4}$. Logarithmic
renormalizations come from the diagrams shown in Fig.~\ref{fig: 1loopS3}(a) and~(a').

One-loop diagrams different from these two do not contribute logarithmically.
This is once more a consequence of symmetry aspects and curvature effects. Discussing
the cubic one-loop vertices, one should also bear in mind that they finally
affect the thermodynamic potential only via the effective quadratic vertex in Fig.~\ref{fig: 1loopS2}(a).
Since the arguments follow the same reasoning as for the quartic interaction in Sec.~\ref{ssec:log_diagrams},
we refrain from an explicit discussion.

\section{Renormalization group}

\label{sec:rg}

With the perturbative analysis from the preceding section, all the relevant logarithmic one-loop diagrams are at hand
and we are ready to apply a one-loop renormalization group (RG) scheme. At the end of the day,
the energy scales above~$T$ will be integrated out of the field theory and we will obtain
the specific
heat in terms of the basic backscattering diagram Fig.~\ref{fig: back}(a) with renormalized coupling constants.

It is important to mention here that the renormalization of the quartic
term~$\cS_4$ differs from the renormalizations of~$\cS_3$ and~$\cS_2$. The former
can be obtained both using the RG scheme and summing ladder diagrams, while
the latter ones do not allow for a study based on simple summations of ladder diagrams.

\subsection{Generalized action}

The actions~$\cS_4$, $\cS_3$, and $\cS_2$, Eqs.~(\ref{2a41})--(\ref{2a44}), contain various subterms which in general
have a specific flow behavior under the RG action. In order to facilitate the RG procedure,
we generalize the action a priori and introduce proper coupling constants. As a result, we write the quartic interaction as
\begin{align}
\mathcal{S}_{4}& =-\frac{1}{4\nu }\int \mathrm{d}\mathbf{n}\mathrm{d}{\tilde{%
\mathbf{n}}}\mathrm{d}\kappa \mathrm{d}{\tilde{\kappa}}\sum_{K{\tilde{K}}Q}u{%
\tilde{u}}\ \Gamma^{c}(\bqbperp) \label{5a01}\\
  &\quad \times \mathrm{tr}\left[ \Psi_{\mathbf{n}}(-K,\kappa )\Psi _{%
\mathbf{n}+\frac{\mathbf{q}_{\perp }}{2p_{F}}}\big(K+Q,\kappa \big)\right]
  \notag \\
 &\quad \times \mathrm{tr}\left[ \Psi _{{\tilde{\mathbf{n}}}}(-\tilde{K},%
\tilde{\kappa})\Psi_{{\tilde{\mathbf{n}}}-\frac{\mathbf{q}_{\tilde{\perp}}}{%
2p_{F}}}\big({\tilde{K}}-Q,\tilde{\kappa}\big)\right]  \notag \\
& \quad -\frac{1}{4\nu }\int \mathrm{d}\mathbf{n}\mathrm{d}{\tilde{\mathbf{n}%
}}\mathrm{d}\kappa \mathrm{d\tilde{\kappa}}\sum_{K\tilde{K}Q}u{\tilde{u}}\
\Gamma^{s}(\bqbperp)  \notag \\
& \quad \times \mathrm{tr}\left[
 \Psi _{\mathbf{n}}(-K,\kappa )\sigma^{k}\Psi_{\mathbf{n}+\frac{\mathbf{q}_{\perp }}{2p_{F}}}\big(K+Q,\kappa\big)\right]  \notag \\
& \quad \times \mathrm{tr}\left[ \Psi _{{\tilde{\mathbf{n}}}}(-{\tilde{K}},{\tilde{\kappa}})\sigma ^{k}\Psi _{{\tilde{\mathbf{n}}}
  -\frac{\mathbf{q}_{\tilde{\perp}}}{2p_{F}}}\big({\tilde{K}}-Q,{\tilde{\kappa}}\big)\right] \;
  \mathnormal{,}  \notag
\end{align}
the cubic action in the form
\begin{align}
\mathcal{S}_{3}&=\frac{1}{2\nu }\int \mathrm{d}\mathbf{n}\mathrm{d}{\tilde{%
\mathbf{n}}}\mathrm{d}\kappa \mathrm{d}{\tilde{\kappa}}\sum_{KQ}u\label{5a02} \\
& \, \times \mathrm{tr}\left[ \Psi _{\mathbf{n}}(-K,\kappa )\Psi _{%
\mathbf{n}+\frac{\mathbf{q}_{\perp }}{2p_{F}}}\big(K+Q,\kappa \big)\right]
\notag \\
& \, \times \left\{{\tilde{u}}\nu v_{F}{\tilde{\theta}}\
\delta\bn\!\cdot\!\B_\perp^c(\bqbperp)+2\mathrm{i}{\tilde{\theta}^{\ast }}\ \B_0^c(\bqbperp)\right\}
\mathrm{tr}\left[ \Psi _{{\tilde{\mathbf{n}}}}(-Q,\tilde{\kappa})\right]
\notag \\
& \, +\frac{1}{2\nu }\int \mathrm{d}\mathbf{n}\mathrm{d}{\tilde{\mathbf{n}%
}}\mathrm{d}\kappa \mathrm{d}{\tilde{\kappa}}\sum_{KQ}u  \notag \\
& \, \times \mathrm{tr}\left[ \Psi _{\mathbf{n}}(-K,\kappa )\sigma
^{k}\Psi _{\mathbf{n}+\frac{\mathbf{q}_{\perp }}{2p_{F}}}\big(K+Q,\kappa %
\big)\right]  \notag \\
& \, \times \left\{{\tilde{u}}\nu v_{F}{\tilde{\theta}}\
\delta\bn\!\cdot\!\B_\perp^s(\bqbperp)+2\mathrm{i}{\tilde{\theta}^{\ast }}\ \B_0^s(\bqbperp)\right\}
\mathrm{tr}\left[ \sigma ^{k}\Psi _{{\tilde{\mathbf{n}}}}(-Q,\tilde{\kappa})%
\right] \komma \notag
\end{align}
and the quadratic action as
\begin{align}
\mathcal{S}_{2} &=-\frac{1}{4\nu }\int \mathrm{d}\mathbf{n}\mathrm{d}{\tilde{%
\mathbf{n}}}\mathrm{d}\kappa \mathrm{d}{\tilde{\kappa}}\sum_{Q} \label{5a03} \\
& \times
  \mathrm{tr}\left[\Psi _{\mathbf{n}}(Q,\kappa )\right]
  \mathrm{tr}\left[-\Psi _{{\tilde{\mathbf{n}}}}(-Q,{\tilde{\kappa}})\right] \nonumber\\
& \times
 \Big\{
  (2\ii \thetax)(2\ii {\tilde{\theta}^{\ast}})\ \D_0^{c}(\bqbperp)
  +
  2(2\ii \thetax) (\ut\nu v_{F}{\tilde{\theta}}\ \delta\bn)\ \D_\perp^{c}(\bqbperp)
  \nonumber\\
& \qquad +
  (u\nu v_{F}\theta\ \delta\bn) (\ut\nu v_{F}{\tilde{\theta}}\ \delta\bn)\ \D_{\perp\perp}^{c}(\bqbperp)
 \Big\}
\notag \\
-&\frac{1}{4\nu }\int \mathrm{d}\mathbf{n}\mathrm{d}{\tilde{\mathbf{n}%
}}\mathrm{d}\kappa \mathrm{d}{\tilde{\kappa}}\sum_{Q}\  \notag \\
& \times
  \mathrm{tr}\left[ \sigma ^{k}\Psi_{\mathbf{n}}(Q,\kappa )\right]
  \mathrm{tr}\left[-\sigma ^{k}\Psi _{{\tilde{\mathbf{n}}}}(-Q,{\tilde{\kappa}})\right]\nonumber\\
&  \times
 \Big\{
  (2\ii \thetax)(2\ii {\tilde{\theta}^{\ast}})\ \D_0^{s}(\bqbperp)
  +
  2(2\ii \thetax) (\ut\nu v_{F}{\tilde{\theta}}\ \delta\bn)\ \D_\perp^{s}(\bqbperp)
  \nonumber\\
& \qquad +
  (u\nu v_{F}\theta\ \delta\bn) (\ut\nu v_{F}{\tilde{\theta}}\ \delta\bn)\ \D_{\perp\perp}^{s}(\bqbperp)
 \Big\}
\punkt  \notag
\end{align}
Equations~(\ref{5a01})--(\ref{5a03}) reduce to the original formulation, Eqs.~(\ref{2a41})--(\ref{2a44}),
if we insert the bare values for the coupling constants, which are given by
\begin{align}
\G^{c/s}(\bqbperp)\Big|_0 = \B_0^{c/s}(\bqbperp)\Big|_0 = \D_0^{c/s}(\bqbperp)\Big|_0 &=
\gamma^{c/s}_{\widehat{\mathbf{n}{\tilde{\mathbf{n}}}}} f(\bqbperp)\komma \nonumber\\
\B_\perp^{c/s}(\bqbperp)\Big|_0 = \D_\perp^{c/s}(\bqbperp)\Big|_0 &= -\bqbperp\ \gamma^{c/s}_{\widehat{\mathbf{n}{\tilde{\mathbf{n}}}}} f(\bqbperp)\komma\nonumber\\
\D_{\perp\perp}^{c/s}(\bqbperp)\Big|_0 &= -\bqbperp^2\ \gamma^{c/s}_{\widehat{\mathbf{n}{\tilde{\mathbf{n}}}}} f(\bqbperp)\punkt
\label{5a04}
\end{align}
The notations for the (one-dimensional) angular variable~$\delta\bn$ and the transverse momentum~$\bqbperp$ are taken from Eqs.~(\ref{3c03}) and~(\ref{3c04}).
Since the coupling constants eventually enter only at backscattering, $\widehat{\mathbf{n}{\tilde{\mathbf{n}}}}= \pi$ or~$\delta\bn\rightarrow 0$, the angular dependence
is not written explicitly.
Note that
in Eqs.~(\ref{5a01})--(\ref{5a04}), the parallel momenta~$\bar{q}_\parallel$ in both the cutoff functions and the prefactors
have been omitted. The former is justified according to the discussion preceding Eq.~(\ref{3c08a0}). In the
prefactors~$(\bn\!\cdot\!\bq)$ of the original actions $\cS_3$ and $\cS_2$, Eqs.~(\ref{2a43}) and~(\ref{2a44}),
the parallel momenta~$\bar{q}_\parallel$ are irrelevant for the non-analyticities following the discussions after Eq.~(\ref{3x05}) and in Appendix~\ref{App_2}. This justifies
the latter.

According to Eq.~(\ref{5a04}), the cutoff functions~$f(\bqbperp)$ are absorbed into the coupling constants.
This is a convenient definition since we have seen in Eq.~(\ref{3c08b}) that the quantities flowing with the RG
are $\gamma_{\widehat{\mathbf{n}{\tilde{\mathbf{n}}}}}^{c/s}f(\bqbperp)$
rather than the interaction constants~$\gamma_{\widehat{\mathbf{n}{\tilde{\mathbf{n}}}}}^{c/s}$ themselves. The coupling constants
with index~``$\perp$'' also contain the transverse momentum~$\bqbperp$ as prefactors. As the flowing coupling constants
in Eq.~(\ref{5a04}) are functions of~$\bqbperp$, we are formally applying a \emph{functional RG} procedure.

\subsection{Renormalization group equations}

We develop an RG scheme using the momentum shell integration.
In one RG step, the large energy cutoff~$\Lambda$ for the
one-dimensional parallel spectrum~$v_F\bar{q}_\parallel$ (and the Matsubara frequencies)
is reduced to a still large but much smaller cutoff~$\Lambda'\ll\Lambda$ by
integrating out the fields with parallel momenta of orders between~$\Lambda'/v_F$ and~$\Lambda/v_F$.
This yields an action at the energy scale~$\Lambda'$ with renormalized coupling constants.
Repeatedly applied RG steps make the coupling constants flow. This RG flow stops at the latest as soon as the cutoff
approaches the order of the temperature~$T$.

In this work, we study the RG flow of the coupling constants in Eq.~(\ref{5a04}), which comprise the
physics of the anomalous low energy behavior of a two-dimensional Fermi liquid, at the leading one-loop order.
This corresponds to a summation of all orders of~$\ln(\Lambda/T)$ at leading order in~$\gamma_\pi^{c/s}$.
The relevant one-loop diagrams have been completely identified in Sec.~\ref{ssec:log_diagrams}.

\subsubsection{RG equations for~$\cS_4$}

Equations~(\ref{3c01}), (\ref{3c07}), and~(\ref{3c08b}) determine how the coupling constant~$\G^c(\bqbperp)$ in the quartic
interaction~$\cS_4$, Eq.~(\ref{5a01}), or its Fourier transform~$\G^c(\brperp)$, cf. Eq.~(\ref{3c08a}),
is renormalized if the energy cutoff~$\Lambda$ is reduced to $\Lambda/b \ll\Lambda$.
Defining
\begin{align}
\label{5a05}
\dt\xi = \frac{4u{\tilde{u}} \nu ^{\ast }}{\nu}\ \ln b\komma
\end{align}
the correction~$\dt\G^c(\brperp)$ to the coupling constant~$\G^c(\brperp)$ in one RG step would be
\begin{align}
\label{5a06}
\dt\G^c(\brperp) &= -\frac{1}{2} \big[\G^c(\brperp)\big]^2\ \dt\xi
\end{align}
if we could neglect the spin channel $\G^s(\brperp)$.

Including the spin channel, we have to examine the spin structure of the one-loop diagram Fig.~\ref{fig: 1loopS4}(a).
Equation~(\ref{5a06}) is the result of attributing both~$\cS_4$-blocks to~$\G^c$. Replacing in one of these blocks~$\G^c$ by~$\G^s$,
reproduces the spin structure of the spin $\G^s$-vertex. If both blocks belong to the spin channel, the algebra of the Pauli
matrices allocates renormalizations to both the $\G^c$ and the $\G^s$ vertices. Then,
Eq.~(\ref{5a06}) should be replaced by the RG
equations which are written as
\begin{align}
\frac{\dt\G^{c}(\brperp)}{\mathrm{d}\xi }
   & =-\frac{1}{2}\big\{\big[\Gamma^{c}(\brperp)\big]^{2}
                      +3\big[\Gamma ^{s}(\brperp)\big]^{2}\big\}\komma  \label{5a07}
   \\
\frac{\dt\G^{s}(\brperp)}{\mathrm{d}\xi }
   & =-\frac{1}{2}\big\{2\Gamma^{c}(\brperp)\ \Gamma^{s}(\brperp)
                -2\big[\Gamma^{s}(\brperp)\big]^{2}\big\}\punkt  \notag
\end{align}
This system of two differential equations is decoupled for the linear combinations
\begin{align}
\G^\romI(\brperp)  &= \G^c(\brperp)-3\G^s(\brperp) \label{5a08}\komma\\
\G^\romII(\brperp) &= \G^c(\brperp)+\G^s(\brperp) \komma
\nonumber
\end{align}
whose bare values are given by
\begin{align}
 \frac{\G^\romI(\brperp)}{f(\brperp)}\Big|_{\xi=0} &=
 \gamma _{\widehat{\mathbf{n}{\tilde{\mathbf{n}}}}}^{\mathrm{I}}
  = \gamma _{\widehat{\mathbf{n}{\tilde{\mathbf{n}}}}}^{c}-3\gamma _{\widehat{\mathbf{n}{\tilde{\mathbf{n}}}}}^{s}\komma  \label{5a_coupling} \\
  \frac{\G^\romII(\brperp)}{f(\brperp)}\Big|_{\xi=0} &=
 \gamma_{\widehat{\mathbf{n}{\tilde{\mathbf{n}}}}}^{\mathrm{II}}
 = \gamma_{\widehat{\mathbf{n}{\tilde{\mathbf{n}}}}}^{c}+\gamma _{\widehat{\mathbf{n}{\tilde{\mathbf{n}}}}}^{s}\punkt  \notag
\end{align}
In terms of~$\G^{\romI/\romII}(\brperp)$, the RG equations~(\ref{5a07}) take the form
\begin{align}
\frac{\dt\G^{\romI/\romII}(\brperp)}{\mathrm{d}\xi }
   & =-\frac{1}{2}\big[\Gamma^{\romI/\romII}(\brperp)\big]^{2}
\punkt  \label{5a07a}
\end{align}
Equation~(\ref{5a07a}) with Eq.~(\ref{5a_coupling}) as boundary condition is easily solved, yielding
\begin{align}
\label{5a09}
\G^\romI(\brperp; \xi) &=
 \frac{\gamma_{\widehat{\mathbf{n}{\tilde{\mathbf{n}}}}}^{\mathrm{I}}f(\brperp)}
     {1+\frac{1}{2}\gamma_{\widehat{\mathbf{n}{\tilde{\mathbf{n}}}}}^{\mathrm{I}}f(\brperp)\ \xi}
     \komma \\
\G^\romII(\brperp; \xi) &=
 \frac{\gamma _{\widehat{\mathbf{n}{\tilde{\mathbf{n}}}}}^{\mathrm{II}}f(\brperp)}
      {1+\frac{1}{2}\gamma _{\widehat{\mathbf{n}{\tilde{\mathbf{n}}}}}^{\mathrm{II}}f(\brperp)\ \xi}
 \komma\notag
\end{align}
where in the relevant backscattering limit~$\delta\bn\rightarrow 0$,
the quantity $\xi$ varies between~$0$ and~$4u{\tilde{u}} (\nu^{\ast}/\nu)\ \ln(\Lambda/T)$.

\begin{figure*}[t]
\includegraphics{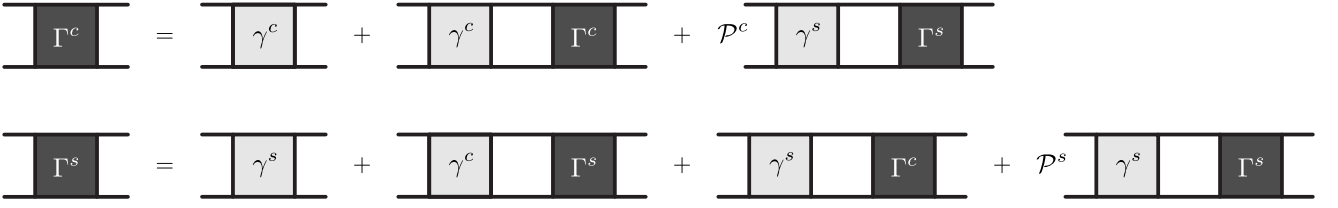}
\caption{Diagrammatical Bethe-Salpether equations for the $\mathcal{S}_{4}$-ladders. Dark gray blocks
stand for the renormalized vertices while light gray blocks stand for the
bare quartic vertices. $\mathcal{P}^{c}$ and $\mathcal{P}^{s}$ are defined
to project a general quartic vertex onto its vertex component of the form $%
[\Psi \Psi ][\Psi \Psi ]$ or $[\Psi \protect\sigma ^{k}\Psi ][\Psi \protect%
\sigma ^{k}\Psi ]$, respectively.}\label{fig: clean}
\end{figure*}

The renormalized coupling constants~$\G^{\romI/\romII}$, Eq.~(\ref{5a09}), can
also be obtained summing the relevant ladder diagrams built of the quartic vertices.
Such a ladder is constructed adding rungs of~$\cS_4$-blocks one by one in the way
corresponding to the diagram in Fig.~\ref{fig: 1loopS4}(a).
Considering the correspondence of the boson diagrams to the conventional fermion ones as discussed in Appendix~\ref{App_3},
one can see that these ladders are related to the usual particle-particle Cooper ladders. The evaluation of the arising geometric series is depicted in form of diagrammatical Bethe-Salpether equations in Fig.~\ref{fig: clean}. The two equations decouple in a complete analogy with Eq.~(\ref{5a08}) and eventually yield the
same result Eq.~(\ref{5a09}) as obtained using the RG equations.

The correspondence between the ladder form of the renormalized quartic vertex and
conventional Cooper ladders suggests that the logarithmic renormalizations of
the coupling constants can be attributed to the superconducting
correlations which in case of an attractive
interaction cause a phase transition toward superconductivity at a critical temperature~$T_c$.
Re-expressing the bare coupling constants~$\gamma_{\widehat{\mathbf{n}{\tilde{\mathbf{n}}}}}^{\romI/\romII}$, Eq.~(\ref{5a09}), in terms
of the original interaction potential, Eq.~(\ref{2a42}), we come at backscattering~$\widehat{\mathbf{n}{\tilde{\mathbf{n}}}}=\pi$
to the relations
\begin{align}
\gamma_{\pi }^{\mathrm{I}}& =\frac{\nu }{2}\left\{ \tilde{V}(0)+\tilde{V}%
(2p_{F})\right\} \;\mathnormal{,}  \label{5a10} \\
\gamma _{\pi }^{\mathrm{II}}& =\frac{\nu }{2}\left\{ \tilde{V}(0)-\tilde{V}%
(2p_{F})\right\} \punkt  \notag
\end{align}
In final results, contributions containing~$\gamma_{\pi }^\romI$ can
be interpreted in terms of a spin singlet while contributions due to~
$\gamma_{\pi }^\romII$, which enter final results with a prefactor of three, can be attributed
to spin triplets. Note that the
latter coupling constant vanishes in models with a contact interaction.

\subsubsection{RG equations for~$\cS_3$}

The relevant one-loop diagrams for the cubic interaction~$\cS_3$, Eq.~(\ref{5a02}), are shown in Fig.~\ref{fig: 1loopS3}(a) and~(a').
In contrast to the quartic interaction~$\cS_4$, the renormalized cubic vertex cannot
be obtained from simple ladder summations and
an RG procedure seems unavoidable.
One can understand this fact from Fig.~\ref{fig: 1loopS3}, which shows that
one has always two different choices
in attaching another quartic block when building higher-order logarithmic diagrams. As a result, a diagram for a general leading vertex correction acquires a rather complicated topology and momentum structure.

Similarly to the RG equations for~$\cS_4$, Eq.~(\ref{5a07}), the RG equations for the cubic
action take the form
\begin{align}
\frac{\dt\B_{0/\perp}^{c}(\brperp)}{\mathrm{d}\xi }
   & =-\big\{\G^c(\brperp)\ \B_{0/\perp}^c(\brperp)
                      +3\G^s(\brperp)\ \B_{0/\perp}^s(\brperp)\big\}\komma  \nonumber
   \\
\frac{\dt\B_{0/\perp}^{s}(\brperp)}{\mathrm{d}\xi }
   & =-\big\{ \Gamma^{c}(\brperp)\ \B_{0/\perp}^s(\brperp)+ \Gamma^{s}(\brperp)\ \B_{0/\perp}^c(\brperp)\nonumber\\
              &\qquad\quad  -2 \G^s(\brperp)\ \B_{0/\perp}^s(\brperp)\big\}\punkt  \label{5a11}
\end{align}
Inserting in analogy with Eq.~(\ref{5a08}) the linear combinations
\begin{align}
\B_{0/\perp}^\romI(\brperp)  &= \B_{0/\perp}^c(\brperp)-3\B_{0/\perp}^s(\brperp) \label{5a12}\komma\\
\B_{0/\perp}^\romII(\brperp) &= \B_{0/\perp}^c(\brperp)+\B_{0/\perp}^s(\brperp)
\nonumber
\end{align}
into the the RG equations in Eq.~(\ref{5a11}) reduces these equations to the form
\begin{align}
\frac{\dt\B_{0/\perp}^{\romI/\romII}(\brperp)}{\mathrm{d}\xi }
   & =-\ \Gamma^{\romI/\romII}(\brperp)\ \B_{0/\perp}^{\romI/\romII}(\brperp)
\punkt  \label{5a13}
\end{align}
With the knowledge of the renormalized coupling constants~$\Gamma^{\romI/\romII}(\brperp)$ of the quartic action, Eq.~(\ref{5a09}),
the RG equation Eq.~(\ref{5a13}) is nothing but a homogeneous linear differential equation for the coupling constant~$\B_{0/\perp}^{\romI/\romII}(\brperp)$.
Using the boundary conditions
\begin{align}
\B_{0}^{\romI/\romII}(\brperp)\Big|_{\xi=0} &= \gamma^{\romI/\romII}_{\widehat{\mathbf{n}{\tilde{\mathbf{n}}}}}f(\brperp)
 \label{5a14}\komma\\
\B_{\perp}^{\romI/\romII}(\brperp)\Big|_{\xi=0} &=
   \gamma^{\romI/\romII}_{\widehat{\mathbf{n}{\tilde{\mathbf{n}}}}}\
   \big[\ii q_0\ \partial_\brperp f(\brperp)\big]\komma
\nonumber
\end{align}
cf. Eq.~(\ref{5a04}), the solutions of the RG equations in Eq.~(\ref{5a13})
are found to be
\begin{align}
\label{5a15}
\B_{0}^{\romI/\romII}(\brperp;\xi) &=
  \frac{\gamma^{\romI/\romII}_{\widehat{\mathbf{n}{\tilde{\mathbf{n}}}}}f(\brperp)}
       {\big[1+\frac{1}{2}\gamma_{\widehat{\mathbf{n}{\tilde{\mathbf{n}}}}}^{\romI/\romII}f(\brperp)\ \xi\big]^2}\komma\\
\B_{\perp}^{\romI/\romII}(\brperp;\xi) &=
  \frac{\gamma^{\romI/\romII}_{\widehat{\mathbf{n}{\tilde{\mathbf{n}}}}}\ \big[\ii q_0\ \partial_\brperp f(\brperp)\big]}
       {\big[1+\frac{1}{2}\gamma_{\widehat{\mathbf{n}{\tilde{\mathbf{n}}}}}^{\romI/\romII}f(\brperp)\ \xi\big]^2}
       \nonumber\punkt
\end{align}
Equipped with the knowledge about quartic and cubic coupling constants at any energy scale, we are now ready to
finally determine the renormalized coupling constants of the quadratic interaction~$\cS_2$.

\subsubsection{RG equations for~$\cS_2$}

The quadratic action~$\cS_2$, Eq.~(\ref{5a03}), is renormalized
according to the one-loop diagrams
in Fig.~\ref{fig: 1loopS2}(a), (b), and~(b'). The spin structure of these diagrams is completely analogous
to the one-loop corrections to the quartic and cubic terms. Introducing corresponding
combinations of the coupling constants in Eq.~(\ref{5a04}) as
\begin{align}
\D_{0/\perp/\perp\perp}^\romI(\brperp)  &= \D_{0/\perp/\perp\perp}^c(\brperp)-3\D_{0/\perp/\perp\perp}^s(\brperp) \label{5a16}\komma\\
\D_{0/\perp/\perp\perp}^\romII(\brperp) &= \D_{0/\perp/\perp\perp}^c(\brperp)+\D_{0/\perp/\perp\perp}^s(\brperp)
\komma\nonumber
\end{align}
we immediately write the decoupled RG equations. These are
\begin{align}
\frac{\dt\D_{0}^{\romI/\romII}(\brperp)}{\mathrm{d}\xi }
   & =-\  \G^{\romI/\romII}(\brperp)\ \D_{0}^{\romI/\romII}(\brperp)
   - \big[
    \B_{0}^{\romI/\romII}(\brperp)
   \big]^2\nonumber\komma\\
\frac{\dt\D_{\perp}^{\romI/\romII}(\brperp)}{\mathrm{d}\xi }
   & =-\  \G^{\romI/\romII}(\brperp)\ \D_{\perp}^{\romI/\romII}(\brperp)
   -  \B_{0}^{\romI/\romII}(\brperp)\ \B_{\perp}^{\romI/\romII}(\brperp)
   \nonumber\komma\\
\frac{\dt\D_{\perp\perp}^{\romI/\romII}(\brperp)}{\mathrm{d}\xi }
   & =-\  \G^{\romI/\romII}(\brperp)\ \D_{\perp\perp}^{\romI/\romII}(\brperp)
      + \big[
       \B_{\perp}^{\romI/\romII}(\brperp)
        \big]^2
\punkt  \label{5a17}
\end{align}
Since~$\G^{\romI/\romII}(\brperp)$, Eq.~(\ref{5a09}), and $\B_{0/\perp}^{\romI/\romII}(\brperp)$, Eq.~(\ref{5a15}), are known functions of~$\xi$, we are once more to solve linear differential equations, which
are now --- in contrast to the RG equations for the cubic action, Eq.~(\ref{5a13}) ---
non-homogeneous.
The boundary conditions for Eq.~(\ref{5a17}) take the form
\begin{align}
\D_{0}^{\romI/\romII}(\brperp)\Big|_{\xi=0} &= \gamma^{\romI/\romII}_{\widehat{\mathbf{n}{\tilde{\mathbf{n}}}}}f(\brperp)
 \nonumber \komma\\
 \label{5a18}
\D_{\perp}^{\romI/\romII}(\brperp)\Big|_{\xi=0} &=
   \gamma^{\romI/\romII}_{\widehat{\mathbf{n}{\tilde{\mathbf{n}}}}}\
   \big[\ii q_0\ \partial_\brperp f(\brperp)\big]\komma\\
\D_{\perp\perp}^{\romI/\romII}(\brperp)\Big|_{\xi=0} &=
   \gamma^{\romI/\romII}_{\widehat{\mathbf{n}{\tilde{\mathbf{n}}}}}\
   \big[q_0^2\ \partial_\brperp^2 f(\brperp)\big]\komma
\nonumber
\end{align}
cf. Eq.~(\ref{5a04}). As a result, the solutions of the RG equations~(\ref{5a17}) are
\begin{align}
\D_{0}^{\romI/\romII}(\brperp;\xi) &= \gamma_{\widehat{\mathbf{n}{\tilde{\mathbf{n}}}}}^{\romI/\romII} f(\brperp)\
  \frac{1-\frac{1}{2}\gamma_{\widehat{\mathbf{n}{\tilde{\mathbf{n}}}}}^{\romI/\romII}f(\brperp)\ \xi}
      {\big[1+\frac{1}{2}\gamma_{\widehat{\mathbf{n}{\tilde{\mathbf{n}}}}}^{\romI/\romII}f(\brperp)\ \xi\big]^3}
 \komma \label{5a19a}\\
\D_{\perp}^{\romI/\romII}(\brperp;\xi) &= \ii q_0\ \frac{\partial}{\partial\brperp}
 \frac{\gamma_{\widehat{\mathbf{n}{\tilde{\mathbf{n}}}}}^{\romI/\romII}f(\brperp)}
      {\big[1+\frac{1}{2}\gamma_{\widehat{\mathbf{n}{\tilde{\mathbf{n}}}}}^{\romI/\romII}f(\brperp)\ \xi\big]^2}
 \komma \label{5a19b}\\
\D_{\perp\perp}^{\romI/\romII}(\brperp;\xi) &= q_0^2\
 \frac{\partial^2}{\partial\br_\perp^2}
\frac{\gamma_{\widehat{\mathbf{n}{\tilde{\mathbf{n}}}}}^{\romI/\romII}f(\brperp)}{1+\frac{1}{2}\gamma_{\widehat{\mathbf{n}{\tilde{\mathbf{n}}}}}^{\romI/\romII}f(\brperp)\ \xi}\punkt\label{5a19c}
\end{align}
Equations~(\ref{5a19a})--(\ref{5a19c}) constitute the final result for the renormalized quadratic vertex and
complete the RG study of our low energy field theory, Eqs.~(\ref{5a01})--(\ref{5a04}). Stopping the RG flow
at~$\xi= 4u{\tilde{u}} (\nu^{\ast}/\nu)\ \ln(\Lambda/T)$ yields the renormalized coupling constants in the relevant
backscattering limit~$\delta\bn\rightarrow 0$, which enter the effective action after integrating out
all superfields at energy scales larger than~$T$. Therefore in low temperature calculations of thermodynamic quantities
based on the renormalized action, the summation of all orders in the large logarithm~$\ln(\Lambda/T)$ is at leading order in~$\gamma_\pi^{\romI/\romII}$ automatically included.

\section{Specific heat}

\label{sec:c}

As a result of the renormalization group procedure developed in the previous section,
we have the full knowledge about the physics of our low energy theory, Eqs.~(\ref{5a01})--(\ref{5a04}),
at any energy scale~$\Lambda' \ll \Lambda$. The relevant energy scale for the calculation
of the specific heat is according to Eqs.~(\ref{3b000}) and~(\ref{3b00}) of the order of the temperature~$T\ll \Lambda$.
So, inserting the quadratic vertex with the renormalized coupling constants, Eqs.~(\ref{5a19a})--(\ref{5a19c}), into
the formula for the thermodynamic potential~$-(T/2)\langle\cS_2^2\rangle$, we obtain instead of Eq.~(\ref{3b01}) the following expression:
\begin{widetext}
\begin{align}
&\quad\delta \Omega(T) =
 \left(T\sum_\om-\int\frac{\dt\omega}{2\pi}\right)
 \int \frac{\mathrm{d}^{2}\mathbf{q}}{(2\pi )^{2}} \
      \mathrm{d}\mathbf{n}\mathrm{d}{\tilde{\mathbf{n}}}\
 \frac{2(v_{F}\delta\bn)^2}
 {[\mathrm{i}\omega -v_{F}(\mathbf{n}\!\cdot \!\mathbf{q})]
  [-\mathrm{i}\omega +v_{F}({\tilde{\mathbf{n}}}\!\cdot \!\mathbf{q})]}\nonumber\\
 &\times  \int_0^1\!\!\int_0^1\dt u\dt\ut\ u\ut\left\{
  \big(\big[\D_{\perp}^{\romI}(\bqbperp;\xi)\big]^2 - \D_{0}^{\romI}(\bqbperp;\xi)\D_{\perp\perp}^{\romI}(\bqbperp;\xi)\big)
  +3\big(\big[\D_{\perp}^{\romII}(\bqbperp;\xi)\big]^2 - \D_{0}^{\romII}(\bqbperp;\xi)\D_{\perp\perp}^{\romII}(\bqbperp;\xi)\big)
  \right\}
  \punkt  \label{5b01}
\end{align}
\end{widetext}
For~$\xi= 4u{\tilde{u}} (\nu^{\ast}/\nu)\ \ln(\Lambda/T)$, Eq.~(\ref{5b01}) yields the anomalous contribution to the thermodynamic potential in all
orders in the logarithm~$\ln(\Lambda/T)\sim\ln(\ve_F/T)$ at leading order in the bare couplings~$\gamma_\pi^{\romI/\romII}$.
This means that in the limit of low temperatures~$T\ll\ve_F$ considered here, Eq.~(\ref{5b01}) gives the full physical result. Explicit
results for the specific heat can be extracted by expansion or specifying the cutoff function~$f(\bqbperp)$ that enters
the renormalized coupling constants, cf. Eqs.~(\ref{5a19a})--(\ref{5a19c}).

An important observation in Eq.~(\ref{5b01}) is that the entire contribution
is expressed as a sum of two contributions. The first one contains only
the bare coupling
constant~$\gamma_\pi^\romI$ while the second contribution contains only the coupling constant~$\gamma_\pi^\romII$. In other words, the fluctuations attributed
to the spin singlet~[$\gamma_\pi^\romI$] are in leading order completely separated from the fluctuations due to the spin triplet~[$\gamma_\pi^\romII$],
cf. the discussion in Secs.~\ref{ssec: fl_corrections} and~\ref{ssec: instabilities}.

\subsection{Low order perturbation theory}

Before evaluating the specific heat in all orders for a certain cutoff function~$f(\brperp)$, let
us explicitly calculate the non-analytic contribution to the specific heat in the third order in the interaction.
This approximation is valid for not very low temperatures~$T$, such that the
quantity $\gamma_\pi^{\romI/\romII} \ln (\Lambda /T)$ is still small, i.e. if
$T\gg \Lambda \exp(-[\gamma_\pi^{\romI/\romII}]^{-1})$ is fulfilled. Results in
this limit are known from conventional perturbation theory\cite{chubukov3}
and therefore, we can check our bosonization approach and see how it works. In Appendix~\ref{App_2},
we recalculate the third order starting from the bosonic diagrams rather than from  Eq.~(\ref{5b01}).

In the second order in~$\gamma_\pi^{\romI/\romII}$, the renormalized vertices in Eqs.~(\ref{5a19a})--(\ref{5a19c}) are reduced to
\begin{align}
\D_{0}^{\romI/\romII}(\brperp) &\simeq \gamma_{\widehat{\mathbf{n}{\tilde{\mathbf{n}}}}}^{\romI/\romII} f(\brperp)
  - 2\big[\gamma_{\widehat{\mathbf{n}{\tilde{\mathbf{n}}}}}^{\romI/\romII} f(\brperp)\big]^2\ \xi
 \komma \label{5b02}\\
\D_{\perp}^{\romI/\romII}(\brperp) &\simeq \ii q_0\frac{\partial}{\partial\brperp}
  \Big\{
\gamma_{\widehat{\mathbf{n}{\tilde{\mathbf{n}}}}}^{\romI/\romII} f(\brperp)
-\big[\gamma_{\widehat{\mathbf{n}{\tilde{\mathbf{n}}}}}^{\romI/\romII} f(\brperp)\big]^2\ \xi
\Big\} \komma \nonumber\\
\D_{\perp\perp}^{\romI/\romII}(\brperp) &\simeq
 q_0^2 \frac{\partial^2}{\partial\br_\perp^2}
 \Big\{
    \gamma_{\widehat{\mathbf{n}{\tilde{\mathbf{n}}}}}^{\romI/\romII}f(\brperp)
    -\frac{1}{2}\big[\gamma_{\widehat{\mathbf{n}{\tilde{\mathbf{n}}}}}^{\romI/\romII}f(\brperp)\big]^2\ \xi
\Big\} \punkt\nonumber
\end{align}
Returning to the momentum representation by Fourier transformation, we find
\begin{align}
\D_{0}^{\romI/\romII}(\bqbperp) &\simeq \gamma_{\widehat{\mathbf{n}{\tilde{\mathbf{n}}}}}^{\romI/\romII} f(\bqbperp)
  - 2\big[\gamma_{\widehat{\mathbf{n}{\tilde{\mathbf{n}}}}}^{\romI/\romII}\big]^2\ [f\!\ast\!f](\bqbperp)\ \xi
 \komma \label{5b03}\\
\D_{\perp}^{\romI/\romII}(\bqbperp) &\simeq -\bqbperp
  \Big\{
\gamma_{\widehat{\mathbf{n}{\tilde{\mathbf{n}}}}}^{\romI/\romII} f(\bqbperp)
-\big[\gamma_{\widehat{\mathbf{n}{\tilde{\mathbf{n}}}}}^{\romI/\romII}\big]^2\ [f\!\ast\!f](\bqbperp)\ \xi
\Big\} \komma \nonumber\\
\D_{\perp\perp}^{\romI/\romII}(\bqbperp) &\simeq
 -\bar{\bq}^2_\perp
 \Big\{
    \gamma_{\widehat{\mathbf{n}{\tilde{\mathbf{n}}}}}^{\romI/\romII}f(\bqbperp)
    -\frac{1}{2}\big[\gamma_{\widehat{\mathbf{n}{\tilde{\mathbf{n}}}}}^{\romI/\romII}\big]^2\ [f\!\ast\!f](\bqbperp)\ \xi
\Big\} \nonumber
\end{align}
with~$[f\!\ast\!f]$ denoting a convolution as in Eq.~(\ref{3c08a0}).

Inserting Eq.~(\ref{5b03}) into Eq.~(\ref{5b01}), using the relation~$\xi= 4u{\tilde{u}} (\nu^{\ast}/\nu)\ \ln(\Lambda/T)$, and
taking the result from the calculation in Appendix~\ref{App_1},
we find the non-analytic thermodynamic potential correction in the third order in the interaction as
\begin{align}
\delta\Omega &\simeq \frac{\zeta (3)T^3}{\pi v_{F}^{2}} \Big\{
 [\gamma_\pi^\romI]^2+3[\gamma_\pi^\romII]^2 \nonumber\\
 -&\frac{4\nu^*}{\nu} [f\!\ast\!f](0)\ \Big(
 [\gamma_\pi^\romI]^3+3[\gamma_\pi^\romII]^3
 \Big)\ \ln\Big(\frac{\Lambda}{T}\Big)
\Big\}\label{5b04}\punkt
\end{align}
Choosing the cutoff function as~$f(\bqbperp)=\Theta(q_0-|\bqbperp|)$,
the prefactor~$[f\!\ast\!f](0)$ just gives unity. In this case, we can also estimate
$\nu^*/\nu \sim 1/2$, following the lines after Eq.~(\ref{3c08}).
Finally applying the thermodynamic relation~$\delta c=-T\partial ^{2}\delta
\Omega /\partial T^{2}$ yields the third order non-analytic specific heat contribution at low temperatures
\begin{align}
\delta c &\simeq -\frac{6\zeta (3)T^2}{\pi v_{F}^{2}} \Big\{
 [\gamma_\pi^\romI]^2+3[\gamma_\pi^\romII]^2 \nonumber\\
 &\qquad\quad-2\ \Big(
 [\gamma_\pi^\romI]^3+3[\gamma_\pi^\romII]^3
 \Big)\ \ln\Big(\frac{\Lambda}{T}\Big)
\Big\}\punkt\label{5b05}
\end{align}
This perturbative result is applicable in the limit of not very low
temperatures such that on one hand $\ln (\Lambda /T)\gg 1$ while on the other hand $\gamma^{\romI/\romII}_\pi\ln (\Lambda /T)\ll 1$.

Equation~(\ref{5b05}) contains the logarithmic contributions from both spin and charge
excitations, which corrects the result of Ref.~\onlinecite{aleiner}, where the
contribution of the charge excitations was missed due to the neglect of the curvature of the Fermi surface. (However, we
emphasize that the results of Ref.~\onlinecite{aleiner} remain correct for one-dimensional systems.)
At the same time, this correction to the specific heat $\delta c$, Eq.~(\ref{5b05}), fully agrees with the result of the later publication
by Chubukov and Maslov\cite{chubukov3}.

To be more specific, the result for the third order non-analytic
contribution to the specific heat at low temperature by Chubukov and Maslov
can be recasted into the form
\begin{align}
\delta c^{(3)} &= \frac{3\zeta (3)}{2\pi v_{F}^{2}} \big\{
(u_{0}\!+\!u_{\pi })\langle
(u_{\vartheta }\!+\!u_{\pi -\vartheta })^{2}\rangle
\nonumber\\
&\quad+3(u_{0}\!-\!u_{\pi
})\langle (u_{\vartheta }\!-\!u_{\pi -\vartheta })^{2}\rangle \big\}\ T^{2}\ln
(\Lambda /T)  \label{5b06}
\end{align}
Here, $\vartheta $ is the scattering angle, $u_{\vartheta
}=\nu \tilde{V}(2p_{F}\sin (\vartheta /2))$, and $\langle g_{\vartheta
}\rangle $ is the angular average for an arbitrary function $g_{\vartheta }$.
Following the decoupling into soft modes, Eq.~(\ref{2a03d}), angular
averages~$\langle g_{\vartheta }\rangle $ are to be replaced by $(g_{0}+g_{\pi })/2$ in our model. Applying this correspondence to Eq.~(\ref{5b06}), we
recover immediately the same logarithmic dependence of the specific heat $\delta c^{(3)}$ at third order as in Eq.~(\ref{5b05}).

Thus, on one hand Eq.~(\ref{5b05}) serves as a good check of our low energy
model, on the other hand we confirm the estimate~$\nu ^{\ast }\sim \nu/2$ discussed after Eq.~(\ref{3c08}).

In the remaining of this analysis, we extend the perturbative
result~(\ref{5b05}) by including the leading in $\gamma_\pi^{\romI/\romII}$ terms of
\emph{all} orders in the logarithm~$\ln (\Lambda /T)$. This calculation shall
complete the picture of the non-analytic corrections to the specific
heat at low temperatures~$T$.

\subsection{Full low temperature result}

In this section, we will extract from Eq.~(\ref{5b01})
the anomalous contribution to the specific heat in all orders in~$\ln(\Lambda/T)$.
As a result, we obtain the full picture of the non-analyticities in the Fermi liquid
thermodynamics at low temperatures~$T$.

In order to accomplish this task, we should choose a model for the cutoff function~$f(\bqbperp)$, which controls the
two soft modes represented by Figs.~\ref{fig: softmodes}(a) and~(b). Following Ref.~\onlinecite{aleiner}, a suitable candidate
is a Lorentzian of the form
\begin{align}
\label{5b07}
f(\bqbperp) &= \frac{1}{1+|\bqbperp|^2/q_0^2}\punkt
\end{align}
This choice implies~$f(0)=1$ so that the result from second order
perturbation theory, Eq.~(\ref{3b11}), remains the same.
We recall that within the low energy theory, it is implied that~$q_0\ll p_F$.
Fourier transforming Eq.~(\ref{5b07}) yields
\begin{align}
\label{5b08}
f(\brperp) &= \frac{\pi}{2}\ \ee^{-|\brperp|}
\end{align}
according to the definition of the Fourier transform in Eq.~(\ref{3c08a}).

Inserting the model function~$f(\brperp)$, Eq.~(\ref{5b08}), into the renormalized quadratic vertices, Eqs.~(\ref{5a19a})--(\ref{5a19c}), we are in a position to Fourier transform them to
the momentum representation, which is needed for formula Eq.~(\ref{5b01}). Since
only the leading quadratic in~$\bqbperp$ term of the expression~$[\D_{\perp}^{\romI/\romII}]^2 - \D_{0}^{\romI/\romII}\D_{\perp\perp}^{\romI/\romII}$ is relevant, we may neglect higher orders in~$\bqbperp$ from the beginning. Thus, the evaluation of the Fourier integrals yields
\begin{align}
\D_{0}^{\romI/\romII}(\bqbperp) &=
  \frac{\gamma_{\widehat{\mathbf{n}{\tilde{\mathbf{n}}}}}^{\romI/\romII}}
     {\big(1+\frac{\pi}{4}\gamma_{\widehat{\mathbf{n}{\tilde{\mathbf{n}}}}}^{\romI/\romII}\ \xi\big)^2}
     \komma\label{5b09a}\\
\D_{\perp}^{\romI/\romII}(\bqbperp) &=  -\bqbperp\
 \frac{\gamma_{\widehat{\mathbf{n}{\tilde{\mathbf{n}}}}}^{\romI/\romII}}
      {1+\frac{\pi}{4}\gamma_{\widehat{\mathbf{n}{\tilde{\mathbf{n}}}}}^{\romI/\romII}\ \xi}
      \komma\label{5b09b}\\
\D_{\perp\perp}^{\romI/\romII}(\bqbperp) &= -\bqbperp^2\
    \frac{4}{\pi\xi}\ln\Big(
     1+\frac{\pi}{4}\gamma_{\widehat{\mathbf{n}{\tilde{\mathbf{n}}}}}^{\romI/\romII}\ \xi
                       \Big)\punkt\label{5b09c}
\end{align}
Applying $\xi = 4 u\ut (\nu^*/\nu)\ln(\Lambda/T)$ to the renormalized
couplings, Eqs.~(\ref{5b09a})--(\ref{5b09c}), inserting them into Eq.~(\ref{5b01}), using the integral
\begin{align*}
\int_0^1\!\!\int_0^1\dt u\dt\ut\ \frac{u\ut + x^{-1}\ln(1+x u\ut)}{(1+x u\ut)^2}
&= \frac{\ln^2(1+x)}{2x^2}\komma
\end{align*}
and adopting the result of Appendix~\ref{App_1} for the remaining integrations, we obtain for
the low temperature non-analytic part of the thermodynamic potential~$\delta\Omega$ the
formula
\begin{align}
\label{5b10}
\delta\Omega &= \frac{\zeta (3)T^3}{\pi v_{F}^{2}}
 \Bigg\{
    \frac{\ln^2(1+\gamma_\pi^\romI L)}{L^2}
    +
    3\ \frac{\ln^2(1+\gamma_\pi^\romII L)}{L^2}
 \Bigg\}\punkt
\end{align}
Herein, the quantity~$L$ is defined as
\begin{align}
\label{5b11}
L &= \frac{\pi\nu^*}{\nu}\ \ln\Big(\frac{\Lambda}{T}\Big)
\end{align}
with $\nu^*$ given by~Eq.~(\ref{3c08}). The bare coupling constants~$\gamma_\pi^{\romI/\romII}$ are expressed in
terms of the original fermion interaction potential~$\tilde{V}$ as
\begin{align}
\gamma_{\pi }^{\romI/\romII}& =\frac{\nu }{2}\left\{ \tilde{V}(0)\pm\tilde{V}%
(2p_{F})\right\}\label{5b12} \komma
\end{align}
cf. Eq.~(\ref{5a10}).

Equation~(\ref{5b10}) constitutes our final result for the non-analyticities of a two-dimensional Fermi gas with repulsive interaction. In the following, we discuss
the corrections to the specific heat of the Fermi liquid and possible instabilities.

\subsection{Corrections to the Fermi liquid}

\label{ssec: fl_corrections}

\begin{figure}[t]
\includegraphics[width=\linewidth]{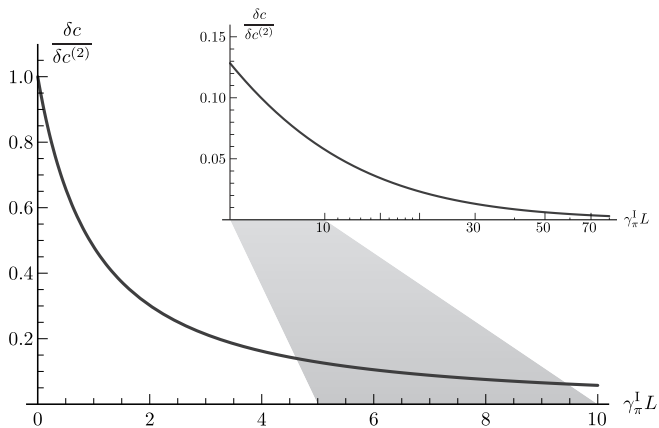}
\caption{
The non-analytic correction to the specific heat of a two-dimensional Fermi liquid for
a repulsive contact interaction as a function of~$\gamma_\pi^\romI L$. The logarithmic-linear
plot in the background illustrates the behavior at large~$\gamma_\pi^\romI L$, i.e. in the asymptotic
limit of very low temperatures~$T$.}
\label{fig: result}
\end{figure}

For the Fermi liquid model, the thermodynamic potential correction~$\delta\Omega$, Eq.~(\ref{5b10}), is a regular function
for all relevant~$\gamma_\pi^{\romI/\romII}$. By means of the formula
\begin{align}
\delta c=-T\frac{\partial^{2}\delta\Omega }{\partial T^{2}}  \label{5c01}
\end{align}
we find the anomalous contribution to the specific heat at low
temperatures~$T$ in the form
\begin{align}
\delta c=-\frac{6\zeta (3)T^{2}}{\pi v_{F}^{2}}
 \Bigg\{
    \frac{\ln^2(1+\gamma_\pi^\romI L)}{L^2}
    +
    3\ \frac{\ln^2(1+\gamma_\pi^\romII L)}{L^2}
 \Bigg\} \label{5c02}\punkt
\end{align}
This formula has been obtained from Eqs.~(\ref{5b10}) and~(\ref{5c01}) by differentiating
only the $T^{3}$ prefactor but not the quantity~$L$, Eq.~(\ref{5b11}). Derivatives of~$L$ would
yield subleading terms in the most interesting low temperature limit.

Equation~(\ref{5c02}) constitutes our main result for the anomalous correction~$\delta c$ to the
specific heat of the two-dimensional Fermi liquid at low temperature in the
weak coupling regime. Figure~\ref{fig: result} shows the plot of~$\delta c$ divided by the second
order result~(\ref{3b11a}) as a function of $\gamma_\pi^\romI L$. For simplicity, only the case~$\gamma_\pi^\romII = 0$ (contact
interaction) is plotted. For a general~$\gamma_\pi^\romII$, a contribution of the same shape as in Fig.~\ref{fig: result}
but dependent on~$\gamma_\pi^\romII L$ has to be added.

With a further decrease of the temperature, the quantity~$L$, Eq.~(\ref{5b11}),
grows such that
the non-analytic part~$\delta c$ of the specific heat of the Fermi liquid asymptotically reaches zero
as $- (\ln^2 L) / L^2$. This asymptotic behavior fully agrees with the estimate by Chubukov
and Maslov\cite{chubukov3} in the limit of small~$\nu^*/\nu$. This estimate was based on the conjectured
relation\cite{chubukov1,chubukov2}
\begin{align}
\delta c = -\frac{3\zeta (3)T^{2}}{2\pi v_{F}^{2}}\ \big[
 f_c^2(\pi) +  3f_s^2(\pi)
\big]\komma
\label{5c03}
\end{align}
where~$f_c(\pi)$ and~$f_s(\pi)$ denote the charge and spin components of the fully renormalized backscattering amplitude.\cite{footnote01}
The asymptotic agreement between our bosonization approach and the conjecture~(\ref{5c03}) supports its validity.
Moreover, full agreement between our result Eq.~(\ref{5c02}) and Eq.~(\ref{5c03})
can be shown in the limit~$\nu^*/\nu\ll 1$ in all orders in~$L$ for a long-range interaction, considering the potential~$V_\bq$
to be non-zero only close to~$\bq=0$.\cite{chubukov_discussion}
We also note that in the supersymmetric approach of Ref.~\onlinecite{aleiner}, where certain effects of the Fermi surface curvature were neglected,
the dependence on~$L$ of the spin contribution to the anomalous specific heat was found to have the same analytic shape as the result of the present work for the contribution due to both charge and spin excitations, Eq.~(\ref{5c02}).

At not very low temperatures, $\gamma_\pi^{\mathrm{I}/\mathrm{II}}\ln \left( \Lambda /T\right) \ll 1$,
the full result for the anomalous specific heat, Eq.~(\ref{5c02}), reduces to the perturbative result
of Eq.~(\ref{5b05}) [except for a prefactor of~$\pi/4$ in front of the third order term which is specific to the choice of the cutoff function, Eq.~(\ref{5b07})].
The logarithmic terms in Eq.~(\ref{5b05}) were attributed to the contribution of superconducting singlet and triplet fluctuations
(fluctuation contribution of $s$- and $p$-wave Cooper pairs).\cite{chubukov3} This is
supported by the form of the coupling constants~$\gamma_\pi^\romI$ and~$\gamma_\pi^\romII$, see Eq.~(\ref{5b12}).
The combination of the interaction amplitudes~$\tilde{V}(0)$ and~$\tilde{V}(2p_F)$
precisely corresponds to what one obtains when summing up the singlet [for~$\gamma_\pi^\romI$] and triplet [for~$\gamma_\pi^\romII$]
Cooper ladders.

Following this line of reasoning, we interpret the first and the second terms
in Eq.~(\ref{5c02}) as contributions coming from singlet and triplet
superconducting fluctuations. Of course, for models with a contact interaction, $\tilde{V}=\mathrm{const}$,
the coupling constant $\gamma _{\pi }^{\mathrm{II}}$ vanishes and only the
singlet Cooper pairs contribute.

It appears that finding the anomalous correction to the specific heat $\delta c$, Eq.~(\ref{5c02}),
from conventional diagrammatic expansions is rather difficult. Identifying the Cooper ``wheels''
as the relevant diagrams and especially identifying the soft modes in these diagrams for an arbitrary
order in the perturbation theory can be rather tricky, which is why earlier works\cite{chubukov3} restricted
themselves to low orders while estimating infinite order results by plausibility.
The bosonization approach presented in this paper allows for a field theory study based
on simple elementary diagrams that can be used as building blocks
for a subsequent renormalization group analysis. As a result,
an explicit derivation of the low temperature non-analytic specific heat of
higher-dimensional Fermi liquids becomes possible in all orders in the large logarithm~$\ln(\Lambda/T)$.

Formula~(\ref{5c02}) for the anomalous specific heat contribution~$\delta c$
has to replace the well-known second order contribution~$\delta c^{(2)}$, Eq.~(\ref{3b11a}), as soon as the logarithm in~$L$,
Eq.~(\ref{5b11}), becomes of the order of~$[\gamma_\pi^{\mathrm{I}/\mathrm{II}}]^{-1}$. The experimental
study on $^3\mathrm{He}$ fluid monolayers  presented in Ref.~\onlinecite{casey}
confirmed the validity of the second order perturbation theory
down to temperatures of order~$1\ \mathrm{mK}$. In this experimental
setting, however, the coupling constant~$\gamma_\pi$ is of order unity,
which is beyond the applicability of our theory as it is based on the assumption of weak
interaction. Nevertheless, provided the logarithmic
renormalization remains valid at least qualitatively also for strong
interactions, the logarithms should be detectable in measurements, although one might need to
investigate a broader temperature interval including temperatures considerably
below the $\mathrm{mK}$~scale.

\subsection{Instabilities}

\label{ssec: instabilities}

According to the main result, Eq.~(\ref{5c02}), discussed in the last section,
the non-analytic corrections to the Fermi liquid thermodynamics at low temperatures
are small as the function~$\delta c(T) /T^2$ decays logarithmically in the limit of $T\rightarrow 0$.
This statement remains valid as long as both singlet and triplet
constants, $\gamma_\pi^\romI\propto \tilde{V}(0)+\tilde{V}(2p_{F})$ and~$\gamma_\pi^\romII\propto \tilde{V}(0)-\tilde{V}(2p_{F})$,
are positive. The assumption of their positivity guarantees that the system
is in the Fermi liquid regime.

If one of these constants is negative, the function~$\ln(1+\gamma_\pi^{\romI/\romII} L)$ in the thermodynamic potential~$\delta\Omega$,
Eq.~(\ref{5b10}), approaches
a pole at $|\gamma_\pi^{\romI/\romII}| L = 1$ when lowering the temperature~$T$, i.e. when boosting
the quantity~$L$, Eq.~(\ref{5b11}). This pole corresponds to the divergency of the geometric
series in the summation of the ladders in Fig.~\ref{fig: clean} for $\gamma_\pi^{\romI/\romII} L = -1$
at $u=\ut=1$ and~$\brperp = 0$. The emergence of this singularity means the breakdown
of the perturbative approach close to and beyond $|\gamma_\pi^{\romI/\romII}| L = 1$.

For a repulsive interaction, the coupling constant~$\gamma _{\pi }^{\mathrm{I}}$
always remains positive. It would be negative if the effective interaction~$\tilde{V}$
were negative, corresponding to attractive interaction. We can identify the resulting singularity in the thermodynamic potential
with the well-known Cooper instability. As a result, one obtains the conventional
$s$-wave superconductivity below a certain transition temperature $T_{c}$
obtained from the condition $\gamma_\pi^{\romI} L = -1$.

So, while $\gamma _{\pi }^{\mathrm{I}}$ is
strictly positive for a repulsive interaction, the triplet coupling constant~$\gamma _{\pi }^{\mathrm{II}%
}$ becomes negative as soon as $\tilde{V}(2p_{F})>\tilde{V}(0)$.
This may actually happen for a repulsive interaction. For instance, for contact
interaction, $\tilde{V}(2p_{F})=\tilde{V}(0)$ and already a small increase of
$\tilde{V}(2p_{F})$ --- for instance due to the closeness to a quantum critical point ---
may make the constant $\gamma _{\pi }^{\mathrm{II}}$ negative. If this
happens, one comes again to an instability in the thermodynamic potential~$\delta\Omega$, Eq.~(\ref{5b10}),
at $\gamma_\pi^{\romII} L = -1$. This instability should correspond to the triplet $p$-wave superconducting
pairing. One can come to this conclusion recalling once more the
superconducting particle-particle ladders in the spin triplet representation. It is easy to
see that for a small angle and backward scattering the combination $\tilde{V}%
(0)-\tilde{V}(2p_{F})$ enters the pre-factor in front of the logarithms,
which confirms our assumption.

The condition for criticality $\gamma_\pi^{\romII} L = -1$ determines the critical temperature~$T_{c}$,
\begin{equation}
T_{c}=\Lambda \ \exp \Big\{-\frac{\nu }{\pi\nu
^{\ast }\ |\gamma _{\pi }^{\mathrm{II}}|}\Big\}\komma  \label{5c04}
\end{equation}%
corresponding to the transition temperature into the triplet
superconducting state. The nature of this transition is similar
to the Kohn-Luttinger transition
towards $p$-wave pairing\cite{kohnluttinger} although in the Kohn-Luttinger
scenario, the inverse coupling
constant~$1/|\gamma _{\pi }^{\mathrm{II}}|$ enters the exponent of the
critical temperature~$T_c$, Eq.~(\ref{5c04}), with a different power.\cite{kohnluttinger,chubukovKL}

Everywhere in this paper, we considered the limit of weak interactions. A more
interesting situation may occur near a quantum critical point (QCP) of a
transition into a magnetic or charge density wave state. In the vicinity of
such a point, the collective mode propagator $\chi \left( \bq,\Omega\right) $
dressed by the electron-hole bubble can be written as
\begin{equation}
\chi \left( \bq,\Omega \right) \sim \frac{\chi _{0}}{\xi ^{-2}+\bq^{2}+\bar{%
\gamma}\left( | \Omega|/|\bq| \right) },  \label{5d01}
\end{equation}%
where $\chi _{0}$ is the susceptibility, $\xi ^{-2}$ determines the
closeness to QCP [$\xi ^{-2}=0$ at QCP], and $\bar{\gamma}$ is
proportional to the interaction. The propagator $\chi \left(\bq,\Omega
\right)$, Eq.~(\ref{5d01}), is written in the limit of small bosonic
Matsubara frequencies $\Omega$. Higher-order
corrections to the propagator $\chi \left(\bq,\Omega\right)$ can be considered but these are highly non-trivial [see, e.g. Refs.~\onlinecite{chubukov4,
dzero,chubukov5,rech,lee}], which invalidates the early conjecture\cite{hertz,millis}
that one can describe a QCP by a conventional $\phi^{4}$-field
theory with the bare propagator $\chi \left(\bq,\Omega\right)$, Eq.~(\ref{5d01}).

Using the conventional fermionic diagrammatic technique for studying the
critical behavior near a QCP is very difficult. At the same time, the bare
bosonic propagator $g_{\mathbf{n}}(K)$, Eq.~(\ref{2a39}), in our bosonization approach describes
directly the electron-hole excitations. Near a QCP, it could be modified and take
a form similar to the one of Eq.~(\ref{5d01}). Then, we would be able to derive a
superfield theory with a modified bare action. Within such a theory, diagrams to be disregarded for
the Fermi liquid like, e.g., diagram Fig.~\ref{fig: 1loopS4}(b) are not necessarily small and
require a special care. This adds to the complexity of the theory with very intriguing consequences.
We hope that our bosonization method can help in studying the QCP problem.

\section{Conclusion}

\label{sec:concl}

Singling out the low energy spin and charge excitations of a higher-dimensional
clean Fermi gas with a repulsive interaction,
we have modified our general bosonization scheme\cite{epm,epm2}
to derive an effective low energy superfield theory. This representation
allows to conveniently calculate thermodynamic quantities in the low temperature limit.
During the derivation, special care has been given to the role of Fermi surface
geometry in higher dimensions. As a result, all curvature effects are
preserved in the final action, correcting the earlier
supersymmetric approach from Ref.~\onlinecite{aleiner}.

The superfields in our low energy theory are anticommuting but periodic in imaginary time.
Consequently, the described excitations obey Bose statistics. These
bosonic excitations include both spin and charge excitations, interacting
in a non-trivial way as described by quartic, cubic, and quadratic terms in the action
of the superfield theory.

A perturbative study of the low energy superfield theory in the backscattering limit yields the well-known
leading non-analyticities in the thermodynamics and also the logarithmic corrections
in any dimension~$d$. In dimensions~$d>1$ however, the class of diagrams producing
logarithms is narrower than in one dimension as well as in the quasi-one-dimensional approximation
of Ref.~\onlinecite{aleiner}, where effects of the curvature of the Fermi surface were neglected. As a result, the renormalization
of the various coupling constants in the theory is for~$d>1$ significantly different from
the one-dimensional scenario. This is reflected by different renormalization group equations and
the observation that in higher dimensions, the renormalization of the quartic part of the action
can be understood also in the framework of ladder diagram summations.

The application of the low energy bosonization approach to the two-dimensional Fermi liquid and
the subsequent renormalization group analysis have yielded an explicit formula for the non-analytic contribution to the specific heat~$\delta
c$, Eq.~(\ref{5c02}). This result is of infinite order in the large logarithm~$\ln(\ve_F/T)$ and leading order
in the coupling constants of the weak interaction. As such, it is valid for an arbitrary low temperature~$T\ll\ve_F$.
The dependence of~$\delta c$ on the logarithm~$\ln(\ve_F/T)$ -- plotted in Fig.~\ref{fig: result} -- indicates
that the function $\delta c(T)/T^2$ decays as $1/\ln^2(\ve_F/T)$ for~$T\rightarrow 0$.
As discussed in Sec.~\ref{ssec: fl_corrections}, our result is in full agreement with asymptotic results of earlier works
based on conventional diagrammatic expansions.

Remarkably, the thermodynamic potential and the specific heat correction, Eqs.~(\ref{5b10}) and~(\ref{5c02}),
consist of two separate terms of an identical analytical form controlled
by two different coupling constants~$\gamma_\pi^\romI$ and~$\gamma_\pi^\romII$. The contribution of the term containing only the coupling constant~$\gamma_\pi^\romII$
comes with a
factor of three as compared to the term with the coupling~$\gamma_\pi^\romI$. In the discussion in Sec.~\ref{ssec: fl_corrections}, we interpret
these two contributions as coming from superconducting fluctuations of spin singlet and spin triplet types. This statement
is supported by comparison of the conventional Cooper ladders with the ladders of the quartic bosonic action and by the
analytical form of the effective coupling constants~$\gamma_\pi^{\romI/\romII}$, Eq.~(\ref{5b12}). For a contact interaction,
$\gamma_\pi^\romII = 0$ and only the singlet superconducting fluctuations contribute.
We note that within our approximation of small fluctuations at backward scattering, we cannot
distinguish between angular harmonics of the same parity, e.g. between $s$- and $d$-pairing, merely distinguishing
between the singlet and triplet excitations.

If one of the coupling constants --- $\gamma_\pi^\romI$ or~$\gamma_\pi^\romII$ --- becomes negative,
the Fermi liquid picture breaks down at a critical temperature~$T_c$ and a superconducting phase transition takes place.
The constant $\gamma_\pi^\romI$ can be negative only for an attractive interaction, leading to the conventional
spin singlet superconducting transition. In contrast, $\gamma_\pi^\romII$ can become negative also for
certain models of repulsive interaction. This scenario of triplet superconductivity, which is similar to the Kohn-Luttinger one, is discussed in Sec.~\ref{ssec: instabilities}.
Since our perturbative approach breaks down close above the critical temperature~$T_c$, Eq.~(\ref{5c04}),
the critical behavior itself should be studied introducing the superconducting order parameter.

A very interesting situation may arise near a quantum phase transition into,
e.g., a ferromagnetic state. Near this point, both superconducting and
paramagnetic excitations are important.\cite{chubukov4,dzero,chubukov5} In the
language of the superfield theory developed here, the non-interacting
paramagnetic excitations should be described by the bare action, Eq.~(\ref{2a37b}),
while the superconducting fluctuations appear as a result of the
interaction between these excitations. Due to the special form of these
excitations, cf. Eq.~(\ref{5d01}), the perturbation
theory is more complicated than the one considered here for the Fermi liquid and
logarithmic contributions may arise in more classes of one-loop
diagrams. Since calculations in this interesting situation
are not simple within the conventional diagrammatic
approaches\cite{chubukov4,dzero,chubukov5,rech,lee}, we hope that the
present bosonization technique will become a helpful analytical tool for future studies
on this topic.

\section*{Acknowledgements}

We are grateful to A.V. Chubukov and D.L. Maslov for invaluable discussions.
H.M. and K.E. acknowledge financial support from the SFB/Transregio~12 of the Deutsche Forschungsgemeinschaft. H.M, C.P., and K.E. acknowledge the hospitality of the International Institute of Physics in Natal where parts of this work were done.


\appendix

\section{Evaluation of the integral in Eq.~(\ref{3b01})}
\label{App_1}


In this appendix, we explicitly evaluate the second order contribution~$\delta\Omega^{(2)}(T)$, Eq.~(\ref{3b01}),
\begin{align}
\label{3x01}
 \delta\Omega^{(2)}(T) &= \left(T\sum_\om-\int\frac{\dt\omega}{2\pi}\right)
             \int\frac{\dt^2\bq}{(2\pi)^2}\dt\bn\dt\bnt\ f^2(\bq)\nonumber\\
             \times&
             \gamma^2_{\widehat{\bn\bnt}}
             \frac{v_F(\bn\!\cdot\!\bq)\ v_F(\bnt\!\cdot\!\bq)}
                  {[\ii\om-v_F(\bn\!\cdot\!\bq)][-\ii\om + v_F(\bnt\!\cdot\!\bq)]}
 \  \punkt
\end{align}
The explicit form of the cutoff function~$f(\bq)$ is not important for the
second order loop. It is sufficient to know that the estimate $|\bq|\lesssim q_0\ll p_F$ holds. Eventually,
the calculation shows that effectively only those~$\bq$ enter $f(\bq)$ which satisfy $|\bq|\ll q_0$. Consequently,
we can effectively put $\bq=0$ in $f(\bq)$ and assume $f(0)=1$.

First, we perform the sum and integral over the frequency~$\om$. The $\coth$-functions resulting
from the finite temperature Matsubara summation are conveniently combined with the sign functions from the zero-temperature
frequency integration using the relation $\coth x = \sgn x [1+2\sum_{l=1}^\infty \exp(-2n|x|)]$. As a result, we obtain
\begin{align}
\label{3x02}
 &\delta\Omega^{(2)}(T) = v_F
  \int\dt\bn\dt\bnt\frac{\dt^2\bq}{(2\pi)^2}\ \gamma^2_{\widehat{\bn\bnt}}\
  \frac{(\bn\!\cdot\!\bq)(\bnt\!\cdot\!\bq)}{(\bn\!\cdot\!\bq)-(\bnt\!\cdot\!\bq)}
  \nonumber\\
  \times & \sum_{l=1}^\infty \Big\{
      \sgn(\bn\!\cdot\!\bq)\exp\Big(-\frac{l v_F|(\bn\!\cdot\!\bq)|}{T}\Big)
  \nonumber\\ &\qquad\qquad
      -\sgn(\bnt\!\cdot\!\bq)\exp\Big(-\frac{l v_F|(\bnt\!\cdot\!\bq)|}{T}\Big)
  \Big\}\punkt
\end{align}
From Eq.~(\ref{3x02}), we understand that relevant ``parallel'' momenta~$(\bn\!\cdot\!\bq)$ or~$(\bnt\!\cdot\!\bq)$
are of order $T/v_F$. It it this observation which makes our entire low energy approach useful.

In order to carry out the remaining integrations, we should transform the variables into a frame that would better reflect
the physics of the scattering processes under consideration.
For this purpose, we introduce new angular variables
\begin{align}
\label{3x03}
\bnb &= \frac{1}{2}(\bn-\bnt)\komma\qquad \delta\bn = \frac{1}{2}(\bn+\bnt)
\end{align}
and the corresponding projections of the momentum vector~$\bq$,
\begin{align}
\label{3x04}
\bar{q}_{\parallel} &= (\bnb\!\cdot\!\bq)\komma\qquad
\bar{\mathbf{q}}_{\perp} = \bq - \bar{q}_{\parallel}\bnb\punkt
\end{align}
The choice of the angular coordinates and the notation in Eq.~(\ref{3x03}) are motivated by the observation
that the most important contributions to $\delta\Omega^{(2)}(T)$, Eq.~(\ref{3x01}), will come from small~$\delta\bn$.
The phase space region around~$\delta\bn=0$ corresponds to backscattering, $\widehat{\bn\bnt}\sim\pi$, and is also exactly the region where the logarithmic renormalizations of the scattering amplitude~$\gamma_{\widehat{\bn\bnt}}$ take place.

Reexpressing Eq.~(\ref{3x02}) with the help of the variables from Eqs.~(\ref{3x03}) and~(\ref{3x04}), we obtain
\begin{align}
\label{3x05}
 &\;\delta\Omega^{(2)}(T) =
  v_F\int\dt\bn\dt\bnt\frac{\dt\bar{\mathbf{q}}_{\perp}}{2\pi}\frac{\dt\bar{q}_{\parallel}}{2\pi}\ \gamma^2_{\widehat{\bn\bnt}}\ \sum_{l=1}^\infty
  \nonumber\\ &\times
  \left\{\frac{|\bar{q}_{\parallel}|^3}
       {\bar{q}_{\parallel}^2-(\delta\bn\!\cdot\!\bar{\mathbf{q}}_{\perp})^2}
       -2|\bar{q}_{\parallel}|
  \right\}
  \exp\Big(-\frac{l v_F|\bar{q}_{\parallel}|}{T}\Big)\punkt
  \end{align}
The second term~$-2|\bar{q}_{\parallel}|$ in the curly brackets produces an insensitive to small~$|\delta\bn|$ and thus purely analytic contribution to the thermodynamic potential. Focussing
on the nonanalytic contributions, we neglect such terms. One can observe at this point that the nonanalytic contributions arise completely
from the term~$(\delta\bn\!\cdot\!\bar{\mathbf{q}}_{\perp})^2$ in the numerator~$(\bn\!\cdot\!\bq)(\bnt\!\cdot\!\bq)=
(\delta\bn\!\cdot\!\bar{\mathbf{q}}_{\perp})^2 - \bar{q}_{\parallel}^2$ of Eq.~(\ref{3x01}). In other words, the momentum components~$\bar{q}_{\parallel}$ of the terms
with $(\bn\!\cdot\!\bq)\theta$ and $(\bnt\!\cdot\!\bq)\tilde{\theta}$ in the actions~$\cS_2$ and~$\cS_3$
are \emph{effectively irrelevant} for the analysis of the nonanalyticities.

We note that the integrand in Eq.~(\ref{3x05}) does not depend on the angular variable~$\bnb$ but only on $\delta\bn$. Therefore, we should transform the integration variables~$\{\bn,\bnt\}$ to~$\{\bnb,\delta\bn\}$ and while we do so, we should already have in mind that the integral will be dominated by small~$\delta\bn$. Let us thus parameterize $\bn=(\cos\phi,\sin\phi)$ and $\bnt=(\cos\tilde{\phi},\sin\tilde{\phi})$ with both $\phi$ and $\tilde{\phi}$ varying between~$0$ and~$2\pi$. Then the normalized integration measure is given by $\dt\bn\dt\bnt = (\dt\phi/2\pi)(\dt\tilde{\phi}/2\pi)$. For a suitable parametrization of the variables~$\bnb$ and~$\delta\bn$, Eq.~(\ref{3x03}), we introduce the angles~$\bar{\phi}$ and~$\delta\phi$ in such a way that $\phi = \bar{\phi}-(\delta\phi+\pi/2)$ and $\tilde{\phi}=\bar{\phi}+(\delta\phi+\pi/2)$ with $\delta\phi$ being small in the backscattering limit. Then, to linear order in $\delta\phi$, we find $\bnb\simeq(\sin\bar{\phi},-\cos\bar{\phi})$ and $\delta\bn \simeq -\delta\phi (\cos\bar{\phi},\sin\bar{\phi})$, which also gives $|\delta\bn| \simeq |\delta\phi|$ in the integrand of Eq.~(\ref{3x05}). The integration measure transforms as $\dt\bn\dt\bnt = (\dt\bar{\phi}/2\pi)\dt\delta\phi/\pi$. The integration over $\bar{\phi}$ is immediately performed over $(0,2\pi)$ and consequently yields unity. After that, the nonanalytic correction~$\delta\Omega^{(2)}(T)$ reads
\begin{align}
\label{3x05a}
  \delta\Omega^{(2)}(T) &=
  \frac{v_F\gamma^2_{\pi}}{\pi}\int\dt\delta\bn\frac{\dt\bar{\mathbf{q}}_{\perp}}{2\pi}\frac{\dt\bar{q}_{\parallel}}{2\pi}\ \nonumber\\
  \qquad &\times \sum_{l=1}^\infty \frac{|\bar{q}_{\parallel}|^3}
       {\bar{q}_{\parallel}^2-(\delta\bn\!\cdot\!\bar{\mathbf{q}}_{\perp})^2}
  \exp\Big(-\frac{l v_F|\bar{q}_{\parallel}|}{T}\Big)\punkt
  \end{align}
Here and in the following, $\delta\bn$ is identified with the one-dimensional variable~$\delta\phi$ and typically takes  small values $|\delta\bn|\ll 1$.

By shifting the integration contour of the parallel momentum~$\bar{q}_{\parallel}$ as $\bar{q}_{\parallel}\mapsto \ii\kappa_\parallel$, Eq.~(\ref{3x05a}) is reduced
to Eq.~(7.17b) of Ref.~\onlinecite{aleiner}. Explicitly, we recast Eq.~(\ref{3x05a}) into the form
\begin{align}
\label{3x06}
&\;\delta\Omega^{(2)}(T) =
  -\frac{2v_F\gamma^2_{\pi}}{\pi}
              \int\dt\delta\bn\frac{\dt\bar{\mathbf{q}}_{\perp}}{2\pi}
              \ \ \sum_{l=1}^\infty \nonumber\\
              \times\ & \mathrm{Re}\left[\int_0^\infty\frac{\dt\kappa_\parallel}{2\pi}\frac{\kappa_\parallel^3}{\kappa_\parallel^2+(\delta\bn\!\cdot\!\bar{\mathbf{q}}_{\perp})^2}\exp\Big(-\frac{\ii l v_F\kappa_\parallel}{T}\Big)
  \right]\punkt
\end{align}
For the integration over~$\delta\bn$,
we indeed notice that the most important contributions come from $|(\delta\bn\!\cdot\!\bar{\mathbf{q}}_{\perp})|/\kappa_\parallel\lesssim 1$ and, since~$\kappa_\parallel\sim T/v_F$ and~$\bar{\mathbf{q}}_{\perp}\sim q_0$, therefore from the backscattering region of small~$\delta\bn$ where
the estimate~$|\delta\bn| \lesssim T/(v_F q_0)\ll 1$ is valid. This also implies that logarithmic renormalizations of the backscattering amplitude~$\gamma_\pi$ become indeed active. Before we come to the~$\delta\bn$-integral, we integrate over the ``imaginary'' momentum~$\kappa_\parallel$ and in order to facilitate that, we recast the rational integrand as a Fourier integral,
\begin{align}
\label{3x07}
  &\quad\frac{1}{\kappa_\parallel^2 + (\delta\bn\!\cdot\!\bar{\mathbf{q}}_{\perp})^2}\nonumber\\
 &= \frac{1}{2q_0|\delta\bn|\kappa_\parallel}\int\ee^{-\kappa_\parallel|\brperp|/(q_0|\delta\bn|)}
 \ \ee^{-\ii\brperp\bar{\mathbf{q}}_{\perp}/q_0}\dt\brperp\punkt
\end{align}
Because of the cutoff~$q_0$ for the momentum~$\bar{\mathbf{q}}_{\perp}$, typical~$|\brperp|$ are of order~$1$. Inserting
Eq.~(\ref{3x07}) into the expression for~$\delta\Omega^{(2)}(T)$, Eq.~(\ref{3x06}), we see that the pre-exponential is just a power of~$\kappa_\parallel$ and the corresponding integration is easily performed,
\begin{align}
\label{3b08}
 \delta\Omega^{(2)}(T) &= -\frac{\gamma_\pi^2T^3}{\pi^2 v_F^2}\int
   \ee^{-\ii\brperp\bar{\mathbf{q}}_{\perp}/q_0}
   \left(
    \frac{v_F q_0 |\delta\bn|}{T|\brperp|}
   \right)^2
   \nonumber\\
   \times\
   \sum_{l=1}^\infty\ \mathrm{Re}&\left[\left\{
    1+\ii l \left(
    \frac{v_F q_0 |\delta\bn|}{T|\brperp|}
   \right)
   \right\}^{-3}\right]
    \frac{v_F q_0\dt\delta\bn}{T|\brperp|}
   \dt\brperp\frac{\dt\bar{\mathbf{q}}_{\perp}}{2\pi q_0}\punkt
\end{align}
According to the estimates discussed in the preceding text,
essential values of the quantity~$\Phi = v_F q_0|\delta\bn|/(T|\brperp|)$ are of order~$1$. Transforming
from the integration variable $\delta\bn$ to~$\Phi$, the integration limits will be of
order $v_F q_0/T \gg 1$, allowing to extend the domain of integration to~$\pm\infty$. Using the integral
\begin{align*}
\int_0^\infty \mathrm{Re}\left[\frac{\Phi^2}{(1+\ii\Phi)^3}
 \right]\ \dt\Phi = -\frac{\pi}{2}\komma
\end{align*}
we obtain
\begin{align}
\label{3b09}
 \delta\Omega^{(2)}(T) &= \frac{ \gamma_\pi^2 T^3}{\pi v_F^2}
 \sum_{l=1}^\infty \frac{1}{l^3}\int
   \ee^{-\ii\brperp\bar{\mathbf{q}}_{\perp}/q_0}
   \dt\brperp\frac{\dt\bar{\mathbf{q}}_{\perp}}{2\pi q_0}\punkt
\end{align}
The sum over~$l$ just gives Ap\'ery's constant~$\zeta(3)$, the remaining integrals are trivial, and
we obtain for~$\delta\Omega^{(2)}(T)$ the result presented in Eq.~(\ref{3b11}).

\section{Third order correction from bosonic diagrams}
\label{App_2}

\begin{figure}[t]
\includegraphics{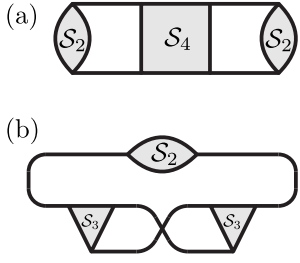}
\caption{Diagrams for the third order backscattering contribution to the
thermodynamic potential.}
\label{fig: 3rd}
\end{figure}

In this appendix, we calculate in the leading logarithmic order the
anomalous specific heat in third order in the interaction by the explicit
evaluation of the bosonic diagrams.

As a result of the analysis in Sec.~\ref{sec:renormalization},
the relevant diagrams are those shown in Fig.~\ref{fig: 3rd}.
Other diagrams are either insensitive to
the backscattering region or their logarithmic divergency is prohibited by
the effects of the curvature of the Fermi surface. Since for the diagrams in
Figs.~\ref{fig: 3rd}(a) and~(b) the curvature terms are not relevant, we will
omit them in the following formulas.

In analytical terms, diagram Fig.~\ref{fig: 3rd}(a) yields the contribution
\begin{align}
& \Delta \Omega ^{(3\mathrm{a})}(T)=
 -\frac{8}{9\nu }\int \mathrm{d}\mathbf{n}\mathrm{d}{\tilde{\mathbf{n}}}\sum_{QQ^{\prime }}
 \nonumber  \\
 & \quad\times
 \left( [\gamma _{\widehat{\mathbf{n}{\tilde{\mathbf{n}}}}}^{\mathrm{I}}]^{3}
      +3[\gamma _{\widehat{\mathbf{n}{\tilde{\mathbf{n}}}}}^{\mathrm{II}}]^{3}\right)
 \
 f(\bq)f(\bq')f(\bq+\bq')
 \nonumber  \\
& \quad \times \left\{ v_{F}^{2}(\mathbf{n}\!\cdot \!\mathbf{q})({\tilde{%
\mathbf{n}}}\!\cdot \!\mathbf{q})-v_{F}^{2}(\mathbf{n}\!\cdot \!\mathbf{q}%
^{\prime })({\tilde{\mathbf{n}}}\!\cdot \!\mathbf{q})\right\}  \notag \\
& \quad \times g_{\mathbf{n}}(Q)g_{{\tilde{\mathbf{n}}}}(-Q)g_{\mathbf{n}}(Q^{\prime }%
)g_{{\tilde{\mathbf{n}}}}(-Q^{\prime }) \label{4a01}
\end{align}%
and the second diagram Fig.~\ref{fig: 3rd}(b) corresponds to
\begin{align}
& \quad \Delta \Omega ^{(3\mathrm{b})}(T)=
 -\frac{8}{9\nu }\int \mathrm{d}\mathbf{n}\mathrm{d}{\tilde{\mathbf{n}}}\sum_{QQ^{\prime }}
 \nonumber\\
 & \times
 \left( [\gamma _{\widehat{\mathbf{n}{\tilde{\mathbf{n}}}}}^{\mathrm{I}}]^{3}+
       3[\gamma _{\widehat{\mathbf{n}{\tilde{\mathbf{n}}}}}^{\mathrm{II}}]^{3}\right)
 \
 f(\bq)f(\bq')f(\bq+\bq')
 \nonumber\\
& \times \left\{ -v_{F}^{2}(\mathbf{n}\!\cdot \!\mathbf{q})({\tilde{\mathbf{n%
}}}\!\cdot \!\mathbf{q})-v_{F}^{2}(\mathbf{n}\!\cdot \!\mathbf{q})({\tilde{%
\mathbf{n}}}\!\cdot \!\mathbf{q}^{\prime })+2v_{F}^{2}(\mathbf{n}\!\cdot \!%
\mathbf{q}^{\prime })({\tilde{\mathbf{n}}}\!\cdot \!\mathbf{q}^{\prime
})\right\}  \notag \\
& \times g_{{\tilde{\mathbf{n}}}}(Q+Q^{\prime })g_{\mathbf{n}}(-Q)
  g_{\mathbf{n}}(Q^{\prime })
  g_{{\tilde{\mathbf{n}}}}(-Q^{\prime })\;%
\mathnormal{.} \label{4a02}
\end{align}%
The coupling constants $\gamma _{\widehat{\mathbf{n}{\tilde{\mathbf{n}}}}}^{%
\mathrm{I}}$ and $\gamma _{\widehat{\mathbf{n}{\tilde{\mathbf{n}}}}}^{%
\mathrm{II}}$ have been introduced in Eq.~(\ref{5a_coupling}).

The third order diagrams for~$\Delta \Omega ^{(3\mathrm{a})}(T)$ and~$\Delta \Omega ^{(3\mathrm{b})}(T)$ contain
two loops with correspondingly two running four-momenta~$Q$ and~$Q'$. Following the idea of Eq.~(\ref{3b000}),
we should subtract the contribution at~$T=0$ and deal with the quantity~$\delta\Omega^{(3)}(T)=\Delta\Omega^{(3)}(T)-\Delta\Omega^{(3)}(0)$ rather than
with $\Delta\Omega^{(3)}(T)$ itself.
Therefore, one four-momentum effectively varies on the scale~$\lesssim T$ while the other one
conversely needs to vary on large scales~$\gg T$ in order to produce the leading logarithmic correction.

Let us begin the explicit evaluation with the expression for diagram~Fig.~\ref{fig: 3rd}(a), Eq.~(\ref{4a01}).
The first term in the curly brackets behaves in a considerably different way for the two cases of small or large~$Q$ --- corresponding
to large or small~$Q'$, respectively. In the case of small~$Q$, the integral over~$Q$ is
calculated analogously to the second order integral Eq.~(\ref{3b01}) while the integral over~$Q'$ is essentially
the logarithmic one-loop integral from Eq.~(\ref{3c02}). As a result, we obtain a correction of relative order $\gamma^{\romI/\romII}_\pi\ln(\Lambda/T)$
to the second order result, Eq.~(\ref{3b11}).

The opposite case of large~$Q$ constitutes an unpleasant divergency at large~$v_F \bqbpar$, which
is only formally cut by the cutoff functions~$f(\bq)$. Fortunately, this ultraviolet divergency is
exactly compensated by an ultraviolet divergency appearing with the opposite
sign in the first term in the curly brackets in the expression for
diagram~Fig.~\ref{fig: 3rd}(b), Eq.~(\ref{4a02}), such that the result is eventually regular. We note once more in this context that
the parallel momentum~$\bqbpar$ of the~$(\bn\cdot\bq)$-terms in the actions~$\cS_3$ and~$\cS_2$ is irrelevant for
thermodynamic quantities, cf. the discussion after Eq.~(\ref{3x05}). Furthermore, since the seeming ultraviolet divergency in~$v_F \bqbpar$
is compensated ---  an observation that is easily generalized to diagrams of arbitrary order ---, it is completely safe to neglect the
$v_F \bqbpar$ part of the~$(\bn\cdot\bq)$-terms in the cubic and quadratic parts of the interaction as done in Eqs.~(\ref{5a01})--(\ref{5a03}).

The second term in the curly brackets of Eq.~(\ref{4a01}) is odd in both~$Q$ and~$Q'$ and for this reason,
one might be tempted to disregard that term. However, because of the presence of the cutoff functions, the
overall integrand is not odd in the perpendicular momenta~$\bar{\bq}_\perp$ and~$\bar{\bq}_\perp'$, cf. Eq.~(\ref{3c04}) for the notation. Explicitly, since
\begin{align}
\int\bar{\bq}_\perp'\ f(\bar{\bq}_\perp')f(\bar{\bq}_\perp+\bar{\bq}_\perp')\ &\frac{\dt\bar{\bq}_\perp'}{2q_0}
 \nonumber\\
=-\frac{\bar{\bq}_\perp}{2}\int f(\bar{\bq}_\perp')&f(\bar{\bq}_\perp+\bar{\bq}_\perp')\ \frac{\dt\bar{\bq}_\perp'}{2q_0}\komma
\label{4a02a}
\end{align}
we observe that the second term of Eq.~(\ref{4a01}) gives the same contribution as the regular part of the first term --- with one half coming
from small~$Q$ and one half coming from small~$Q'$.

Now, let us turn our attention to the expression for~$\Delta \Omega
^{(3\mathrm{b})}$, Eq.~(\ref{4a02}). The presence of~$Q'$ in three denominators
implies that $Q$ is necessarily the large four-momentum. The first term in the curly brackets of
Eq.~(\ref{4a02}) consequently does nothing more than neutralize the ultraviolet divergency in~$\Delta \Omega
^{(3\mathrm{a})}$ as discussed above. The second term is treated in complete analogy with the second term in Eq.~(\ref{4a01})
while, finally, the third term is effectively
of the same form as the first term of the diagram Fig.~\ref{fig: 3rd}(a) in
the limit of small~$Q$.

As to the cutoff functions~$f(\bq)$, they have played an important role in understanding the seemingly odd terms in Eqs.~(\ref{4a01}) and~(\ref{4a02}).
After applying Eq.~(\ref{4a02a}), all relevant terms have the form of the first term in Eq.~(\ref{4a01}) at small~$Q$.
The remaining integration of~$Q$ is completely equivalent to the second order integral presented in Appendix~\ref{App_1}.
There, we learned that the transverse momentum
of the small four-momentum could be safely put to zero in the cutoff functions. Thus, the integral of the ``large'' transverse momentum
over the cutoff functions yields a prefactor of $f(0)\int f^2(\bar{\bq}_\perp) [\dt\bar{\bq}_\perp/2q_0]$. For the choice
$f(\bar{\bq}_\perp)=\Theta(q_0-|\bar{\bq}_\perp|)$, this prefactor is just unity.

Collecting all the terms, we find that diagram Fig.~\ref{fig: 3rd}(a)
gives $4/9$ and diagram~(b) $5/9$ of the third order
correction $\delta \Omega ^{(3)}(T)$ to the thermodynamic potential, which can be written as
\begin{equation}
\delta \Omega ^{(3)}(T)=-\frac{4\nu ^{\ast }}{\nu }\frac{\zeta (3)}{\pi
v_{F}^{2}}\left( [\gamma _{\pi }^{\mathrm{I}}]^{3}+3[\gamma _{\pi }^{\mathrm{%
II}}]^{3}\right) T^{3}\ln \left( \frac{\Lambda }{T}\right) \punkt
\nonumber
\end{equation}
This result clearly agrees with the one obtained from the low order expansion of the renormalized coupling constants, Eq.~(\ref{5b04}).

\section{Boson model versus fermion picture}
\label{App_3}

\begin{table}[t]
\caption{\label{tbl: bosonvsfermion}
Vertices of the bosonic excitations and related diagrammatical structures
of the Hartree and Fock soft modes in conventional fermion language.}
\includegraphics[width=0.9\linewidth]{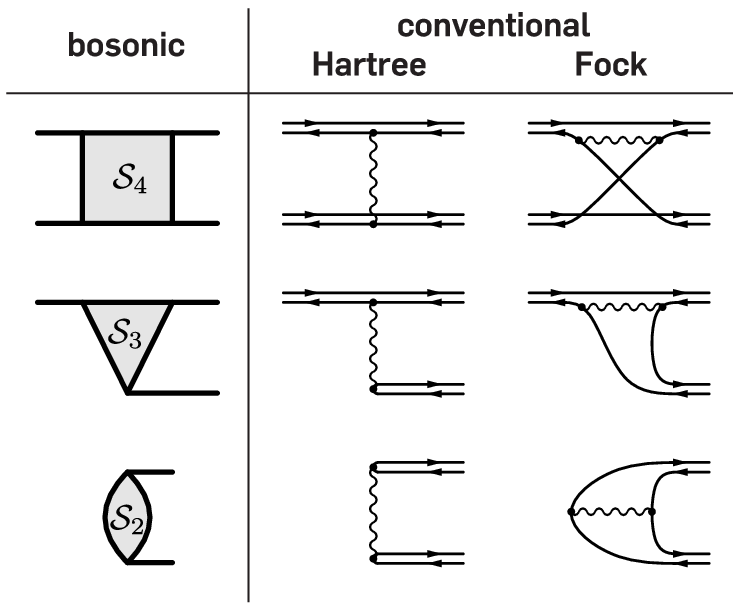}
\end{table}
In Sec.~III of Ref.~\onlinecite{epm2}, it was checked in the second order
in the interaction that our method of bosonization allows for, in principle, an exact reformulation
of the original fermion model in terms of bosonic excitations and reproduces exactly each single contribution from
the fermionic diagrammatics. The choice of a proper diagrammatical representation allowed
us to identify the bosonic contributions with the ones of the fermionic picture already on the level
of diagrams --- before explicitly evaluating the analytical expressions.

Following the decoupling into soft modes in Sec.~\ref{sec:model}, the vertices in the bosonic theory collect
at the same time the fermionic Hartree vertices with a momentum transfer close to zero and Fock vertices
transferring momenta of order~$2p_F$. As discussed in Ref.~\onlinecite{epm2}, where
an exact Hartree-like decoupling scheme has been applied, the bosonic propagator corresponds to the
propagation of a particle-hole pair in the fermion picture, which will be reflected diagrammatically
by opposite oriented double-lines. Following the derivation of the exact supersymmetric representation
in Ref.~\onlinecite{epm2} for eachwise the Hartree and the Fock decoupling schemes, we obtain
the diagrammatical representation in Table~\ref{tbl: bosonvsfermion}. Table~\ref{tbl: bosonvsfermion}
can be understood as a dictionary translating diagrams in the boson picture into corresponding standard fermionic diagrams.

\begin{figure}[t]
\includegraphics[width=0.8\linewidth]{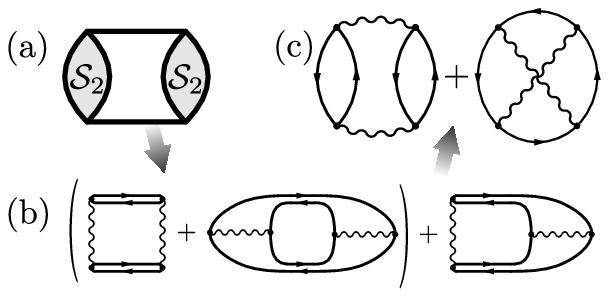}
\caption{\label{fig: S2_conv}
Correspondence between the bosonic low energy representation and conventional
diagrammatics: (a) the bare backscattering diagram from Fig.~\ref{fig: back}(a),
(b) its representation in Hartree and Fock soft modes according to Table~\ref{tbl: bosonvsfermion},
and finally (c) the conventional fermionic diagrams.}
\end{figure}
As an example, let us consider the diagram for the bare anomalous contribution, Fig.~\ref{fig: S2_conv}(a). Its evaluation
in Sec.~\ref{ssec:backscattering} returns Eq.~(\ref{3b11}) for the correction to~$\Omega$, which yields
the leading anomalous $T^2$-term in the specific heat~$c$.

Redrawing the bosonic quadratic vertices with the help of Table~\ref{tbl: bosonvsfermion} in all possible ways that the soft Hartree
and Fock vertices may enter, we obtain the diagrams shown in Fig.~\ref{fig: S2_conv}(b). Finally, ``literally'' interpreting the bosonic propagators as pairs of opposite
directed fermion ones, we identify the corresponding conventional diagrams, Fig.~\ref{fig: S2_conv}(c),
which share the same low energy physical content with Fig.~\ref{fig: S2_conv}(a). Indeed, standard fermion
perturbation theory\cite{chubukov1} yields exactly Eq.~(\ref{3b11}) as the anomalous contribution to~$\Omega$,
which exclusively comes from the conventional second order diagrams in Fig.~\ref{fig: S2_conv}(c).

\subsection*{One-loop diagrams in the fermion picture}

\begin{figure}[t]
\includegraphics[scale=1.2]{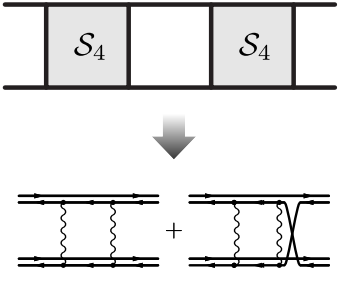}
\caption{\label{fig: 1loopS4_conv_1}
The bosonic one-loop diagram~Fig.~\ref{fig: 1loopS4}(a) and its relation to
conventional diagrams. The latter form particle-particle ladders, giving logarithms
independently from the dimension~$d$ of the system.}
\end{figure}

\begin{figure}[t]
\includegraphics[scale=1.2]{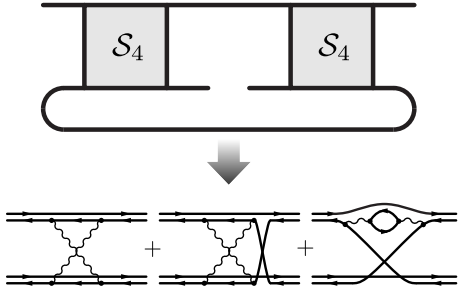}
\caption{\label{fig: 1loopS4_conv}
The bosonic one-loop diagram~Fig.~\ref{fig: 1loopS4}(b) and its related
conventional diagrams. The latter are of particle-hole ladder and polarization bubble type, reflecting
the absence of a logarithmic divergency in $d>1$ dimensions.}
\end{figure}

Figures~\ref{fig: 1loopS4_conv_1} and~\ref{fig: 1loopS4_conv} constitute the correspondence between the one-loop quartic vertex
corrections~$\delta\cS_4^{\mathrm{(a)}}$ and~$\delta\cS_4^{\mathrm{(b)}}$, Fig.~\ref{fig: 1loopS4}(a) and~(b), and their contribution
in the fermionic picture according to Table~\ref{tbl: bosonvsfermion}.

Two out of the four fermion lines are seemingly free while the remaining two lines
 interact with an effective renormalized interaction. Figure~\ref{fig: 1loopS4_conv_1} shows that~$\delta\cS_4^{\mathrm{(a)}}$ corresponds
to a particle-particle ladder in conventional diagrams. Particle-particle ladders are known to give rise to logarithmic divergencies independently from the dimension~$d$,
which is in agreement with the analytical result for~$\delta\cS_4^{\mathrm{(a)}}$, Eqs.~(\ref{3c01}), (\ref{3c08b}), and~(\ref{3c07}).
As to Fig.~\ref{fig: 1loopS4_conv}, depending on whether at least one of the two $\cS_4$-blocks is recasted into a Hartree vertex or not, this renormalization
appears in form of either a particle-hole ladder or of a polarization bubble. Both of them share the feature of being logarithmically divergent in $d=1$ dimension
but convergent in $d>1$. Thus once more, the graphical correspondence reflects the analytical result we have found in the discussion of the bosonic diagram, namely
the suppression of the logarithmic divergency due to curvature effects in higher dimensions, cf.~Eq.~(\ref{3c11}).


\begin{thebibliography}{99}
\bibitem{footnote00} Instead of Eq.~(\ref{2a36c}), one could also define BRST symmetry
by the variation~$\delta\Psi = \eta\ \partial_\theta \Psi$. This transformation
leaves the first term in~$\cS_S$ and the action~$\cS_B$ invariant while symmetry
would be broken by the second term in~$\cS_S$.
\bibitem{footnote01} Notwithstanding the overall agreement of Eqs.~(\ref{5c02}) and~(\ref{5c03}) in the limits considered,
the separation into charge and spin components in Eq.~(\ref{5c03}) is to be distinguished from the separation
into the spin singlet [$\gamma_\pi^\romI$] and triplet [$\gamma_\pi^\romII$] components in our result Eq.~(\ref{5c02}).
\end{thebibliography}
\end{document}